\documentclass[a4paper]{spie}  %>>> use this instead for A4 paper
 \def\ksec{kilo-seconds}
 \def\micron{$\mu $m}
\usepackage{amsmath,amsfonts,amssymb}
\usepackage{graphicx}
\usepackage[colorlinks=true, allcolors=blue]{hyperref}

\title{The Athena X-ray Integral Field Unit (X-IFU)}
\author[a]{Didier Barret}
\author[b]{Thien Lam Trong}
\author[c]{Jan-Willem den Herder}
\author[d]{Luigi Piro}
\author[e]{Xavier Barcons}
\author[f]{Juhani Huovelin}
\author[g]{Richard Kelley}
\author[h]{J. Miguel Mas-Hesse}
\author[i]{Kazuhisa Mitsuda}
\author[j]{St\'ephane Paltani}
\author[k]{Gregor Rauw}
\author[l]{Agata Rozanska}
\author[m]{Joern Wilms}
\author[n]{Marco Barbera}
\author[j]{Enrico Bozzo}
\author[e]{Maria Teresa Ceballos}
\author[o]{Ivan Charles}
\author[p]{Anne Decourchelle}
\author[c]{Roland den Hartog}
\author[o]{Jean-Marc Duval}
\author[q]{Fabrizio Fiore}
\author[r]{Flavio Gatti}
\author[s]{Andrea Goldwurm}
\author[c]{Brian Jackson}
\author[c]{Peter Jonker}
\author[g]{Caroline Kilbourne}
\author[d]{Claudio Macculi}
\author[t]{Mariano Mendez}
\author[u]{Silvano Molendi}
\author[v]{Piotr Orleanski}
\author[a]{Fran\c cois Pajot}
\author[a]{Etienne Pointecouteau}
\author[g]{Frederick Porter}
\author[p]{Gabriel W. Pratt}
\author[w]{Damien Pr\^ele}
\author[a]{Laurent Ravera}
\author[x]{Etienne Renotte}
\author[y]{Joop Schaye}
\author[z]{Keisuke Shinozaki}
\author[A]{Luca Valenziano}
\author[B]{Jacco Vink}
\author[a]{Natalie Webb}
\author[i]{Noriko Yamasaki}
\author[b]{Fran\c coise Delcelier-Douchin}
\author[b]{Michel Le Du}
\author[b]{Jean-Michel Mesnager}
\author[b]{Alice Pradines}
\author[C]{Graziella Branduardi-Raymont}
\author[A]{Mauro Dadina}
\author[f]{Alexis Finoguenov}
\author[D]{Yasushi Fukazawa}
\author[E]{Agnieszka Janiuk}
\author[F]{Jon Miller}
\author[k]{Ya\"el Naz\'e}
\author[q]{Fabrizio Nicastro}
\author[G]{Salvatore Sciortino}
\author[H]{Jose Miguel Torrejon}
\author[b]{Herv\'e Geoffray}
\author[b]{Isabelle Hernandez}
\author[b]{Laure Luno}
\author[a]{Philippe Peille}
\author[b]{J\'er\^ome Andr\'e}
\author[b]{Christophe Daniel}
\author[b]{Christophe Etcheverry}
\author[b]{Emilie Gloaguen}
\author[b]{J\'er\'emie Hassin}
\author[b]{Gilles Hervet}
\author[b]{Irwin Maussang}
\author[b]{J\'er\^ome Moueza}
\author[b]{Alexis Paillet}
\author[b]{Bruno Vella}
\author[b]{Gonzalo Campos Garrido}
\author[b]{Jean-Charles Damery}
\author[b]{Chantal Panem}
\author[b]{Johan Panh}
\author[g]{Simon Bandler}
\author[b]{Jean-Marc Biffi}
\author[g]{Kevin Boyce}
\author[a]{Antoine Cl\'enet}
\author[g]{Michael DiPirro}
\author[x]{Pierre Jamotton}
\author[d]{Simone Lotti}
\author[b]{Denis Schwander}
\author[g]{Stephen Smith}
\author[c]{Bert-Joost van Leeuwen}
\author[c]{Henk van Weers}
\author[m]{Thorsten Brand}
\author[e]{Beatriz Cobo}
\author[m]{Thomas Dauser}
\author[c]{Jelle de Plaa}
\author[a]{Edoardo Cucchetti}

\affil[a]{IRAP CNRS, 9 Av. colonel Roche, BP 44346, F-31028 Toulouse cedex 4, France and Universit\'e de Toulouse III Paul Sabatier / OMP, Toulouse, France}
\affil[b]{Centre National d'Etudes Spatiales, Centre spatial de Toulouse, 18 avenue Edouard Belin, 31401 Toulouse Cedex 9, France}
\affil[c]{SRON, Netherlands Institute for Space Research, Sorbonnelaan 2, 3584 CA Utrecht, The Netherlands}
\affil[d]{INAF/Istituto di Astrofisica e Planetologia Spaziali, Via Fosso del Cavaliere 100, 00133, Roma, Italy}
\affil[e]{Instituto de F\`isica de Cantabria (CSIC-UC), E-39005 Santander, Cantabria, Spain}
\affil[f]{Department of Physics, Division of Geophysics and Astronomy, P.O. Box 48, FI-00014, University of Helsinki, Finland}
\affil[g]{NASA/Goddard Space Flight Center, 8800 Greenbelt Rd, Greenbelt, MD 20771, United States}
\affil[h]{Centro de Astrobiolog\`ia, CSIC / INTA, Ctra de Torrej\'on a Ajalvir, 4 km,  28850 Torrej\'on de Ardoz, Madrid, Spain}
\affil[i]{Institute of Space and Astronautical Science (ISAS) \& Japan Aerospace Exploration Agency (JAXA), 3-1-1 Yoshinodai, Chuo-ku, Sagamihara, 252-5210, Japan}
\affil[j]{Department of Astronomy, University of Geneva, Chemin d'Ecogia 16, CH-1290 Versoix, Switzerland}
\affil[k]{University of Li\`ege, Institute for Astrophysics \& Geophysics, All\'ee du 6 Ao\^ut 19c, B-4000 Li\`ege, Belgium}
\affil[l]{Nicolaus Copernicus Astronomical Centre  of the Polish Academy of Sciences,  ul. Bartycka 18,  00-716 Warsaw, Poland}
\affil[m]{ECAP, University of Erlangen-N\"uremberg, Sternwartstr. 7, 96049 Bamberg, Germany}
\affil[n]{Universit\`a degli Studi di Palermo, Dipartimento di Fisica e Chimica, Via Archirafi 36, 90123 Palermo, Italy and INAF/Osservatorio Astronomico di Palermo G.S.Vaiana, Piazza del Parlamento 1, 90134 Palermo, Italy}
\affil[o]{Univ. Grenoble Alpes, CEA INAC-SBT, 38000 Grenoble, France}
\affil[p]{Laboratoire AIM, UMR 7158, CEA/CNRS/Universit\'e Paris Diderot, CEA DRF/IRFU/SAp, F-91191 Gif sur Yvette , France}
\affil[q]{INAF-Osservatorio Astronomico di Roma, Via Frascati, 33 - 00078, Monte Porzio Catone, Italy}
\affil[r]{University of Genova, Dept. of Physics, Via Dodecaneso 33, 16146, Genova, Italy}
\affil[s]{APC - Astroparticule et Cosmologie, Universit\'e Paris Diderot, 10 rue A. Domon et L. Duquet, 75205 Paris cedex 13, France and Service d'Astrophysique IRFU/DRF/CEA Saclay, F-91191 Gif sur Yvette cedex, France }
\affil[t]{University of Groningen, Landleven 12, 9747 AD Groningen, The Netherlands }
\affil[u]{INAF - IASF Milano, Via E. Bassini 15, I-20133 Milano, Italy}
\affil[v]{Centrum Badan Kosmicznych, Polish Academy of Science, Bartycka 18a, 00-716 Warszawa, Poland}
\affil[w]{APC - Astroparticule et Cosmologie, Universit\'e Paris Diderot, 10 rue A. Domon et L. Duquet, 75205 Paris cedex 13, France }
\affil[x]{CSL - Centre Spatial de Li\`ege, Avenue du Pr\'e-AiIly (B29), B-4031 Angleur (Liege), Belgium }
\affil[y]{Leiden Observatory, Leiden University, PO Box 9513, NL-2300 RA Leiden, The Netherlands}
\affil[z]{Japan Aerospace Exploration Agency, Research Unit II (U2), Research and Development Directorate, 305-8505 2-11, Sengen, Tsukuba, Ibaraki, Japan}
\affil[A]{INAF-IASF  Istituto di Astrofisica Spaziale e Fisica Cosmica, Area della Ricerca - via Piero Gobetti, 101 - 40129, Bologna, Italy}
\affil[B]{Anton Pannekoek Institute/GRAPPA, University of Amsterdam, PO Box 94249, NL-1090 GE Amsterdam, The Netherlands}
\affil[C]{University College London - Mullard Space Science Laboratory, Holmbury St. Mary Dorking, Surrey, RH5 6NT, United Kingdom}
\affil[D]{Hiroshima University, High Energy Astrophysics Group, Department of Physical Sciences,1-3-1 Kagamiyama, Higashi-Hiroshima, Hiroshima 739-8526, Japan}
\affil[E]{Center for Theoretical Physics, Polish Academy of Sciences, Al. Lotnikow 32/46, 02-668 Warsaw, Poland}
\affil[F]{University of Michigan Department of Astronomy, Ann Arbor, 1085 South University Avenue, MI 48109-1107, United States }
\affil[G]{INAF/Osservatorio Astronomico di Palermo G.S.Vaiana, Piazza del Parlamento 1, 90134 Palermo, Italy}
\affil[H]{IUFACyT, Universidad de Alicante, Campus de San Vicente del Raspeig, E03690, Alicante, Spain}

\authorinfo{Further author information: (Send correspondence to Didier Barret: dbarret@irap.omp.eu)}

\begin{document} 
\maketitle
\newpage
\begin{abstract}
The X-ray Integral Field Unit (X-IFU) on board the Advanced Telescope for High-ENergy Astrophysics (Athena) will provide spatially resolved high-resolution X-ray spectroscopy from 0.2 to 12 keV, with $\sim 5$'' pixels over a field of view of 5 arc minute equivalent diameter and a spectral resolution of 2.5 eV up to 7 keV. In this paper, we first review the core scientific objectives of Athena, driving the main performance parameters of the X-IFU, namely the spectral resolution, the field of view, the effective area, the count rate capabilities, the instrumental background. We also illustrate the breakthrough potential of the X-IFU for some observatory science goals. Then we briefly describe the X-IFU design as  defined at the time of the mission consolidation review concluded in May 2016, and report on its predicted performance. Finally, we discuss some options to improve the instrument performance while not increasing its complexity and resource demands (e.g. count rate capability, spectral resolution).

{\it The X-IFU will be provided by an international consortium led by France, The Netherlands and Italy, with further ESA member state contributions from Belgium, Finland, Germany, Poland, Spain, Switzerland and two international partners from the United States and Japan.}
\end{abstract}

% Include a list of keywords after the abstract 
\keywords{Athena, Instrumentation, Space telescopes, X-ray spectroscopy, X-ray Integral Field Unit}

% introduction
%
%
\section{Introduction}

ESA's Athena X-ray observatory mission was selected in 2014 to address the Hot and Energetic Universe science theme (Ref.~\citenum{NandraWP2013}). The Hot Universe refers to the baryons in the Universe at temperatures above $10^{5-6}$K and that amount to about half of the total baryonic content. There are as many baryons at $T >10^7$~K trapped in dark-matter potential wells which trace the large-scale structures of the Universe as there are locked into stars. Gas at lower temperatures (down to $T\sim 10^{5}$ K) is expected to reside in filamentary structures pervading the intergalactic medium. The hot gas distributed on large-scales is strongly influenced by phenomena occurring in the immediate vicinity of black holes (the Energetic Universe) through a poorly understood process called Cosmic Feedback. As X-rays are copiously produced by hot gas and accretion around black holes, the best observational handle on the Hot and Energetic Universe is through X-ray observations. In addition X-rays can escape relatively unimpeded from significantly obscured environments and are only weakly contaminated by phenomena other than those mentioned. To address these and other science objectives Athena is conceived as a large observatory offering an unprecedented combination of sensitive X-ray imaging, timing and high-resolution spectroscopy.   

The X-IFU is a cryogenic imaging spectrometer, offering spatially-resolved high-spectral resolution X-ray spectroscopy over a 5 arc minute equivalent diameter field of view (see Table \ref{tab_performance} for the performance requirements). The breadth of the science affordable with the X-IFU encompasses key scientific issues of the Hot and Energetic Universe science theme and beyond. In a nutshell, the X-IFU will provide: 
\begin{itemize}
\item 3D integral field spectroscopic mapping of hot cosmic plasmas, enabling measurements of gas bulk motions and turbulence, chemical abundances and the spatial distribution of these and other physical parameters. {\it This drives the X-IFU field of view and spatial resolution, particle background, spectral resolution and calibration accuracy.}
\item Weak spectroscopic line detection, enabling the detection of unresolved absorption and emission lines from Warm and Hot Intergalactic Medium filaments and weak spectral features produced by unusual ion species or states. {\it This drives the X-IFU spectral resolution, calibration and throughput.}
\item Physical characterization of the Hot and Energetic Universe, including plasma diagnostics using emission line multiplets, AGN reverberation and black hole spin measurements, winds in galactic sources in outburst, AGN winds and outflows, stellar outflows, solar wind charge exchange, etc. {\it This drives the X-IFU spectral resolution, calibration and high-count rate capability.}
\end{itemize}

\begin{figure}[!t]
\begin{center}
\begin{tabular}{c}
\includegraphics[scale=0.3]{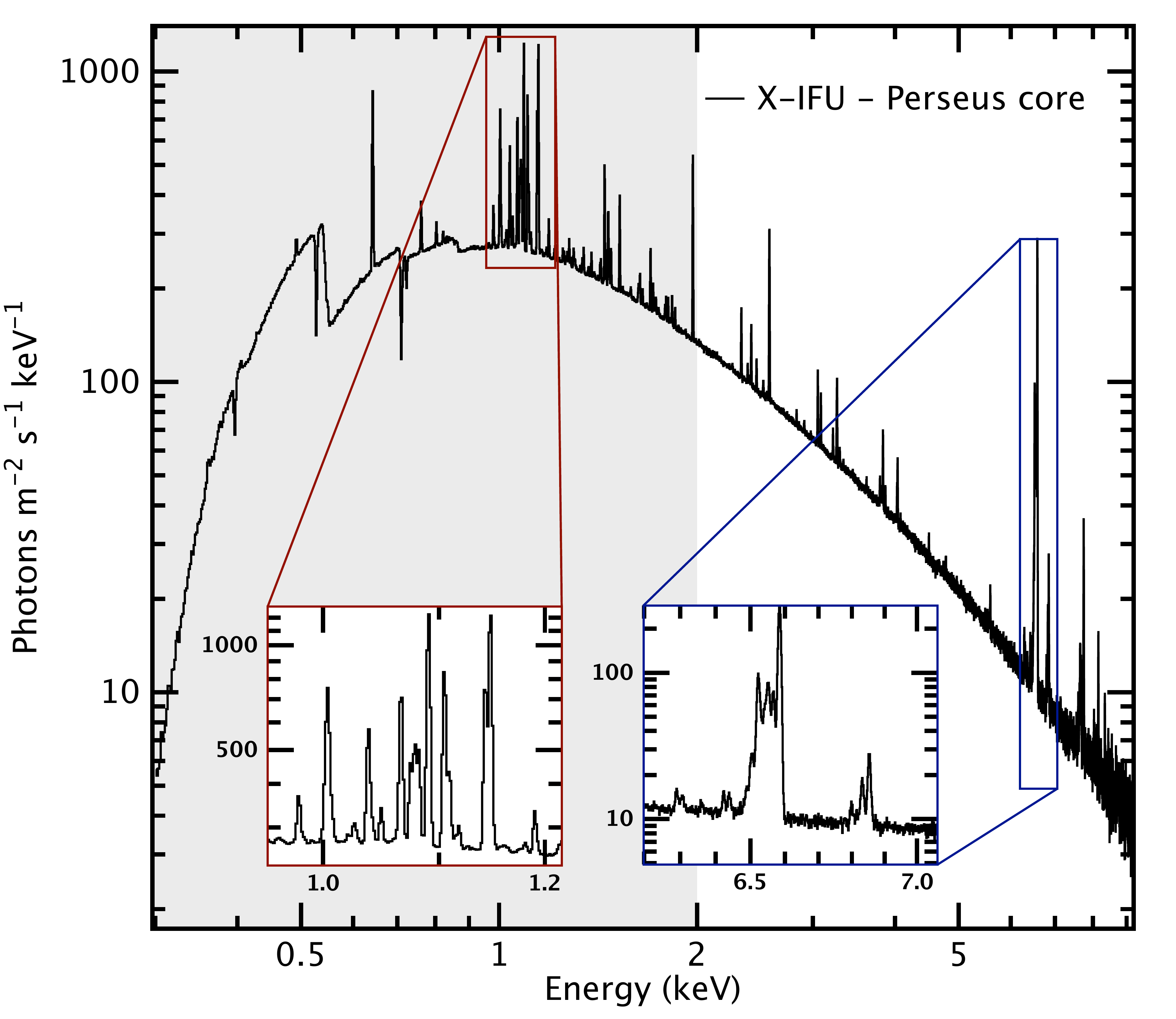}
\end{tabular}
\caption{\label{fig:perseus}The simulated X-IFU spectrum of the core of the Perseus cluster, based on {\it Hitomi SXS} observations  (Ref.~\citenum{Hitomi2016}, Courtesy of A.C. Fabian and C. Pinto). The exposure time is 100 \ksec. The grey area indicates the region not fully explored by SXS, due to the fact that the {\it Hitomi SXS} Perseus observation was performed with gate valve closed (blocking most X-rays below 2 keV). The insets show the region around the iron L and K complexes. The wealth of information provided by such a spectrum, that will be measured on sub-arc minute scales enables in depth studies of the physical properties of the hot cluster gas (e.g. temperature, density, turbulence, bulk motion, abundance, \ldots) (see Section \ref{s:hu}).}
\end{center}
\end{figure}

The breakthrough capabilities of the X-IFU are illustrated in Figure \ref{fig:perseus}, which shows the expected X-IFU spectrum of the core of the Perseus cluster, based on very recent results obtained with the Hitomi Soft X-ray Spectrometer (Ref.~\citenum{Hitomi2016}).

In this paper we will first review the core scientific objectives of Athena as driving the X-IFU performance requirements: this will cover the Hot and Energetic Universe as well as the Observatory science (Section \ref{sec:drivers}). Next we will describe the  baseline instrument configuration (Section \ref{current_instrument_configuration}) and present the current instrument performance in Section \ref{current_instrument_performance}. In Section \ref{performance_optimization}, we list the different options that are being studied for optimizing the instrument performance. This precedes a description of the short term plan and conclusions (Section \ref{conclusions}).
% science
%
%
\section{SPECTROSCOPY OF THE HOT AND ENERGETIC UNIVERSE}
\label{sec:drivers}  % \label{} allows reference to this section
Let us now review the core X-IFU related scientific objectives of Athena and describe some of the observatory science goals relevant for the X-IFU.
\subsection{The Hot Universe}
\label{s:hu}

The X-IFU on board Athena will be vital to characterise the Hot Universe, by  measuring the mechanical energy stored in gas bulk motions and turbulence in groups and clusters of galaxies, the distribution of its metal abundances across cosmic time, the effect of Active Galactic Nuclei (AGN) in the intra-cluster medium (ICM) and by obtaining the distribution and properties of the warm/hot baryonic filaments in the intergalactic medium.

\subsubsection{Cluster bulk motions and turbulence}
\label{s:bul}
The hierarchical growth of large-scale structures happens through continuous accretion of material  and successive merging events (Refs.~\citenum{ber98,pla14,vog14}). These processes heat the gas  filling the massive halos  by adiabatic compression and by countless shocks they generate at all scales (e.g., Ref. \citenum{mar07}). 

Part of the gravitational energy released at the formation of a halo is channeled through bulk motions and turbulent flows. It cascades down to smaller scales where it is dissipated, thereby contributing to the virialisation  of their hot gaseous atmosphere (Refs. \citenum{mcc07,nag13,lau13}). Turbulence in the ICM is also related to the weak magnetic field bathing the cluster gas and further links to the viscosity, convection and conduction of the gas (Ref.~\citenum{sch07}). The connection between these micro-scale physical processes, as well as their impact on the larger scales is still to be unveiled in order to fully understand the  overarching  process of assembling large scale structures.

To date this has been investigated mostly by establishing a connection between the statistics of surface brightness fluctuations and of  the turbulent velocity field (e.g., Refs.~\citenum{chu12,zhu14,kha16}). Direct measurement of the turbulent velocity and bulk motions can be obtained from the respective measurement of the broadening and shift they induce on the atomic emission lines from the ICM. X-ray  grating spectroscopy has only provided upper limits (e.g., Refs.~\citenum{san13,pin15}). Thanks to the capabilities of the SXS calorimeter spectrometer (Ref.~\citenum{Mitsuda2014}) {Hitomi} (Ref.~\citenum{Takahashi2014}) has provided an unprecedented view of the Perseus cluster (Ref.~\citenum{Hitomi2016}), showing for the first time what spatially-resolved high-resolution X-ray spectroscopy can deliver. These unique observations have shown that the level of turbulence close to the core of Perseus is rather modest ($< 164\pm 10 \, {\rm km}\, {\rm s}^{-1}$), despite the highly structured spatial distribution of hot gas resulting from AGN feedback. This is only a small glimpse of what X-IFU will be able to do, but on 12 times smaller spatial scales and with a spectroscopic throughput which will be 25 times larger than that of Hitomi/SXS.

The direct measurement of the bulk motions and turbulence velocity field thus requires spatially-resolved, high resolution spectroscopy. The X-IFU on board the Athena observatory will be the first integral field unit operating at X-ray wavelengths that will allow, thanks to  its joint spatial ($5''$) and spectral (2.5~eV) resolution, to map the velocity field of the hot gas in groups and clusters of galaxies down to a precision of $10-20\, {\rm km}\, {\rm s}^{-1}$ for velocities ranging between $100-1000\, {\rm km}\, {\rm s}^{-1}$ (see Fig.~\ref{fig:yields}).

The joint high spatial and spectral resolution of the X-IFU shall also allow to resolve line complexes (e.g., Iron around 6.7~keV) and measure line ratios to further constrain thermodynamics or ionisation state of the gas far out into the clusters outskirts (see e.g. Ref. \citenum{molendi_2016}). All these diagnostics will provide us with a comprehensive view of how the dark and baryonic matter assemble and evolve into large scale potential wells.

\subsubsection{Chemical enrichment}
\label{s:che}

With masses up to and exceeding $10^{15}\, {\rm M}_{\odot}$, the deep potential wells of galaxy clusters retain all the information regarding the chemical enrichment of their intracluster medium (ICM) across cosmic time (this is not the case for galaxies, which lose their gaseous haloes and eject metals into the inter-galactic medium through stellar winds and stellar explosions).  Just 40 years ago a feature corresponding to Fe\,\textsc{XXV} and Fe\,\textsc{XXVI} transitions was discovered in the X-ray spectrum of the Perseus cluster (Ref.~\citenum{mit76}). Since then, it has been recognised that the hot gas of the ICM is continuously enriched with heavy elements generated in type Ia (SN$\rm{1a}$)  and core-collapse supernova (SNcc) explosions in the cluster member galaxies. Elements from O to Si and S are mainly produced in massive stars and ejected in SNcc at the end of their lifetime. White dwarfs in binary systems give rise to SNIa explosions that produce elements up to Si, Fe and Ni. X-ray observations of emission lines from highly-ionized elements are the only way to access information on the abundances of the hot gas, its evolution to high redshift, and the processes by which heavy elements  are redistributed into the surrounding ICM.

\begin{figure}
\begin{center}
\begin{tabular}{c}
\includegraphics[height=7.5cm]{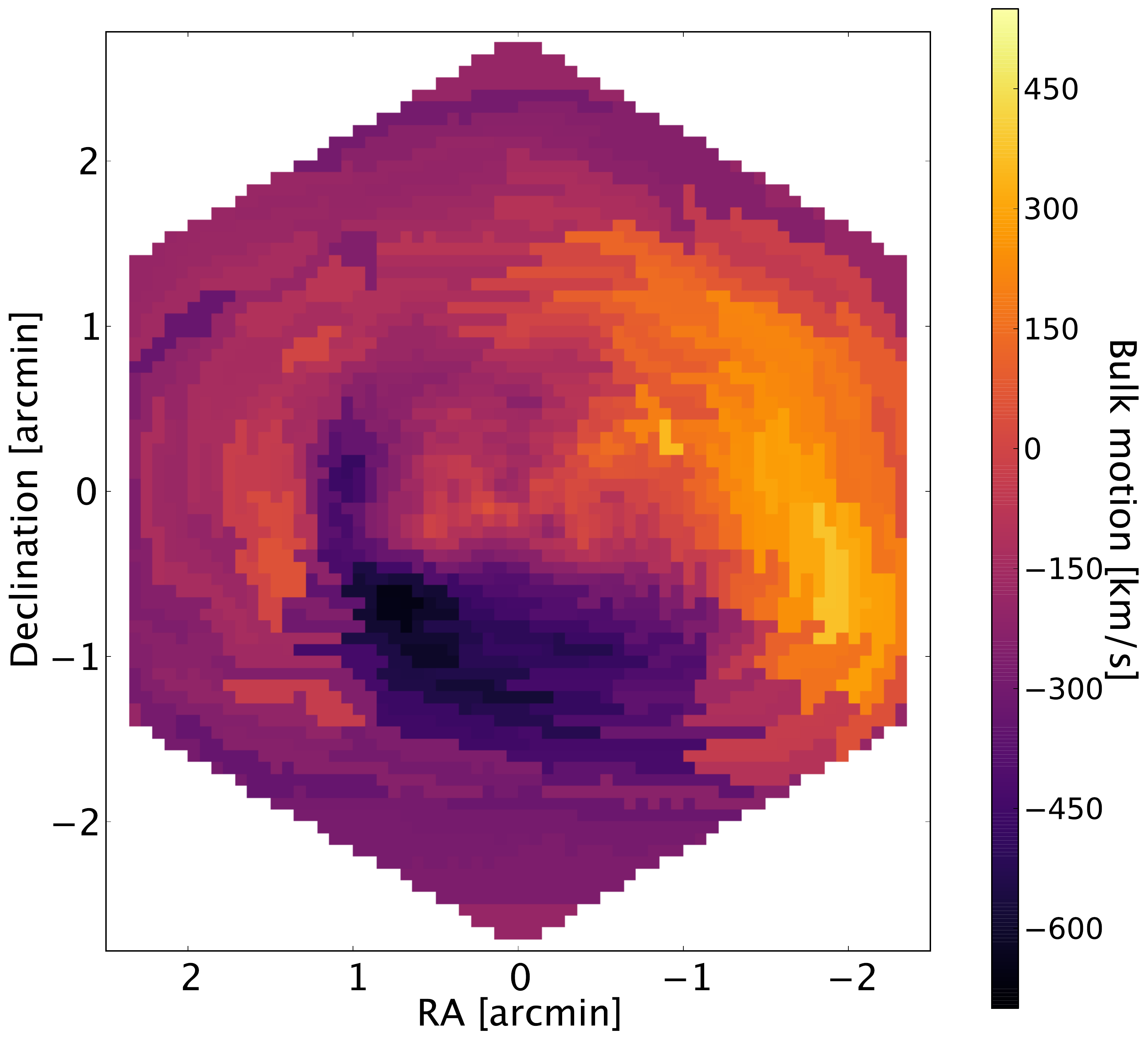}\includegraphics[height=7.5cm]{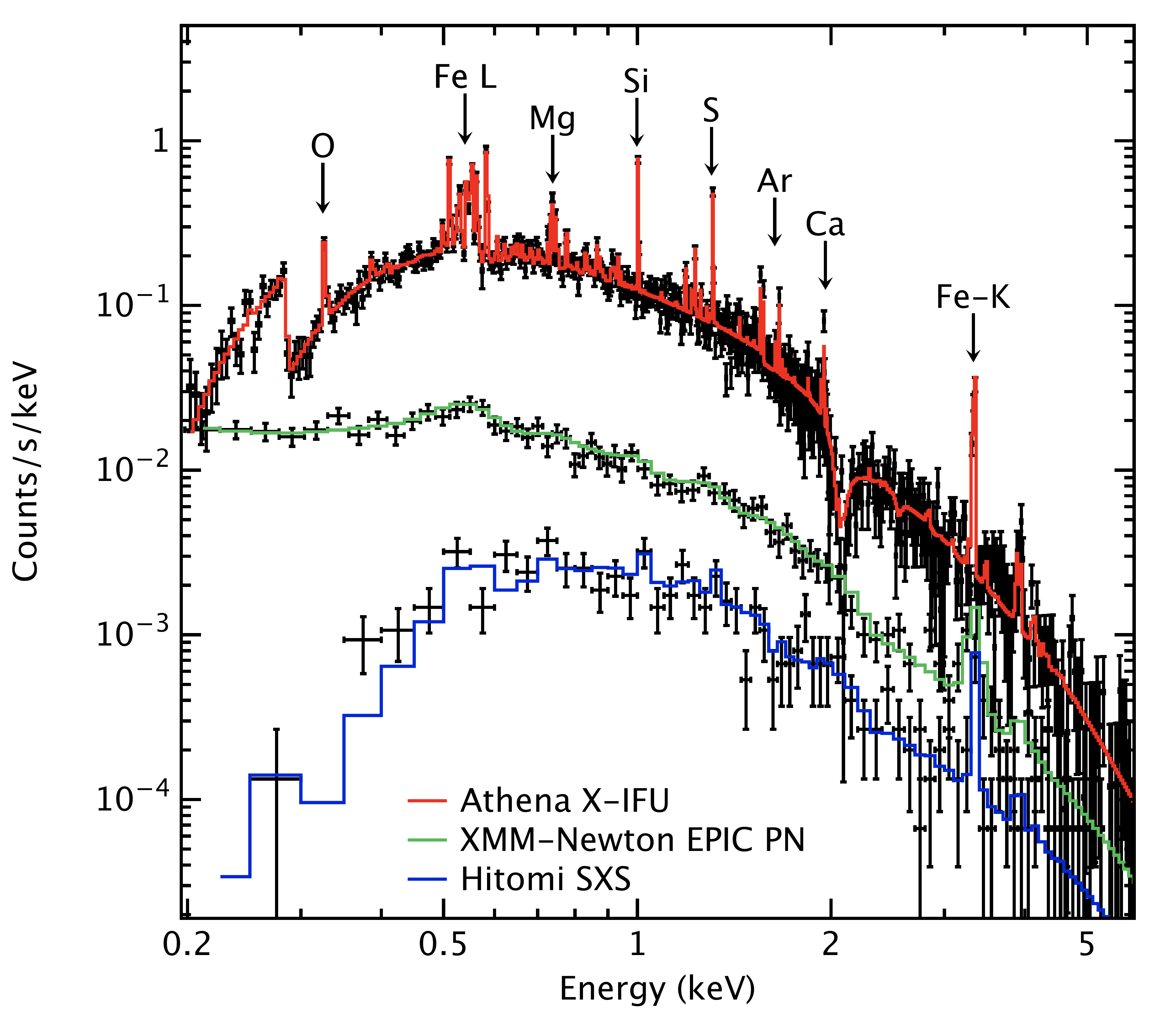}
\end{tabular}
\end{center}
\caption 
{ \label{fig:yields}
Left: Reconstructed bulk motion induced velocity field (in km/s) of the hot intra-cluster gas for a 50 \ksec\ X-IFU observation of the central parts of a Perseus like cluster from the numerical simulations in Ref. \citenum{rasia15}. The cluster has the luminosity of Perseus but is considered at a redshift of 0.1. Right: Simulated X-IFU spectrum of a $z=1$ galaxy group with $kT=3~{\rm keV}$ and $L_\mathrm{X}=1\times 10^{44}\, {\rm erg}\, {\rm s}^{-1}$ for 50~ks. Emission lines from elements which are key to understand chemical evolution can be clearly seen.
} 
\end{figure}  

The combination of {Athena}'s large effective area and the 2.5~eV spectral resolution of the X-IFU will allow the abundances of many common heavy elements to be measured to unprecedented precision. 
Abundance ratios are a powerful method for constraining the contribution of SN$\rm{1a}$, SNcc and AGB stars to the total heavy element abundance, as each source produces different heavy element yields. Information on the Initial Mass Function (IMF), the stellar populations, and the star formation history of the galaxies in the cluster can also be gleaned from the evolution of abundance ratios across time. 

%evolution
The exceptional spectral resolution of the X-IFU will allow accurate abundance ratios to be determined to high redshift ($z>1$) for the first time (see Fig.~\ref{fig:yields}). An example application is the measurement of the evolution of the ratios of O/Fe and Si/Fe for an ensemble of clusters. This will allow discrimination between models where recent enrichment is caused by  SNIa ejection, which produces relatively more Fe than O and Si at low redshift, and models where the enrichment is due to stripping of already pre-enriched member galaxies, which do not show evolution in redshift.

The heavy elements are ejected and redistributed into the ICM by a number of processes, including outflows and jets from active galactic nuclei (e.g., Ref.~\citenum{kir11}), galactic winds and starbusts (e.g., Ref.~ \citenum{str00}), and ram-pressure stripping of the galaxies (e.g., Ref.~\citenum{sch08}); in addition, it is also possible that intracluster stars may also contribute to the ICM enrichment (e.g., Ref.~\citenum{gal03}). 
Spatially-resolved measurements such as abundance profiles provide insight into the different enrichment mechanisms and their spatial distribution, their timescales, and how the gas is mixed by gas-dynamical processes.  Detailed abundance mapping  is a powerful tracer of the jet energy distribution and can supply constraints on entrainment of enriched gas by the jets, the jet power itself (which is correlated with the radial range of the metal-enriched outflows). It can also put strong constraints on gas physics through  ram-pressure stripping of enriched plasma from infalling sub-clumps. The X-IFU will enable all of these measurements to be obtained to unprecedented precision.

\subsubsection{AGN feedback on cluster scales}
\label{s:agn}
Active galactic nuclei (AGN) at the centres of galaxy groups and clusters play a critical role in shaping the properties of the central galaxy and the surrounding ICM. Mechanical feedback from AGN jets is thought to be one of the best candidates for suppression of star formation in the massive central galaxies and for heating the gas inside and beyond the cluster core. Observations of X-ray cavities surrounding radio lobes in systems from massive elliptical galaxies to galaxy clusters (e.g., Refs.~\citenum{boe93,mcn05}) provide observational support for the existence of feedback from AGN jets at all scales.

In the centre of bright, nearby clusters, X-IFU measurements of the X-ray line profiles and variations of the line centroid will allow estimation of the characteristic spatial scales of the turbulent motions induced by AGN jets on scales of tens of kpc, and mapping of the velocity field of the hot gas to an accuracy of $\sim 20\, {\rm km}\, {\rm s}^{-1}$. This will give unprecedented insights into how power from the initially highly collimated jets is distributed into the surrounding ICM. The thermal and non-thermal energy content of the X-ray cavities will be measured accurately for the first time, helping to establish their composition. X-IFU will help to detect directly the shocked gas surrounding expanding radio lobes, with a spectral resolution sufficient to resolve shock expansion speeds for the first time.

The dynamics of the hot gas in the vicinity of cool filaments will yield essential clues to the cooling and mixing process, and enable for the first time a measurement of the amount of material cooling out of the hot phase and relate this to the fuel available to power the AGN. The X-IFU will enable understanding of the entire cycle of heating and cooling in the cores of nearby clusters and groups. Robust jet power measurements for large samples can be compared to accretion rates of hot and cold material, enabling insights into the accretion process and black hole growth to be obtained.

\subsubsection{The missing baryons and the Warm-Hot Intergalactic Medium}
\label{s:whim}

The number of visible baryons in the local ($z<2$) Universe (stars, cold atomic gas and molecular gas in galaxies) adds up to only about 15\% of the total number of baryons inferred through a number of independent measurements of cosmological parameters, and recent radio observations have shown the evidence that these âinvisibleâ baryons mostly reside in the gaseous phase. However, not only the detailed state of the baryons remains unclear, but at least half of these baryons are still elusive, and thought to lie in a tenuous web of warm-hot intergalactic medium (the WHIM). X-IFU studies of WHIM will provide unprecedented information on the hot phase of the baryons in large scale structures, complementing the COS/HST observations which are sensitive to the luke-warm phase at $10^{(5-5.5)}\, {\rm K}$.

X-IFU will pursue three independent approaches to the problem:
\begin{itemize}
\item Studies of WHIM in absorption against bright AGN at $z<0.5$, probing low-z LSS and their interplay with the surrounding intergalactic medium;  
\item Studies of WHIM in absorption (and simultaneously in emission at $z<0.1$) against Gamma-Ray Burst (GRB) afterglows, probing the WHIM up to redshift of $\sim 2$ or higher and strongly constraining the physics (i.e., density and temperature) and kinematics (turbulence and bulk motions) of the WHIM for those filaments detected both in emission and absorption;  
\item WHIM in emission in the outskirts of galaxy concentrations, probing the kinematics of the warm-hot gas near large structures. 
\end{itemize}

\begin{figure}[!h]
\begin{center}
\begin{tabular}{c}
\includegraphics[scale=0.225]{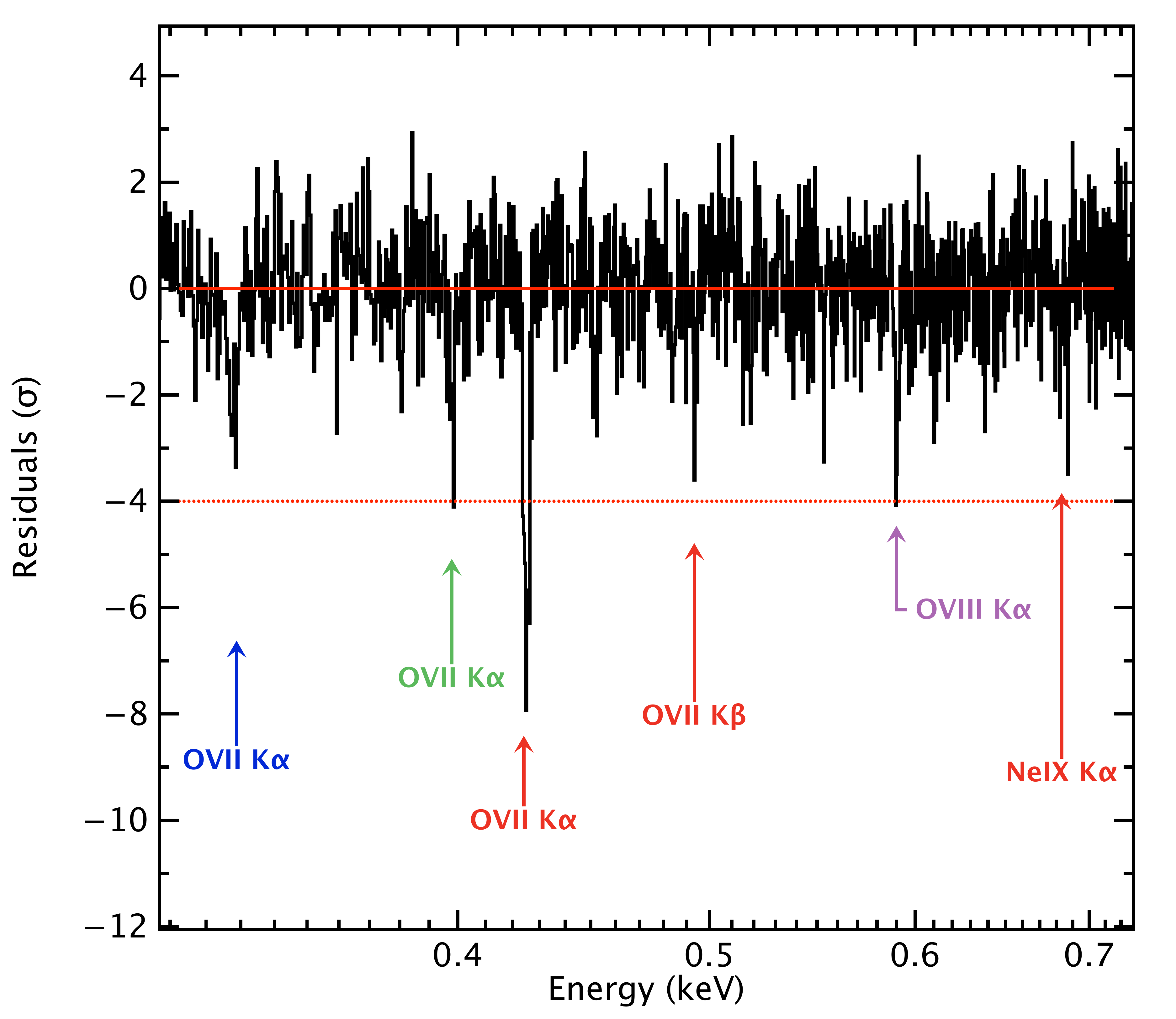}\includegraphics[scale=0.225]{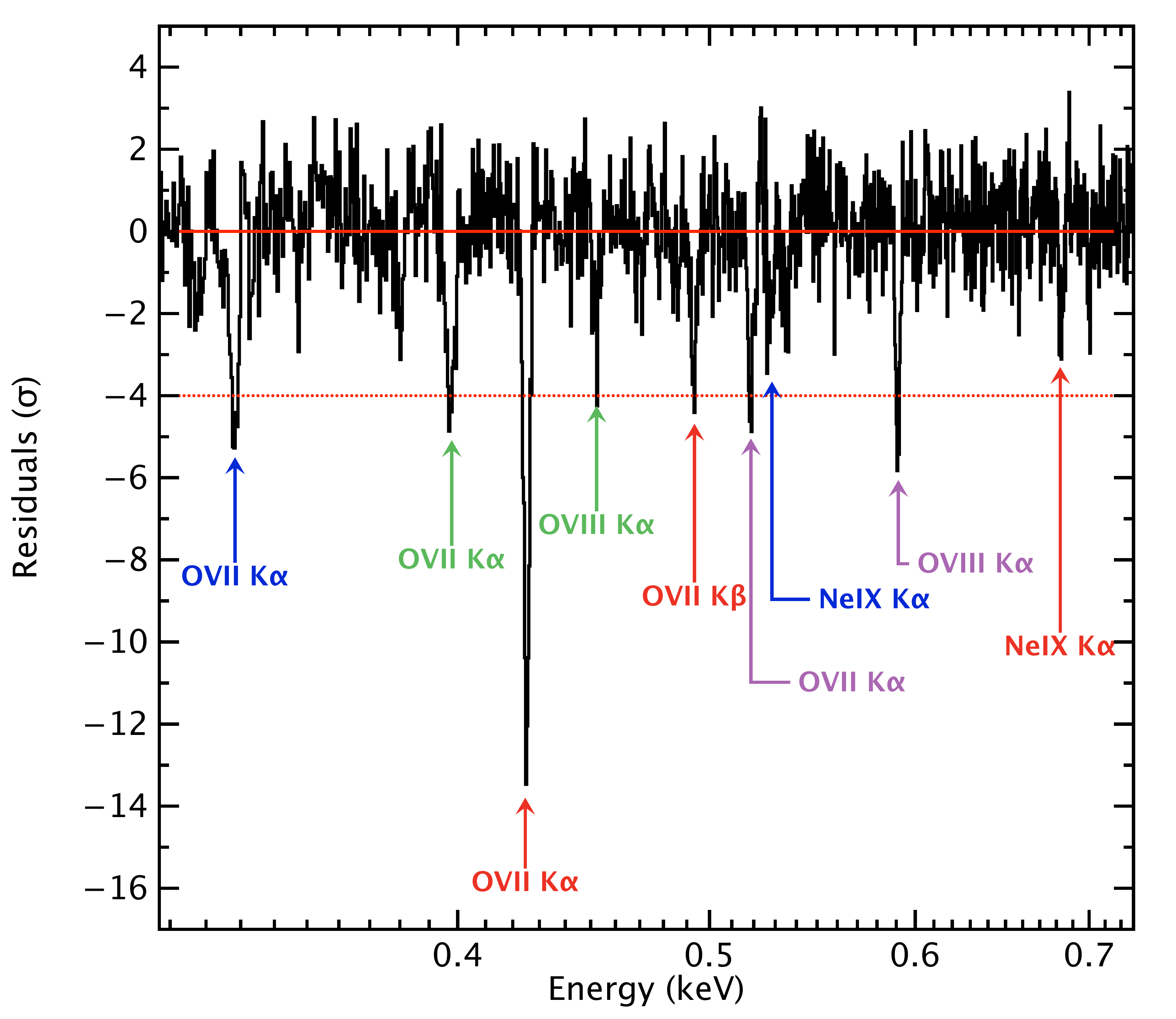}
\end{tabular}
\end{center}
\caption 
{ \label{fig:WHIM}
The two panels show a simulated X-IFU spectrum of a bright background source at $z>0.8$ crossing the same random patch of the WHIM as predicted by the simulations in Ref.~\citenum{CenOstriker2006}. Four WHIM filaments are clearly detected in both spectra, in some cases through several absorption lines, enabling direct measurements of the gas temperature and even turbulence. Left: The background source is a GRB afterglow at $z>0.8$ with an effective intrinsic column density of $10^{22}\, {\rm cm}^{-2}$ yielding about $1.5 \times 10^6$ counts in the 0.3-2~keV band. This corresponds to the brightest 10\% of the GRB afterglows expected in the sky. Right: The background spectrum is that of the brightest known BL Lac at $z>0.8$ 3C454.3 ($z=0.86$), with a flux $F (0.3-2\, {\mathrm keV})= 2.7 \times 10^{-11}\, {\rm erg}\, {\rm cm}^{-2}\, {\rm s}^{-1}$ and integrated for 100 ks yielding about $7\times 10^6$ counts.} 
\end{figure}

The main advances compared to previous instrumentation are the unique combination of large collecting area and spectral resolution. Accurate simulations based on theoretical predictions, show that the X-IFU will be able to fully understand the WHIM baryon budget with a set of Athena observations of bright nearby ($z<0.5$) AGN and more distant ($z<2$) GRB afterglows. Fig.~\ref{fig:WHIM} (left) displays a simulated spectrum of a GRB afterglow among the 10\% brightest in the sky, going through a random WHIM line of sight extracted from the hydrodynamical simulations in Ref.~\citenum{CenOstriker2006}\ (spanning the redshift range $z=0-0.85$). Four WHIM filaments are detected at redshifts 0.108, 0.350, 0.444 and 0.753, one of them with 3 lines. Taking the Swift/BAT GRB monitor as a reference, there are $\sim 25$ such GRB afterglows per year, and therefore with a moderate ToO efficiency several such targets could be observed each year. Obtaining higher quality spectra of high-z objects will be rather infrequent in terms of GRB afterglow availability. Alternatively (Fig.~\ref{fig:WHIM}, right) a bright BL Lac like 3C 454.3, ($z=0.86$) could provide a high-quality spectrum along the same simulated line of sight. The challenge in that approach to sample sufficiently long lines of sight is the scarcity of such bright objects at $z>0.5$. An optimised combination of both approaches as a function of the various performance parameters will be necessary.
    
\subsection{The Energetic Universe}
\label{s:Energetic}
X-IFU observations will measure the energy released by accretion onto black holes by winds and outflows, all the way up from the local Universe to $z\sim 3$, providing solid ground to understand AGN feedback. In nearby galaxies, the amount of gas, energy and metals blown into the circum-galactic medium by both AGN and starbursts will be mapped. By targeting distant GRB afterglows, X-IFU observations will characterise the ambient interstellar medium of high-z galaxies and constrain their prevailing stellar populations. Deep X-IFU observations of obscured distant AGN will be able to unveil their redshift via the Fe K emission line. 

\subsubsection{High-z Gamma-ray Bursts: early metal enrichment of the Universe}
\label{s:GRB}
X-IFU observations of gamma ray bursts can play a unique role in the study of metal enrichment as GRBs are the brightest light sources at all redshifts and, for long duration events (LGRBs), occur in star-forming regions. As LGRB progenitors are short-lived massive stars, they provide an ideal probe of the effect of stellar evolution on galaxy chemical enrichment across cosmic time. Beginning with metal free (Population III) stars, the cycle of metal enrichment started when their final explosive stages injected the first elements beyond Hydrogen and Helium into their pristine surroundings, quickly enriching the gas. These ejecta created the seeds for the next generation of stars (population II). Finding and mapping the earliest star formation sites (population III/II stars) is one of the top priorities for future astrophysical observatories. Tracing the first generation of stars is crucial for understanding cosmic re-ionization, the formation of the first seed Super-Massive Black Holes (SMBH), and the dissemination of the first metals in the Universe. Photons from Pop III stars and radiation generated from accretion onto the first SMBH initiate the Cosmic Dawn. 
\begin{figure}
\begin{center}
\begin{tabular}{c}
\includegraphics[scale=0.3]{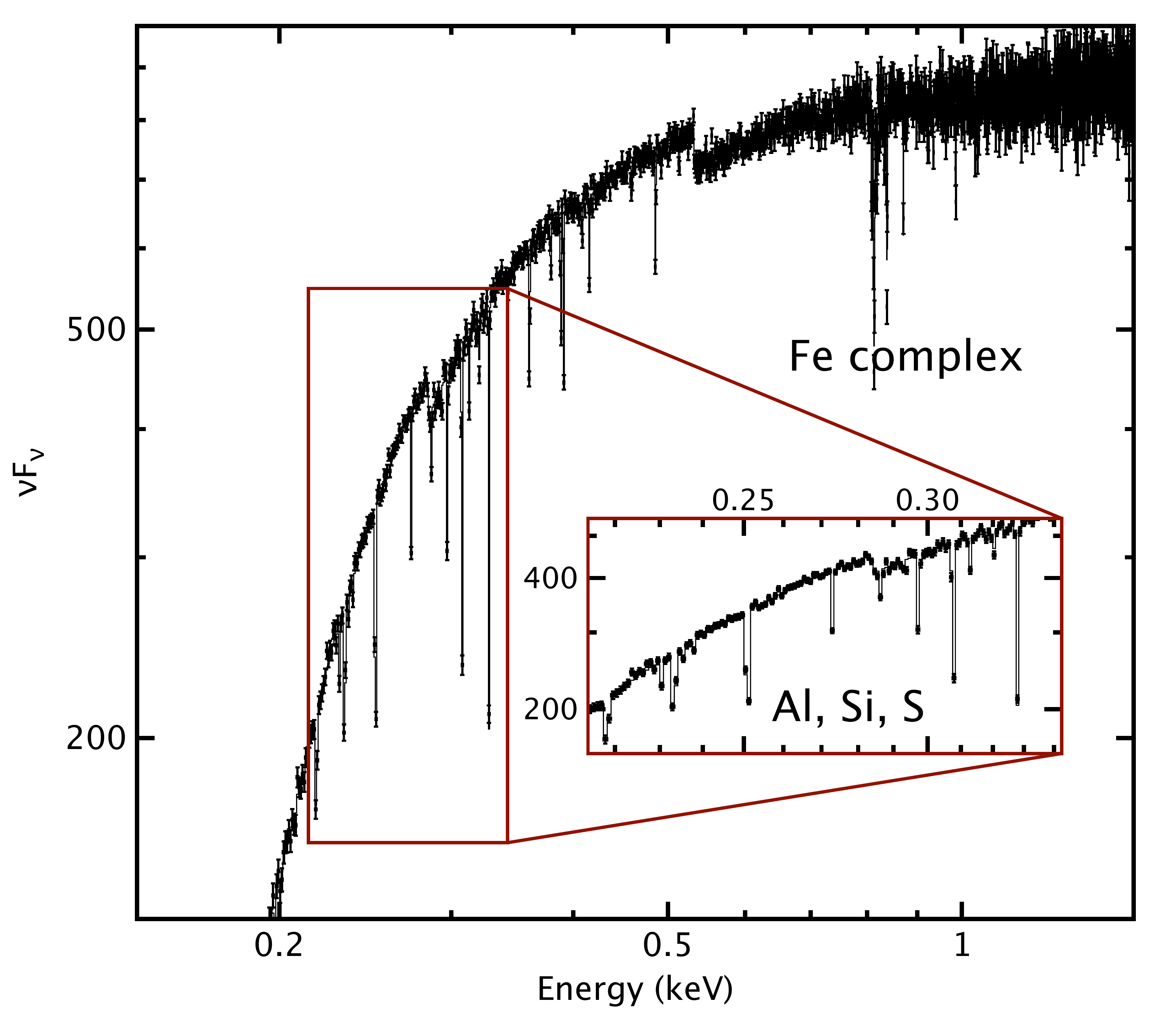}
\end{tabular}
\end{center}
\caption 
{ \label{fig:highzGRB}
A simulated X-IFU X-ray spectrum of a medium bright (fluence$=0.4\times 10^{-6}\, {\rm erg}\, {\rm cm}^{-2}$) afterglow at $z=7$, characterized by deep narrow resonant lines of Fe, Si, S, Ar, Mg, from the gas in the environment of the GRB. An effective intrinsic column density of $2\times 10^{22}\, {\rm cm}^{-2}$ has been adopted.
} 
\end{figure} 

The chemical fingerprint of Pop III star explosions is distinct from that of later generations, opening the possibility to probe the Initial Mass Function (IMF) of the Universe. Stellar evolution studies show that the nucleosynthetic yields of Pop III and Pop II explosions differ significantly. The convolution of these yields with an IMF directly translates to abundance patterns, which can differ up to an order of magnitude depending on the characteristic mass scale of the IMF (Ref. \citenum{HegerWoosley2010}). The X-IFU will be able to measure metal abundance patterns for a variety of ions (e.g., S, Si, Fe) for at least 10 medium-bright X-ray afterglows per year with H equivalent column densities as small as $10^{21} {\rm cm}^{-2}$ and gas metallicities as low as 1\% of solar for the denser regions expected in early star-forming zones; in even denser regions the accuracy will be further improved (Fig.~\ref{fig:highzGRB}). Measuring these patterns using GRBs, combined with Athena studies of AGN sightlines, galaxies and supernovae, will enable us to determine the typical masses of early stars, thereby testing whether or not the primordial IMF is indeed top heavy. For more information see Ref.~\citenum{JonkerSP2013}.

\subsubsection{AGN and Star-formation driven winds and outflows}
\label{s:outflows}
The measurement of tight correlations between the mass of galaxy bulges, or of the velocity dispersion of their stellar content, and the mass of the SMBH hosted in all galaxies (Refs.~\citenum{KormendyRichstone1995,Magorrian1998}) strongly indicate that some feedback mechanism must have acted between these components during the galaxy formation and evolution phases. In the last fifteen years, thanks to X-ray observations of quasars and nearby Seyfert galaxies, we have been able to identify AGN driven ultra-fast outflows (UFOs) as one of the plausible mechanisms (Refs.~\citenum{Chartas2002,Pounds2003,Tombesi2010a}). UFOs manifest themselves as blushifted resonant absorption lines due to highly ionized iron. They are detectable in the 7-10 keV band (Refs.~\citenum{Tombesi2010a,Tombesi2010b}) and are thought to be related to winds that are energetic enough to quench the star formation of the host galaxies. This provides the link between the tiny SMBH and the huge galaxy bulges (Ref.~\citenum{Tombesi2015}, see Ref.~\citenum{KingPounds2015} for a review on this topic). However, we still lack some fundamental pieces of evidence to fully verify this scenario: 1)  we do not know what is the launching mechanism of the UFOs, how they deposit their energy into the interstellar medium and how  they interact with star forming regions; 2) we have only poor knowledge of the physical conditions of the AGN engine and of the importance of UFOs at high-z, i.e., in epochs where both QSO and starburst activity were at their highest (Refs.~\citenum{Boyle1988,Madau1996,BrandtHasinger2005} and references therein).

\begin{figure}
\begin{center}
\begin{tabular}{c}
\includegraphics[scale=0.3]{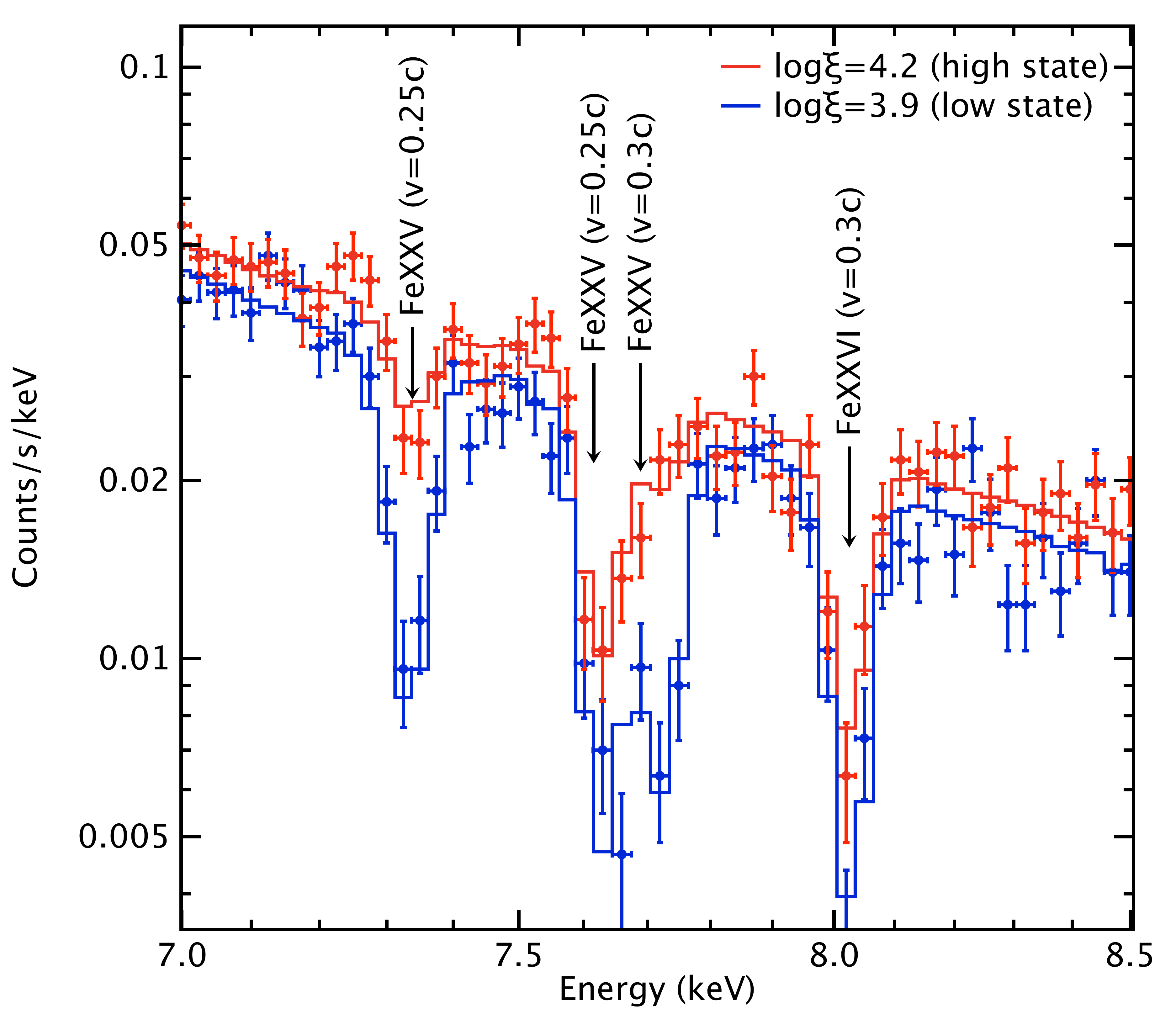}
\end{tabular}
\end{center}
\caption 
{ \label{fig:Winds}
X-IFU simulated spectrum (100 ks) of the QSO PDS456 showing the UFO components. The simulation is for two different spectral states (high ionisation in red, low ionisation in black).   
} 
\end{figure} 

To investigate the first of the above mentioned questions, we should obtain a detailed characterization of the physical properties of UFOs (column density, ionization state, outflow velocity, location, geometry, covering factor, etc.) and check how they evolve with respect to the distance from the source, maximum outflow velocity, and X-ray or UV luminosity (see Fig.~\ref{fig:Winds}). We must also consider that UFOs are known to be variable (Ref.~\citenum{Tombesi2015}) and so we must be able to investigate the above mentioned properties on time-scales of $\approx 10-100$~ks (corresponding to the dynamical time-scale at $10\, R_g$ for a SMBH with $M\sim 10^{7-8} M_{\odot}$). The coupling of the large collecting area and of the exquisite energy resolution of X-IFU provides the capability to perform such kind of studies (Ref.~\citenum{CappiSP2013}) on a sample of at least $\sim 50-80$ nearby ($z<0.01$) and bright ($F_{2-10\, {\rm keV}} > 10^{-11}\,  {\rm erg}\, {\rm cm}^{-2}\, {\rm s}^{-1}$) Seyfert galaxies with the accuracy needed to disentangle the right scenario among the radiation-driven (e.g., Refs.~\citenum{ProgaKallman2004,Sim2010}), momentum-driven (Ref.~\citenum{King2010}), and magnetically-driven (Ref.~\citenum{Fukumura2010}) accretion disc wind models. Moreover, the ability of the X-IFU to simultaneously obtain high-resolution X-ray spectra and good quality images will allow us to investigate where and how the AGN driven winds interact with the interstellar medium, what is the interplay among the AGN- and the starburst-driven winds, and how these winds enrich the intergalactic matter. In this respect it will be important to study nearby starburst and Luminous Infrared Galaxies (LIRG)/Ultra Luminous Infrared Galaxies (ULIRG) with the main goal to map, from good quality X-ray spectra, regions dominated by collisionally-ionized plasmas (hot gas), non-equilibrium ionization (shocks) and photo-ionized plasmas. Within its nominal life, X-IFU will be able to obtain such information for at least 30 objects thus allowing, for the very first time, to have a clear vision of such phenomena in the nearby Universe. 

To fully understand if and how these winds acted along cosmic time to shape the SMBH-galaxy co-evolution, we must finally couple the knowledge on the launch and SMBH-host galaxy interaction mechanisms obtained at low redshift with a reliable census of the UFOs and of their properties at high-z, i.e., we must answer  the second of the above mentioned questions. Up to now, UFOs have been measured only in a handful of mostly lensed (i.e., brightness enhanced) QSO at high redshift (Refs.~\citenum{Chartas2002,Chartas2003, Chartas2007,Lanzuisi2012,Vignali2015}). This clearly demonstrates that the capability to collect photons is fundamental to obtain the information we need. The Athena collecting area will enable investigations of the properties of ionized and outflowing absorbers in large samples of QSO ($L_\mathrm{X}\geq 10^{44}\, {\rm erg}\, {\rm s}^{-1}$) up to redshift $z\sim 4$ (Ref.~\citenum{GeorgakakisSP2013}).  Moreover, the wide angle surveys performed using WFI should produce rich samples of hundreds of QSO suitable for detailed spectroscopic studies with the X-IFU. These studies are expected to allow reliable estimates of the physical parameters of UFOs at the knee ($L^*$) of the X-ray luminosity function, which dominate the growth of SMBH at $z\sim 1-4$. We expect to be able to fully test the scenarios that assume the AGN driven winds as the mechanism to self-regulate the star-formation and SMBH growth along cosmic times.

\subsubsection{SMBH spins}
\label{s:SMBH}
Thanks to its high energy resolution and sensitivity, the X-IFU will allow the measurement of SMBH spins with unprecedented accuracy in a large number of AGN even beyond the local Universe.

The angular momentum is, in addition to mass, the other fundamental parameter that characterizes astrophysical black holes and therefore a proper census of SMBH in the Universe requires the measurement of their spins. This is also of fundamental importance in order to understand the black hole growth history (Ref.~\citenum{BertiVolonteri2008}), particularly the relative roles of mergers and chaotic accretion, that tend to reduce the spin, versus prolonged accretion, that generally increases it. Moreover a systematic study of SMBH spins would shed light on the relation between SMBH rotation and its outflow power in the form of relativistic jets. Since the influence of the spin is felt only up to a few gravitational radii, X-ray observations, probing the innermost regions of SMBH systems, are the main tool for such measurements.

The simplest and most widely applicable way of measuring SMBH spins is via time-averaged spectral fitting of relativistically broadened Fe K$\alpha$ lines (Ref.~\citenum{Fabian2000}). The larger the spin, the closer to the horizon will be the innermost stable circular orbit (ISCO) of a prograde accretion disk resulting in a larger broadening of the reflected lines. Broad lines are present in at least $\sim 30-40$\% of bright nearby type 1 AGN (Refs.~\citenum{Nandra2007,delaCalle2010}) and reliable estimates of SMBH spins have already been made in about 20 AGN with this technique (e.g., Refs.~ \citenum{Reynolds2014,Brenneman2013,Risaliti2013,Walton2013}). The Fe line is always accompanied by a reflection continuum in hard X-rays and, if the reflecting matter is at least partly ionized, also in the soft X-rays. With its large effective area over a broad energy range X-IFU will permit the simultaneous use of the iron line and of the soft X-ray reflection continuum to measure black hole spins also at intermediate redshifts. As an example, the spin of a maximally rotating black hole spin in PB5062, a luminous ($L_\mathrm{X} \sim 3\times 10^{46} \, {\rm erg}\, {\rm s}^{-1}$) QSO at  $z=1.77$ can be recovered with a precision of 20\% in a 100 ks observation.

One of the difficulties in these measurements is to disentangle the different AGN spectral components that come from regions at different distances from the SMBH. With its excellent energy resolution the X-IFU will easily separate the broad lines from the narrow features, which are ubiquitous in AGN and originate from more distant matter (Ref.~\citenum{YaqoobPadmanabhan2004}) and this will again allow to increase drastically the sample of measured SMBH spins.

X-IFU will also be able to map the inner regions of the accretion disks in the time-energy plane. Any deviation from axial symmetry in the disk emissivity (e.g., associated with hot spots) will lead to a characteristic variability of the iron line (Ref.~\citenum{Dovciak2004}), with ÒarcsÓ being traced out on the time-energy plane (Ref.~\citenum{ArmitageReynolds2003}). Evidence of hot spots were found in XMM-Newton data (Refs.~\citenum{Iwasawa2004,deMarco2009}) and they are of great diagnostic power for tracing the inner turbulent flow of the disk in the strong gravity environment. General Relativity makes specific predictions for the arc forms and from a fit of these features one can derive SMBH mass and spin as well as the disk view inclination.

\subsubsection{Accretion physics}
\label{s:BHB}

Outbursts from Galactic stellar-mass black holes and neutron stars span orders of magnitude in mass accretion rate, and evolve over days, weeks, and months (Ref. \citenum{RemillardMcClintock2006,Dunn2010,ReynoldsMiller2013}). In contrast, the same dynamic range in AGN is not accessible on human timescales. The high flux observed from Galactic sources ensures very high sensitivity in the crucial Fe K band, wherein the most highly ionized gas --likely tied to the region closest to the black hole-- is observed. Discoveries made in the Fe K band in stellar-mass black holes help to direct and sharpen subsequent observations of AGN. In short, Galactic compact objects represent rapidly evolving, high-flux proxies that are vital to understanding the much larger classes of Seyfert galaxies and quasars. 

\begin{figure}
\begin{center}
\begin{tabular}{c}
\includegraphics[scale=0.3]{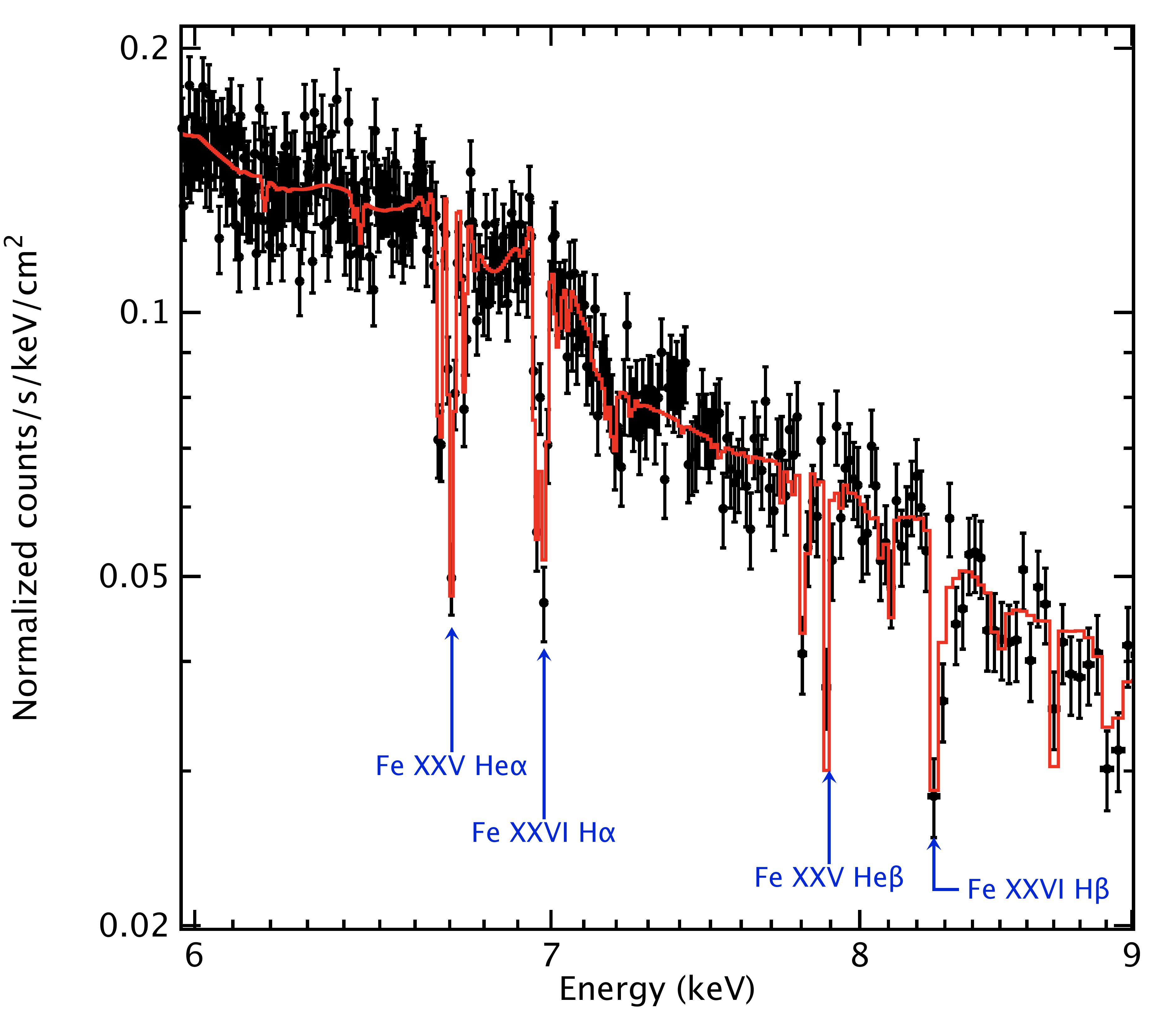}
\end{tabular}
\end{center}
\caption 
{ \label{fig:BHB}
X-IFU simulated observation lasting only 100 seconds of the Black Hole binary GRS1915$+$105. A disk wind, as reported in Ref.~\citenum{Miller2016} has been simulated. Strong spectral features can be clearly seen in the spectrum, including Fe\,\textsc{XXV} K$\alpha$ ($6.65-6.75$~keV), the resolved doublet Fe\,\textsc{XXVI} K$\alpha$ ($6.95-7.00$~keV), Fe\,\textsc{XXVI} K$\alpha$ shifted by $0.01$c ($7.00-7.05$~keV), Fe\,\textsc{XXVI} K$\alpha$ shifted by $0.03$c, Fe\,\textsc{XXV} K$\beta$ ($7.65-7.90$~keV) and Fe\,\textsc{XXVI} K$\beta$ ($8.20-8.35$~keV). This rich set of features will enable unprecedented studies of the structure of the disk winds.
} 
\end{figure} 

Theoretical studies demand that black holes must be fuelled by accretion disks wherein magnetic processes mediate the transfer of mass and angular momentum (Ref. \citenum{ShakuraSunyaev1973,BlandfordPayne1982,BalbusHawley1991}). In addition, the angular momentum can be transported throughout the accretion flow via the non-axisymmetric waves and shocks, related to the self-gravitating parts of disks in active galactic nuclei, or to the regions tidally excited by the companion star in black hole binaries (Ref.~ \citenum{FragileBlaes2008}).

For decades, one were only able to observe the effects of the fundamental disk physics (thermal spectra, jet ejection episodes), but unable to probe the underlying process that fuels black holes in Seyfert and quasar accretion modes. Very new observations may now indicate that disk winds can be used to constrain the emergent magnetic field of accretion disks (Ref.~\citenum{Miller2016}). This opens a long-awaited window on the fundamental physics of disk accretion and enables connections to numerical simulations. This is also a window on mechanical feedback from black holes, since feedback modes may depend on the disk magnetic field strength and configuration (Ref. \citenum{Livio2003,Begelman2015}). 

Equipped with the X-IFU, Athena will have the power to reveal the fundamental physics that drives accretion onto black holes. The keys are spectral resolution, and sensitivity. Whereas current telescopes can only offer initial constraints on disk physics in 100 \ksec, the X-IFU will be able to constrain magnetic fields via winds in just hundreds of seconds - the dynamical time scale of such winds. The evolution of the outflow properties (mass outflow rate, kinetic power) and disk fields will finally be accessible on their natural time scale (see Fig.~\ref{fig:BHB}). 

Jets may be the main agents of feedback from black holes into environments as large as galaxy clusters.
In both stellar-mass black holes and AGN, momentary dips in the X-ray flux have been associated with relativistic jet ejection events (Ref. \citenum{Belloni1997,Chatterjee2009,Chatterjee2011}). These dips last 100s of seconds in stellar-mass black holes. Whereas prior missions have only been able to observe continuum variations that do not reveal velocities or accelerations, X-IFU will be able to obtain sensitive line spectra as mass is transferred from the disk into the jet, providing a new and unparalleled view of the jet launching process.

For black holes that accrete at rates more than a few per cent of the Eddington limit, the inner region of the flow will be dominated by radiation rather than gas pressure (Ref.~\citenum{Lightman1974}). Observationally, two microquasars have shown spectacular X-ray variability on timescales of order of 50-100 s, that confirm the limit-cycle nature of the underlying process (Refs.~\citenum{Belloni2000, Altamirano2011}). Moreover, the observed interplay between the wind outflow launched in some states of the sources and such 'heartbeat' variations has shed some light on the plausible disk stabilizing force (Ref.~\citenum{Janiuk2015}). The magnetic fields, or, in general, the intrinsically stochastic nature of the turbulent dissipation must be an important agent here (Ref.~\citenum{Janiuk2012}). X-IFU observations of these and other targets will reveal to exquisite detail the structure of these winds and allow a detailed study of disk wind launching mechanisms and open a window into the study of fundamental disk physics. 

A remarkable case that links the stellar BH to the SMBH of AGN is the massive BH at the center of our own galaxy, Sgr A*. With its $4\times 10^6\, {\rm M_{\odot}}$ mass and consequently a time scale of $\sim 30$~min at the ISCO, its clean close environment due to the low accretion rate and the fact that it is relatively nearby, Sgr A* offers the opportunity to study the details of its steady accretion flow and the non-thermal flares that take place within a few Schwarzschild radii from the horizon. Even if challenging because of the confusion in the inner $5"$ of the galaxy, the X-IFU will allow to study the spectral lines of the Sgr A* accreting plasma (Ref.~\citenum{Wang2013}) providing insights in the physical conditions and dynamics of the flow. X-IFU will also be able to explore in detail the variable fluorescent spectral lines from the molecular clouds of the central zone which are reflecting radiation from ancient outbursts of the BH (Refs.~ \citenum{Ponti2013,Clavel2013}). With its excellent spectral-imaging performances the X-IFU will obtain Molecular Cloud line diagnostics unreachable to present instruments and crucial for the reconstruction of Sgr A* light curve in the past 1000 yr, probing the link between dormant SMBH and their past active phases.

\subsection{Observatory science}
\label{s:Obs}
While focusing on the Hot and Energetic Universe, the breakthrough capabilities (e.g., excellent spectral resolution combined with high throughput and fast timing capabilities) of the X-IFU will enable a wealth of new science investigations to be performed for a wide range of astrophysical sources of great interest to the broader astronomical community. This so-called observatory science covers very different topics and we highlight here only a few of them. We also highlight the potential of X-IFU observations in unforeseen discoveries thanks to the Athena fast Target of Opportunity observation capability, in particular in science triggered by new Gravitational Wave detectors.

\subsubsection{Solar system and planetary science}
\begin{figure}
\begin{center}
\begin{tabular}{c}
\includegraphics[scale=0.3]{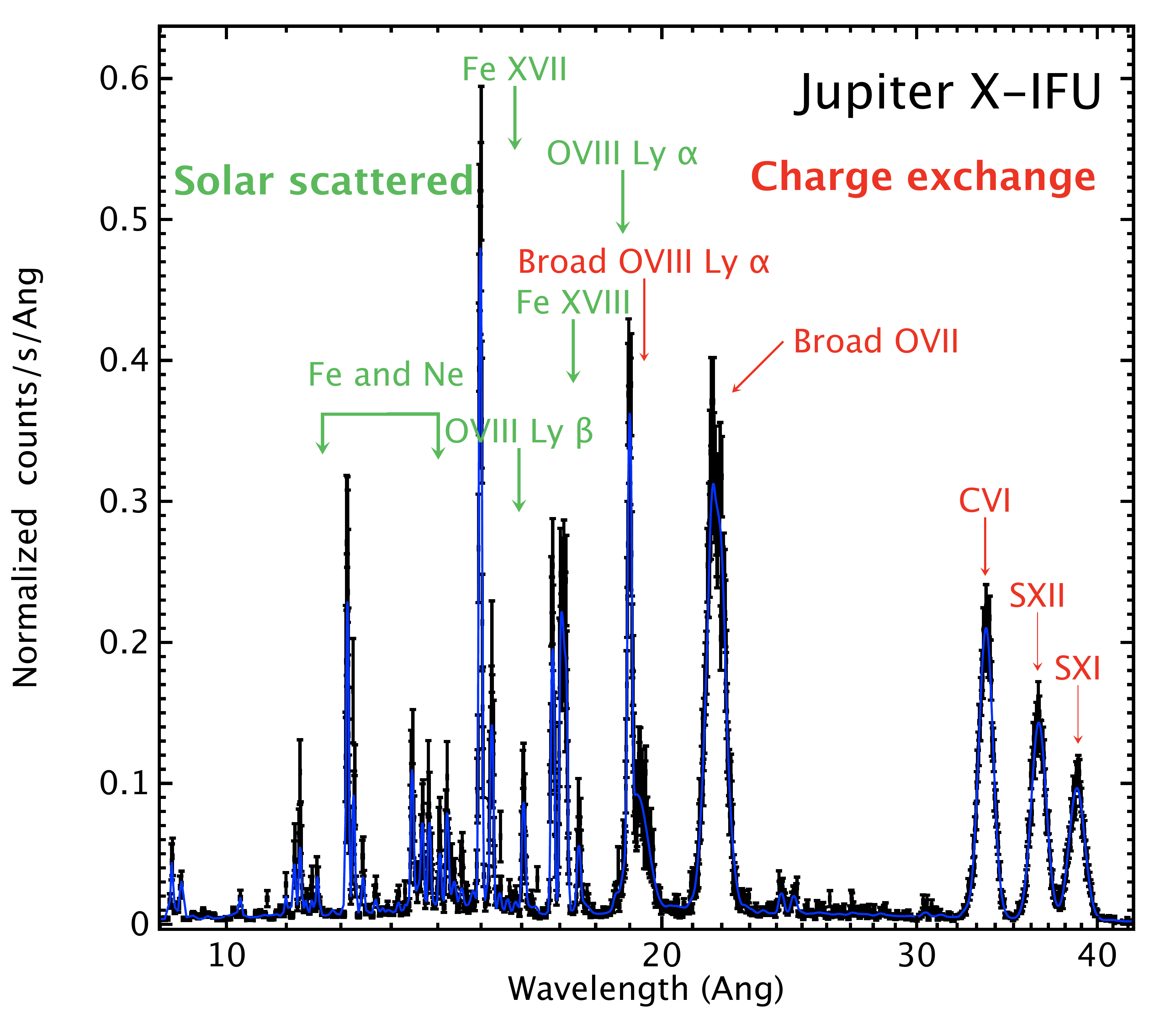}
\end{tabular}
\end{center}
\caption 
{ \label{fig:SWCX}
Simulated Jupiter's spectrum for a 20 \ksec\ Athena X-IFU observation, showing clearly the emission lines produced by charge exchange between solar wind  particles and Jupiter's atmosphere.  
} 
\end{figure}

Athena investigations of the solar system will answer questions still open following the pioneering work carried out with Chandra and XMM--Newton, and will add enormously to our understanding of the interactions of the solar wind with planetary bodies, and between space plasmas and magnetic fields. The X-IFU will determine the species, and thus the origin (solar wind or Io's volcanoes), of the ions responsible for Jupiter soft X-ray aurora, and will test theories of ion acceleration in the planet magnetosphere through line broadening velocity measurements (see Fig.~\ref{fig:SWCX}). High sensitivity observations of X-ray fluorescence from the Galilean moons will allow surface composition measurements, and studies of the Io plasma torus will shed light on the yet unknown mechanisms energizing its X-ray emission. A search for X-ray aurorae on Saturn with the X-IFU will reach much greater depth than possible so far. The X-IFU will spectroscopically map Mars extended exosphere through differing solar wind conditions and seasons, as well as the very extended comae of comets transiting in the Sun neighbourhood.

The X-IFU will drastically improve our knowledge of the consequences of X-ray incidence on exoplanets, a crucial element in order to understand the effects of atmospheric mass loss and, more generally, of the chemical and physical evolution of planet atmospheres especially in the early evolutionary stages.  In a few selected nearby known planetary systems hosting hot Jupiters,  X-IFU will search for ingress/eclipse/egress effects during planetary orbits. In a wider sample of planetary systems X-IFU can confirm/improve the statistical evidence of Star-Planet Interactions and search for those variability features that are imprints of such interactions. Athena may also discover unexpected spectral signatures (and their orbital modulation) of planetary atmospheres due to the host stars high energy radiation and particle emission. 

\subsubsection{Massive stars}
\label{s:massivestars}

The strong stellar winds of massive stars make them key players in feedback processes within galaxies, whatever their redshift. However, large discrepancies remain in the evaluation of the wind properties, impeding a proper understanding of massive star evolution and feedback. X-rays form inside the stellar winds, so they are a sensitive probe of their physical properties. Time resolved high-resolution spectroscopy, collected by X-IFU, will thus provide major breakthroughs in this field. In single stars, small-scale structures directly linked to the degree of wind inhomogeneity will be probed by examining short-term variations as well as line profile shapes in a large sample of objects, which is not within reach of current X-ray spectrometers. In addition, detailed Doppler mapping will pinpoint the properties of larger-scale structures, notably linked to magnetic confinement, pulsational activity, or co-rotating features, whose presence could just be hinted at in the best datasets currently at hand. In massive binary systems, where the two stellar winds collide, precise line profile studies, in particular of the Fe-K line complex at $6.7$~keV, will for the first time be performed, yielding the immediate post-shock conditions in the wind interaction zone while Doppler tomography will directly map it. X-IFU will thus dramatically increase our understanding of the plasma physics in the most important stellar feedback drivers.

\subsubsection{Cool stars}
\label{s:coolstars}

High-energy irradiation of circumstellar disks during star formation and early stellar evolution are crucial for disk evolution and, eventually, resulting planetary system formation (Ref. \citenum{Glassgold2000}). The radiation associated with magnetospheric accretion onto CTTS (Classical T Tauri Stars) and young brown Dwarfs is believed to originate in localized structures (accretion streams and hot spots). Therefore, the observed emission changes throughout the stellar rotation cycle and the X-IFU will probe the X-ray line emission from the heated plasma in shocks forming upon impact of the accreting matter on the stellar surface, while accompanying simultaneous optical studies will monitor the line emissions produced in magnetic accretion channels. The viewing geometry crucially determines what we see in X-rays (cf. TW Hya Ref., \citenum{Brickhouse2010} and V2129 Oph, Ref. \citenum{Argiroffi2011}) and the detailed mapping of the accretion geometry, therefore, requires simultaneous optical and X-ray monitoring for at least one rotation cycle, i.e., typically a few days. Spectro-polarimetric optical monitoring (e.g., with E-ELT/HIRES) simultaneous (e.g for at least one night) with X-IFU observations provides both time-resolved emission line fluxes tracing accretion columns and maps of the large-scale magnetic field structure.  With X-IFU for the brightest ($F_X\sim 3\times 10^{-13}\, {\rm erg}\, {\rm cm}^{-2}\, {\rm s}{^-1}$) CTTSs high resolution time-resolved spectroscopy down to 3 \ksec\ will allow us i) to explore the variability of the accretion process (on predicted time scales of hours) and/or the modulation due to accretion stream shadowing, ii) to constrain with Doppler line shifts down to $100-400\, {\rm km}\, {\rm s}^{-1}$ the bulk velocity of accreting material, iii) to investigate from simultaneous observations of many density sensitive triplets the controversial issue of density stratification of accreting material, and iv) to address the open issue of the excitation mechanism of the Fe K$\alpha$ 6.4 line emitted from the circumstellar disk (Ref. \citenum{Giardino2007}). 

Strong flares (peak $L_\mathrm{X}\sim 10^{32}\, {\rm erg}\, {\rm s}^{-1}$, peak $T\sim  2\times 10^8\, {\rm K}$) on young active stars and the likely associated  huge  Coronal Mass Ejection are likely a major perturbation source either of circumstellar disk evolution and/or of the early planetary system formation (Ref. \citenum{OstenWolk2015,SanzForcada2011,SanzForcada2014}).  The X-IFU will allow us to investigate the initial phase of the intense flares including mass motions as well as their influence on circumstellar disks and/or early planetary evolution.

Structures on the surface of late-type stars such as dark star spots or bright plages can be expected to produce corresponding signatures in the outermost atmosphere, the corona, which is traced by X-ray emission. Yet, a clear rotational modulation of the X-ray luminosity has been observed only in few cases (eg., the M star AB Dor, Ref. \citenum{Hussain2005}). Joint X-IFU and optical spectro-polarimetric monitoring (e.g., with E-ELT/HIRES) will yield unique constraints on the magnetic field structure in active stars, by probing different temperature regimes separately and, consequently, different structures on the stellar surface.

\subsubsection{Supernova Remnants}
\label{s:SNR}
% New text by Anne

The  X-IFU will provide the spatially-resolved spectral capabilities that have long been wished for in the study of supernova remnants (SNRs) (See Ref. \citenum{decourchelle_sp} and references therein). Like clusters of galaxies, SNRs are extended sources, ranging in angular scale from less than 0.5 arc minutes (in external galaxies) to, in extreme cases, several degrees in our Galaxy. X-IFU will be able to spatially resolve the spectra of SNRs with unprecedented spectral resolution. This is of great importance to obtain information on the supernova event and explosion mechanism by providing detailed abundance ratios for all elements with Z=6-28 (carbon to nickel) and new insights into the dynamics of the explosion by mapping Doppler shifts/broadening of the ejecta in young SNRs, as well as for understanding the physics of the hot non-equilibrium plasmas in SNRs, its evolution and impact on the interstellar medium.

To start with the abundance patterns in SNRs, it should be noted that SNRs originate from two different types of explosions: core collapse supernovae (spectroscopy classes Type II, Ib/c) and thermonuclear supernovae (Type Ia). There are still major uncertainties regarding these explosions. Core collapse supernovae are caused by the implosion of the cores of massive stars, leading to the formation of a neutron star or black hole. How the energy liberated leads to the ejection of the outer envelope is very uncertain, issues being the roles of neutrinos, instabilities, rotation and magnetic field. Thermonuclear supernovae are caused by CO white dwarf which explodes if its masses approaches the Chandrasekhar limit ($1.4\, M_{\odot}$). Here a main issue is the supernova progenitor system, whether the evolution towards this limit is caused by accretion from a normal stellar companion, or due to merging of two white dwarfs. For both types of supernovae much can be learned from the resulting nucleosynthesis products. With the current instruments the more abundant even elements have been reasonably well observed in young SNRs, but it is only with the X-IFU that their 3D distribution can be mapped  through line doppler shift and broadening measurements, and their spectral properties along the line of sight determined. This is crucial information for the models of Supernova explosions. Indeed the level of asymmetry of the ejecta  is closely related to the explosion mechanism itself, leading to relatively symmetrical explosion for Type Ia compared to core collapse events, and leading to different nucleosynthesis yields.

Different explosion mechanisms leave also distinct patterns in the abundance of the odd elements, like F, Na, Al, P, Sc, Ti, Cr and Mn. These rare elements reveal in core collapse how much the exposure was to intense neutrino radiation, and for thermonuclear supernovae it will inform us about the initial pre-main sequence abundances of the progenitor, but also deviations from nuclear statistical equilibrium reactions, which can be used to distinguish among different types of explosions. X-IFU will be able to detect these weaker lines, among the wealth of lines from more abundant elements and to map their spatial distribution, expected to be in different regions of the SNR, often grouped according to the layer of the supernova in which they were synthesized. 

As for the physics of SNRs, the hot plasmas in SNRs are often out of both ionisation equilibrium and thermal equilibrium, as the plasmas often did not have time enough to relax to equilibrium. Ionisation non-equilibrium in young SNRs results in line emission from lower ionisation species than one would expect given the temperature. Low ionisation species can also be enhanced by dust sputtering, releasing new atoms into the hot plasma from broken up dust particles. Interestingly, some older SNRs have plasmas for which the ionisation degree is higher than expected given the electron temperatures. This could arise by rapid adiabatic cooling of the electrons, whereas the recombination rate is too low to keep up with the cooling. With X-IFU, this can be studied in detail, for individual ions, but also as a function of location and will lead to new insights on the progenitor properties (stellar wind, shell) and interstellar environment.

Temperature non-equilibrium means that the electron temperature can be cooler than the ion temperature. This has important consequences as the electron temperature is easier to measure, but the internal energy is dominated by the ions. With X-IFU the ion temperatures can be measured through the thermal line broadening at the edges of some SNRs, as in the center line broadening will be dominated by different velocities from different parts of the shell. The ion temperatures are also interesting as efficient cosmic-ray acceleration by SNRs shocks may lead to lower ion temperatures (as the internal energy will now be divided between hot plasma and cosmic rays). On the topic of cosmic-ray acceleration, some relatively young SNRs are dominated by X-ray synchrotron radiation. This radiation itself does not need X-IFU to characterise its spectrum, however, X-IFU is important to find signatures of thermal emission, which can be more easily picked up by high resolution spectroscopy. The thermal emission will be used to estimate the plasma densities, which is notably needed to model the gamma-ray emission from these young SNRs

\subsubsection{Discovery science through fast ToOs}
\label{s:GW}

Athena will have a fast Target of Opportunity (ToO) observation capability enabling observations of transient phenomena within hours of the trigger. Transients will be very much at the focus of astrophysics by the late 2020s, thanks to facilities like the LSST in the optical, SKA in radio or the all-sky Gamma-ray monitors like SVOM or others. X-IFU observations of some of these triggers will reveal critical astrophysical information on these sources (e.g., on high-z GRBs as explained in Sec.~\ref{s:GRB}), extending into the high-energy domain the observations at longer wavelengths.

Likewise, at the end of the next decade, one may expect that gravitational wave sources will be located to an accuracy that will enable follow-up observations with Athena. Coalescing compact objects in binaries (involving at least a neutron star, with a stellar mass black hole or another neutron star) are expected, during their runaway orbital decay due to gravitational radiation, to produce electromagnetic radiation related to the energetic outflows generated. A relativistic jet may form and produce a short gamma-ray burst, followed by an X-ray to radio afterglow lasting from hours to days. So far no such X-ray counterpart has been found, despite extensive searches (Ref. \citenum{abbott_2016} for a review of the follow-up observations of GW150914).  Likewise for high redshift GRBs, X-IFU observations of the X-ray counterparts of compact binary gravitational wave sources would shed light on the nature and properties of their progenitors, their energy output in electromagnetic form,  as well as on the properties of their host galaxies and circumstellar environments (see Ref. \citenum{berger_2014} for a review).

\section{The X-IFU instrument}
\label{current_instrument_configuration}
\subsection{Top level X-IFU performance specifications}
To meet the scientific objectives described above (Section \ref{sec:drivers}), the derived X-IFU performance requirements are listed in Table \ref{tab_performance}.
\begin{table}[!h]
\begin{center}       
\begin{tabular}{|l|l|}
\hline
Parameters &	Requirements \\
\hline
Energy range &		$0.2-12$ keV \\
Energy resolution$^{1)}$: E $<$ 7 keV &		2.5 eV \\
Energy resolution: E $>$ 7 keV &		E/$\rm \Delta E$ = 2800\\
Field of View	 &	5$'$ (equivalent diameter) \\
Effective area @ 0.3 keV &		$>1500$ cm$^2$\\
Effective area @ 1.0 keV &		$>15000$ cm$^2$\\
Effective area @ 7.0 keV &		$>1600$ cm$^2$\\
Gain calibration error (RMS, 7 keV)	 &	0.4 eV \\
Count rate capability $-$ nominally bright point sources$^{2)}$	 &	1 mCrab ($>80$\% high-resolution events)\\
Count rate capability $-$ brightest point sources	 &	1 Crab ($>30$\% throughput)\\
Time resolution	 &	10 $\mu$s\\
Non X-ray background (2-10 keV)	 &	$< 5\times 10^{-3}$ counts/s/cm$^2$/keV (80\% of the time)\\
\hline
\end{tabular}
\end{center}
\caption{Baseline X-IFU top level performance requirements.$^{1)}$ The goal energy resolution is 1.5 eV. $^{2)}$ The goal point source count rate capability is 10 mCrab ($>80$\% high-resolution events).}
\label{tab_performance}
\end{table}
Those performance requirements can be achieved with a large format array of actively cooled X-ray absorbers thermally coupled to Transition Edge Sensors (TES) (3840 TES with a 249 $\mu$m absorber pitch) operating at $\sim 90$ mK, routinely calibrated, and  shielded by an active cryogenic anti-coincidence system. 
%
% The X-IFU instrument 
%
\subsection{Functional block diagram}
\begin{figure}[!t]
\begin{center}
\begin{tabular}{c}
\includegraphics[scale=0.125,clip=true]{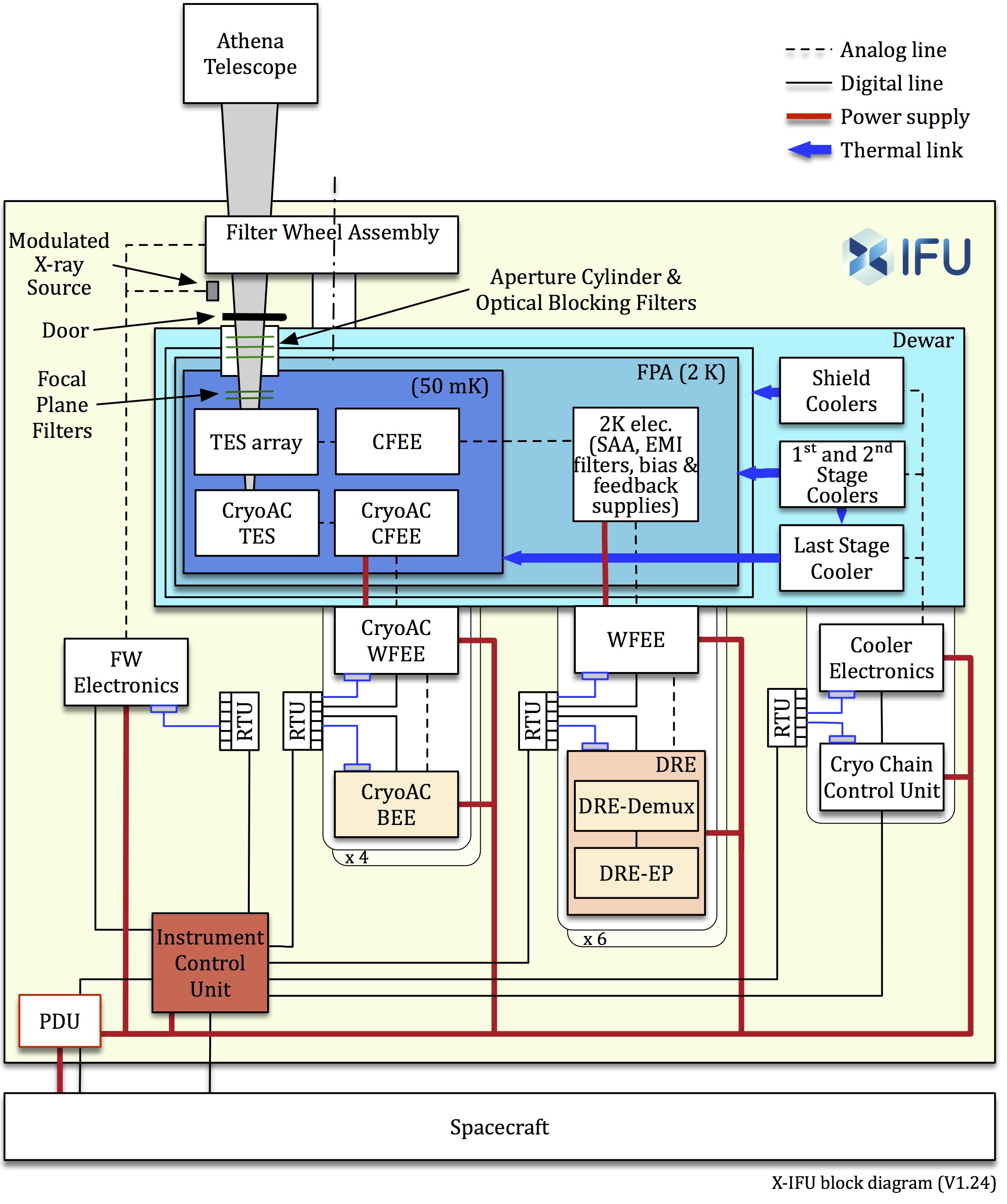} 
\end{tabular}
\caption{\label{fig:db_spie_2016_fbd}The X-IFU functional diagram, showing the main components of the instrument. FPA stands for Focal Plane Assembly, FW for Filter Wheel, CFEE for Cold Front End Electronics, WFEE for Warm Front End Electronics, DRE for Digital Readout Electronics, Demux for Demultiplexing, EP for Event Processor, CryoAC-BEE for Cryo-AC Back-End Electronics, RTU for Remote Terminal Unit, ICU for Instrument Control Unit, PDU for Power Distribution Unit. }
\end{center}
\end{figure}

The X-IFU functional block diagram is shown in Figure \ref{fig:db_spie_2016_fbd}. Here we will briefly describe each component of the diagram, starting from the top left.

\subsubsection{Filter wheel assembly} The filter wheel assembly and control electronics are described in Ref. \citenum{bozzo_spie_2016}. The filter wheel may include up to eight different filter positions, with the aim of providing: 1) Protection of the X-IFU focal plane detectors against micro meteorites and contamination; 2) Reduction of the optical load from stars and solar system bodies to limit the energy resolution degradation; 3) Possibility to optimize the throughput in case of bright X-ray targets; 4) Capability to separate the sky contributions from the intrinsic instrumental background by offering on-off measurements with a closed position; 5) Provision of a radioactive calibration source as a backup to the Modulated X-ray Sources (MXS).

%Pending on a number of trade-off studies, the filter wheel may include:
%\begin{enumerate}
%\item One closed position (Molybdenum filter), to be used during launch for the protection of the focal plan detectors and in-flight for the evaluation of the internal detectors background (the filter will be opaque to X-rays from the telescope).
%\item One open position (no filter) that can be used for the observations of faint X-ray sources where no contamination from bright optical targets is expected. 
%\item Two Beryllium filters of different thickness that are able to suppress to different extents the X-ray fluxes of celestial sources at energies $< 3$ keV, where the bulk of the X-ray photons are usually emitted. These two filters are suitable for particularly soft and bright X-ray sources that are characterized by fluxes as high as $\sim 50$ mCrab and for which the nominal spectral resolution should be preserved (the brightness limit is still being assessed by proper simulations).
%\item A single neutral density blocking filter that can be used to induce an overall decrease of the rate of X-ray photons coming from particularly bright sources onto the X-IFU focal plane by a factor that could be as high as 100. This filter may allow  observations of X-ray sources with fluxes of 100 mCrab or brighter.
%\item Two optical blocking filters that are needed to limit the optical load (and thus the degradation of the instrument energy resolution) from the bright UV/V counterparts of the X-ray sources.
%\end{enumerate}
\subsubsection{Modulated X-ray Sources} Due to the high sensitivity of TES to their environment, changes in the energy scale throughout an observation are expected. These deviations shall be regularly corrected using on-board modulated X-ray sources (MXS) illuminating the TES array with X-rays emitted at known energies. Modulated X-ray sources similar to the one developed for Hitomi SXS are baselined for X-IFU. The MXS can produce short pulses of X-rays (tens of $\mu$s) at multiple energies (e.g. 5.4 keV and 8.0 keV). Its intensity and duty factor will be tuned to give a sufficient number of high spectral resolution events over the typical duration of gain drifts in the electronics (10 minutes). By selection of the appropriate on and off times, the number of events per read-out channel can be tuned to reduce the effect of cross talk (see Ref. \citenum{denhartog_spie_2016} for a description of the various cross-talks components expected). The MXS will be driven by the same electronics as the filter wheel.
%In a MXS, a LED illuminates a photocathode, releasing electrons. Those electrons are then accelerated by an electrical field to hit the anode (a Be window) coated with relevant materials (Cu and Cr). Fluorescent X-ray photons are thus produced in a non-collimated way. 
\subsubsection{Dewar Door} The cryostat is expected to be launched in vacuum at ambient temperature and this requires a door to close its entrance window at the interface with the aperture cylinder. Based on ASTRO-H/SXS heritage, which was able to observe with the gate valve closed, the X-IFU Dewar door should be transparent to X-rays above a few keV. 
\subsubsection{Thermal filters and aperture cylinder}
Thermal Filters (TF) for the X-IFU are described in Ref. \citenum{barbera_2016_LTD} (see also Ref. \citenum{sciortino_spie_2016} for a report on their on-going characterization). TF should maximize the soft X-ray transmission, while providing attenuation of the visible/infrared/UV radiation from the warm surfaces seen from the TES array as to limit the radiation heat load on the cold stage and prevent the degradation of the energy resolution due to  photon shot noise. In addition TF must also provide radio-frequency attenuation, as TES and associated readout are very sensitive to electromagnetic perturbations. Finally TF should also protect the detector from molecular contamination, which would lead to a loss of effective area at soft energies (de-icing of the outer filter has to be considered in the design).

The thermal filter baseline design presently implemented in the X-IFU response matrix (based on the Athena original proposal) consists of five filters for a total of 280 nm Polyimide, 210 nm Aluminum and a Polyimide mesh 10 \micron\ thick 93\% open area on the two larger filters. The metallic aluminum blocks the visible and infrared radiation mainly by reflection and the Polyimide blocks the UV radiation by absorption. Such filter design does not allow to fully achieve the effective area requirements at low energies (E $< 1$ keV) reported in Table \ref{tab_performance} (see Section \ref{s:effective_area}). Section \ref{s:low_energy_response}  discusses possible solutions under study to improve the low energy response.

The aperture cylinder holds the TF located at the interface of the thermal and structural shields of the cryostat. 

%
%The optical load from the bright UV/VIS counterparts of X-ray sources such as massive stars, AGN's in outburst, Mars, etc., can degrade the XIFU energy resolution by additional photon shot noise. Two additional filters will be placed on the filter wheel: the thin FW optical blocking filter (2000 \AA~polyimide + 400 \AA~aluminum)  allows to observe hot stars fainter than m$_V$ $\sim 8$ with no significant degradation of energy resolution, while the use of the thick filter wheel optical blocking filter (2000 \AA~polyimide + 800 \AA~aluminum) allows to observe hot stars as bright as m$_V$ = 2.
\subsubsection{The focal plane assembly (FPA)}
The FPA is presented in detail in Ref. \citenum{jackson_spie_2016}. The FPA includes all of the cryogenic components of the X-IFU detector mounted on the (nominally) 2 K stage of the cryostat, including the main TES array (Ref. \citenum{smith_spie_2016}) and its cold front end  electronics (CFEE) and the active cryogenic anti-coincidence TES (CryoAC TES) and its cold front end electronics (CryoAC CFEE). The CryoAC enables to flag on the ground and remove events corresponding to background particles depositing energy in both TES arrays (see Ref. \citenum{macculi_spie_2016,lotti_spie_2016}). The TES and their SQUID readout electronics are sensitive to disturbances from their environment. As such, the FPA includes magnetic and straylight shielding, EMC shielding and filtering to allow the TES to reach their required performance within the expected environment of the cryostat, during both ground tests and in-flight operations. The inner two filters of the aperture cylinder are related to the FPA. The FPA also contains a thermal suspension system to insulate the TES, operating at a bath temperature of $\sim 50$ mK, from the FPA 2 K mounting interface in the cryostat.
\def\micron{$\mu $m}
\subsubsection{The TES array} The core of the focal plane assembly is a large-format Transition Edge Sensor (TES) microcalorimeter array. In its baseline configuration (LPA 1 configuration in Figure \ref{fig:array_configurations}), the TES array contains 3840 single size pixels (249 \micron\ pitch) combining a TES thermistor thermally coupled to a Au/Bi X-ray absorber, with the combination of thermistor and absorber thermally isolated from the thermal bath of the silicon wafer frame by a thin (1 \micron) SiN membrane (see Ref. \citenum{smith_spie_2016} for details).  It has been designed to meet the instrument energy resolution, count-rate, and field-of-view requirements in a single pixel design. Operating at an assumed bath temperature of $\sim 50$ mK with a MoAu bilayer transition temperature of $\sim 90 $ mK, the raw pixel energy resolution for energies less than 7 keV is $\sim 2.0$ eV, leading to a total instrument energy resolution of 2.5 eV, once noise contributions from the readout electronics, disturbances from the environment (FPA, cooling chain), a finite event record length, gain calibration uncertainty  are all taken into account in the spectral resolution budget of the X-IFU.

The Au/Bi absorbers form a filled-array to provide a high quantum efficiency, with the combined thickness of the Au and Bi layers defined to produce the required stopping power for 7 keV X-rays, and the relative thicknesses of the Au and Bi layers adjusted to provide the required heat-capacity and thermal conductivity to the TES thermistors.
\subsubsection{Cold front end electronics and 2 K electronics}The main TES array is operated using frequency domain multiplexed SQUID readout electronics. SQUID amplifiers are used to provide low-noise, low-dissipation readout of the low-impedance TES pixels. An individual SQUID amplifier chain is used to operate (nominally) 40 pixels in a frequency division multiplexed readout scheme. In this scheme, a comb of 40 frequency carriers in the 1-5 MHz range is generated in warm electronics and pass into the cryostat via a single twisted-pair. This comb is divided over 40 pixels by an array of 40 resonant superconducting LC filters, such that each pixel is voltage biased by a single AC carrier. The bias (signal) currents of those 40 pixels are summed at the output of the LC filter chip, which is connected to the input of a SQUID amplifier chain that combines a low-dissipation first-stage SQUID and a high output-power SQUID amplifier array at 2 K. The 2 K electronics assembly will also combine bias dividers and filter networks for all electric lines entering the FPA.

%Impedance-matching of the  pixels, LC filters, and first-stage SQUID amplifier is performed using a combination of the electrical designs of the LC filter, the input circuit of the SQUID amplifier, the normal-state resistance of the TES thermistor, and an impedance transformer between the  pixel and the LC filter chip. 
\subsubsection{Cryo-AC and associated electronics}The CryoAC detector is a 4-pixel TES array whose physically large pixels are over-dimensioned relative to the active area of the sensor array to intercept the majority of the primary cosmic rays passing through the sensor array, including events at large angles of incidence (Ref. \citenum{macculi_spie_2016}). The CryoAC pixels are micromachined out of a silicon wafer to produce physically large pixels with a substantial stopping power for minimum ionizing particles (MIPs), thermally isolated from their frame by thin Si support legs. The energy deposited in a CryoAC pixel is detected by a parallel array of $\sim 100$ TES thermistors patterned on one surface of the CryoAC detector. The energy deposition of a MIP event allows high signal-to-noise detections by means of an appropriate TES covering of the large-area silicon absorber pixels (Ref. \citenum{macculi_spie_2016}). The 4 CryoAC pixels will be read-out using 4 DC-biased SQUID amplifiers operating at $\sim 50$ mK.
\subsubsection{The Dewar assembly}
The Dewar assembly consists of a thermo-mechanical assembly providing support structure, thermal insulation, vacuum tight enclosure and entrance baffle to the FPA, between room temperature and FPA mounting interface temperature (2K). It includes external supports, outer vessel, door/valve, radiative shields, cryogenic supports and structures, thermal hardware (thermal insulation, thermal straps) and the aperture cylinder. The Dewar assembly includes all active coolers and their control electronics as well as all cryogenic harnesses. The FPA and WFEE are also included in the Dewar assembly and must also provide EMC/EMI shielding and filtering.
\subsubsection{The reference X-IFU cyrogenic chain}
\begin{figure}[!t]
\begin{center}
\begin{tabular}{c}
\includegraphics[scale=0.35,clip=true]{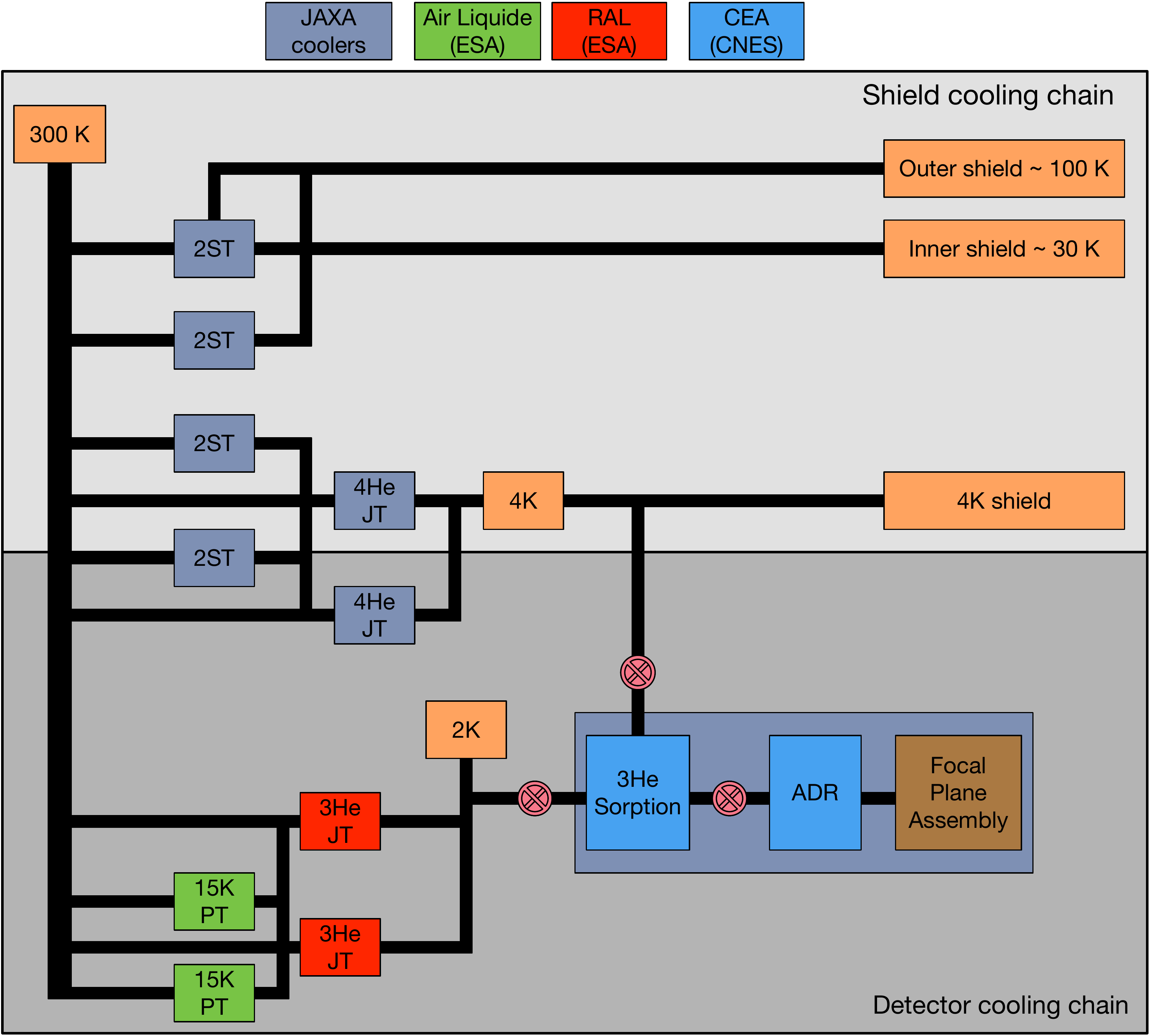} 
\end{tabular}
\caption{\label{fig:db_spie_2016_cyrochain}The reference cryogenic chain configuration for the X-IFU, split in two parts: the shield cooling chain and the detector cooling chain. The Dewar assembly, including the full cyrogenic chain, is currently under optimization, with the objective of consolidating its thermal and mass budgets.}
\end{center}
\end{figure}

The X-IFU cyrogenic chain, and its potential evolution is presented in Ref. \citenum{charles_spie_2016}, based on an initial configuration derived from the Athena Concurrent Design Facility study of Athena. To satisfy the 5 year mission lifetime requirement (10 years as a goal), the X-IFU cryogenic chain is based only on mechanical coolers. The main assumptions used are 1) the use of a 4K stage to damp some of the parasitic losses (harnesses, straps) that should have reached directly the 2 K stage and to operate the subK cooler, 2) a fully redundant chain, accepting only the sub-K cooler as a single point of failure. As shown in Figure \ref{fig:db_spie_2016_cyrochain}, the X-IFU cooling chain is divided in two parts: a shield cooling chain and a detector cooling chain. Its components are 1) 2 two-stage Stirling coolers provided by JAXA are foreseen to cool the inner and outer shields, 2) a 4 K Joule-Thomson, pre-cooled by 3 two stage Stirling coolers, both provided by JAXA, 3) 2 K Joule-Thomson cooler developed by RAL, with a pre-cooling provided by a 15 K pulse Tube developed by Air Liquide, 4) and finally a hybrid Sorption He3-Adiabatic Demagnetisation Refrigerator (ADR) developed by CEA-SBT to reach the $\sim 50$ mK bath temperature.  As discussed in Ref. \citenum{charles_spie_2016}, further optimization of the cooling chain is required (e.g. number of coolers), in particular in view of consolidating the thermal budgets, and reducing the mass budget of the whole Dewar assembly. Moving the first stage SQUIDs at $\sim 300$ mK, to reduce the heat load at the $\sim 50$ mK stage is an option that will be studied, assuming that it is demonstrated as not impacting the spectral resolution budget.

\subsubsection{Warm Front End Electronics (WFEE)}
The presentation of the WFEE for the X-IFU is addressed in Ref. \citenum{prele_spie_2016}. The WFEE is composed of 96 channels. Each channel biases and adjusts the offset of the 2 SQUID stages. The last SQUID stage is amplified by the Low Noise Amplifier just before the Analogue to Digital Converter (ADC) of the DRE. 
\subsubsection{The Digital Readout electronics (DRE)}
%Moreover, the high dynamic range in the combined bias (signal) currents of 40  pixels, which would be much too large to be matched to a typical low-power SQUID amplifier, is compensated by the application of feedback at the input to the SQUID. A 40-pixel comb of feedback signals, generated at baseband in the warm electronics, is used to suppress the 40 carrier signals, plus a fraction of the signal from an X-ray event in one of the 40 pixels, allowing the use of a low-dissipation first-stage SQUID. The readout of the X-IFU FPA is multiplexed in frequency domain. With a frequency range of $\sim 1$ to 5 MHz and an AC-carrier separation of 100 kHz (40 pixels are multiplexed in a single readout channel and 96 channels are considered). 
%
In the CFEE the first stage SQUID needs to be linearized with a high gain feedback loop. For each readout chain the DRE-DEMUX computes the AC signal to voltage-bias the 40 TES. It digitizes the signal amplified by the CFEE and the WFEE. Then it de-modulates the digital signal to reconstruct the data stream of each pixel and it computes the feedback signal for the first stage SQUID. The DRE implements a so-called base-band feedback (BBFB) technique to ensure that the feedback signal carriers are properly phased with the TES signal carriers at the SQUID input (accounting for delays). After de-multiplexing, the pixels data streams are transmitted to the DRE Event Processor (DRE-EP), which includes two major functions: event triggering and pulse processing. The DRE-EP delivers the scientific data combing the arrival time and energy of each event. 

The instrument requirements for a 2.5 eV energy resolution at 7 keV, operation up to 12 keV, and a pixel speed of $436~\mu$s (given here as the electro-thermal decay constant corresponding to a critically damped time constant of $160~\mu$s , see Ref. \citenum{smith_spie_2016}) to reach 1 mCrab point source count rates  drive a large dynamic range requirement along the full detection chain. In Frequency Domain Multiplexing, the DAC components used to generate the AC carriers and the feedback signals must produce high-amplitude signals (sum of 40 AC carriers) with very low noise levels (minimizing the contribution to the energy resolution budget). A preliminary evaluation indicates that a DAC with a dynamic range of $\sim 161.5$ dB per unit frequency is needed. Assuming a sampling frequency of 20 MHz, this implies an effective number of bits of 15 (Ref. \citenum{denhartog_spie_2014}). A DAC with such requirements is under study, but appears very challenging. As such, it is essential to optimize the DAC requirements by considering tradeoffs in the detection chain. This includes: optimizing the intrinsic dynamic range of the sensor pixels by relaxing the pixel speed requirement; considering optimized application-specific DAC configurations, including the use of multiple DACs per frequency comb to optimize the dynamic range and power dissipation; locating the AC carriers on a frequency grid to suppress intermodulation products; and considering alternative readout schemes using Time Domain or Code Domain Multiplexing. The implications of all these tradeoffs at system level must also be considered.

\begin{figure}[!t]
\begin{center}
\begin{tabular}{c}
\includegraphics[scale=0.45,clip=true]{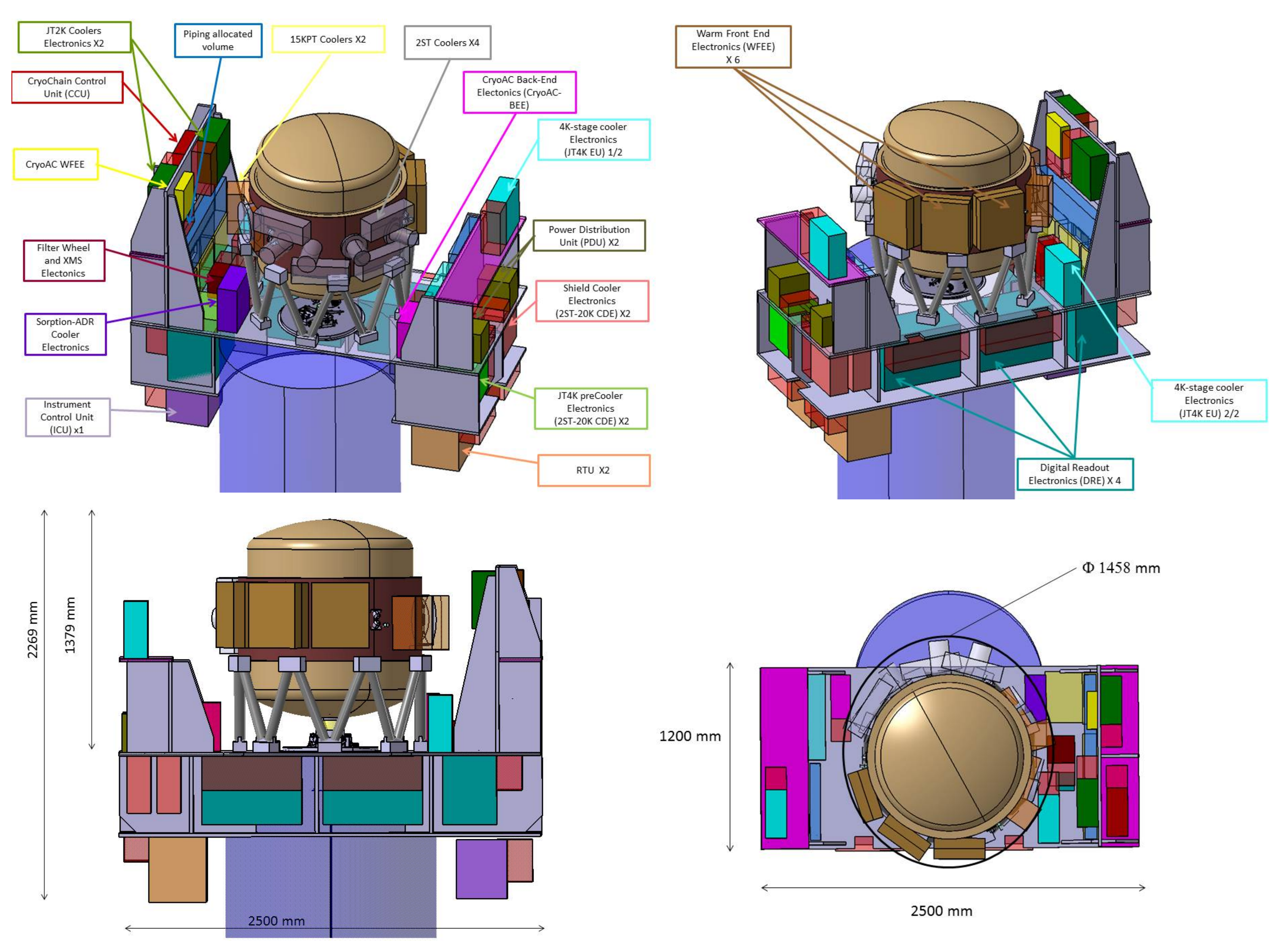} 
\end{tabular}
\caption{\label{fig:db_spie_2016_mechanical_design}The X-IFU mechanical design presented at the Mission Consolidation Review (May 2016, mid-Phase A). This accommodation does not take into account any potential optimization of the placement of the different electronic boxes on the focal plane module.}
\end{center}

\end{figure}

\subsubsection{Instrument Control Unit (ICU)}
The X-IFU Instrument Control Unit is presented in Ref. \citenum{Corcione_spie_2016}. The ICU functions include the telecommand \& telemetry( TC/TM) management with the spacecraft, science data formatting and transmission to the spacecraft mass memory, housekeeping data handling, time distribution for synchronous operations and the management of the X-IFU components (i.e. Coolers, Filter Wheel, Detector Readout Electronics Event Processor, Power Distribution Unit).
%Each of the two can manage the full instrument. 
%Only one section can be on at a given time. 
%The present ICU concept is based on the Euclid NISP unit, in which the two sections are mounted in the same box. This configuration is to be assessed and presented at the PDR. Each section is composed by 3 double eurocard PCB: a processor board, a power supply board (DC/DC converters), a data acquisition (HK) and controller board.
%The CPU in the baseline design is based on a Leon 2/3 Fault Tolerant processor, which at present has the required computing power, being the X-IFU subsystems controlled by independent electronics. In other terms, ICU is commanding the controlled units by sending TCs and receiving TM. Each unit, especially in the cryo-chain and detector/cryo AC, is providing internal controlling functions. 
\subsubsection{Remote Terminal Unit (RTUs)}
To better deal with the multiple instrument interfaces, to reduce harness lengths, mass and volume by reducing the distance between the I/O boards and the units, to decrease connections and connectors, Remote Terminal Units (RTUs) are proposed. A RTU is a flexible, miniaturized interface and control unit with support for various protocols and analogue/digital input and output control interface. TC/TM  and thermal regulation functions are integrated in the RTUs.
\subsection{The X-IFU mechanical design}
The X-IFU mechanical design as presented at the Mission Consolidation Review is shown in Figure \ref{fig:db_spie_2016_mechanical_design}. This design will evolve for two main reasons: First, since the overall mass of the X-IFU exceeds the mass allocation from the satellite, consolidation of the X-IFU mass budget is a priority. This will likely impact the design, most particularly the Dewar assembly (e.g. number of shields, cryostat compactness, number of coolers \ldots), and the digital electronics as being the two largest contributors to the mass budget. Second, the X-IFU accommodation will have to be optimized as being the driving component of the Athena focal plane module, hosting the two instruments. This will likely result in a less compact and integrated design of the instrument, with electronics boxes more distributed around the two instruments. Nevertheless, Figure \ref{fig:db_spie_2016_mechanical_design} provides a good view on the X-IFU system complexity and size. In terms of mass and power budgets, at the MCR, the X-IFU weighted 860 kg (margins included) and required up to $\sim 2.3$ kW (in observation and regeneration mode).

%current performance
%
%
\section{X-IFU Current performance}
\label{current_instrument_performance}
In this section, we will present the anticipated scientific performance  of the X-IFU, based on current design and knowledge. In table \ref{tab_dependences}, we list the various sub-system/components presented in Figure \ref{fig:db_spie_2016_fbd} that are involved in the main performance parameters of the X-IFU. 
\begin{table}[!h]
\begin{center}       
\begin{tabular}{|l|l|}
\hline
Parameters &	Component or subsystem \\
\hline
Energy range &		TES array, blocking filters, filter wheel\\
Energy resolution: E $<$ 7 keV &		TES array, blocking filters, CFEE and 2K electronics,\\
 &		 WFEE, DRE, cryogenic chain\\
Energy resolution: E $>$ 7 keV &		TES array, blocking filters, CFEE and 2K electronics,\\
 &		 WFEE, DRE, cryogenic chain\\
Field of View	 &	TES array, FPA, aperture cylinder, blocking filters,\\
 &		 dewar, MXS, filter wheel\\
Detector quantum efficiency @ 0.3, 1 keV & TES array, blocking filters\\
Detector quantum efficiency @ 7 keV	 & TES array, blocking filters\\
Gain error (RMS, 7 keV)	 & TES array, cryogenic chain, MXS\\
Count rate capability $-$ nominal bright sources	 &TES array, blocking filters, CFEE and 2K electronics,\\
 &		 WFEE, DRE\\
Count rate capability $-$ brightest sources	 & TES array, blocking filters, CFEE and 2K electronics,\\
 &		 WFEE, DRE\\
 Time resolution	 &	TES array, DRE, ICU\\
Non X-ray background	 &	Cryo-AC, FPA, dewar\\
\hline
\end{tabular}
\end{center}
\caption{X-IFU performance parameters and subsystem dependencies. The subsystems are presented in Figure \ref{fig:db_spie_2016_fbd}. }
\label{tab_dependences}
\end{table}

\subsection{Effective area}
\label{s:effective_area}
\begin{figure}[!ht]
\begin{center}
\begin{tabular}{c}
\includegraphics[scale=0.3]{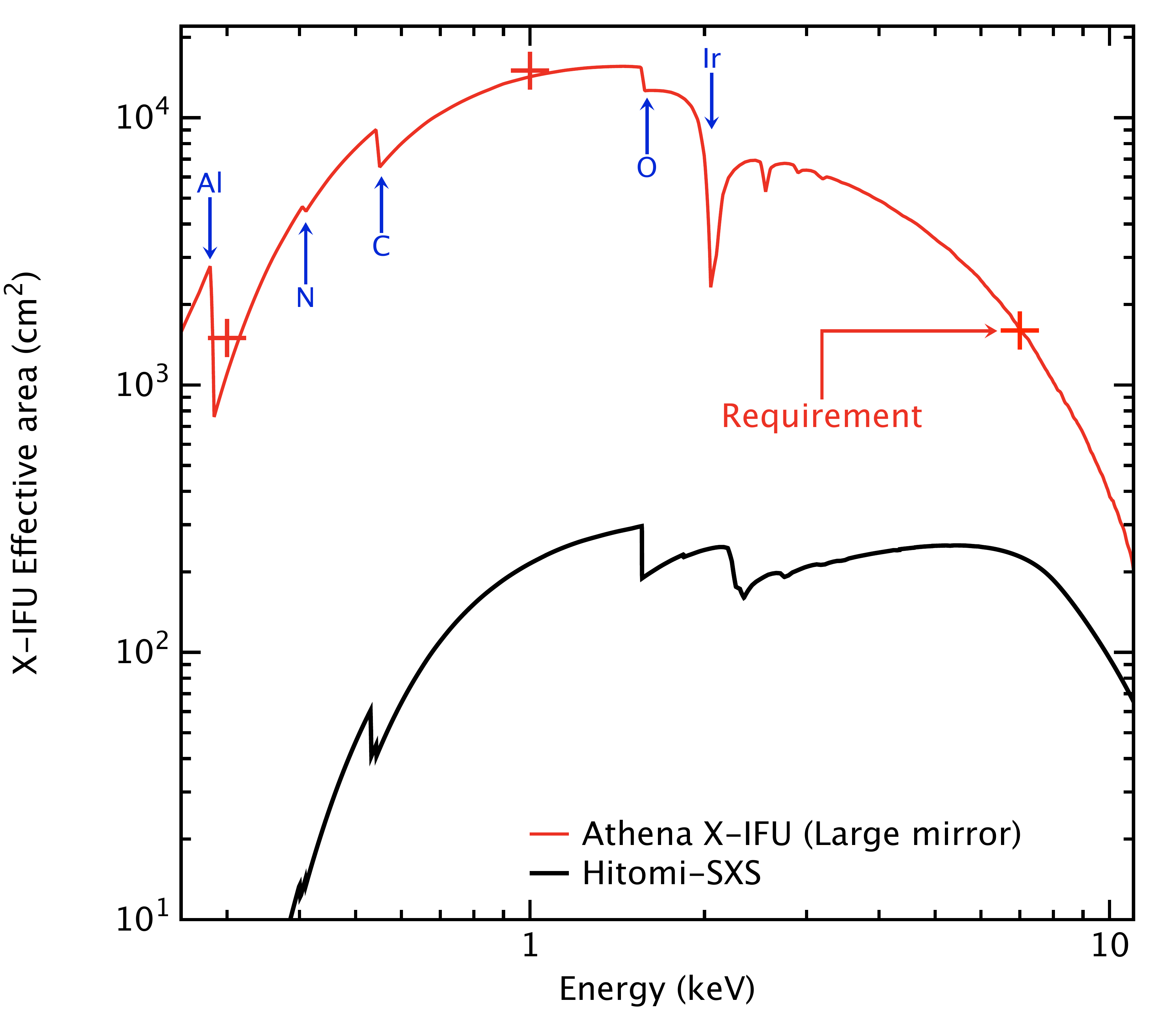}
\end{tabular}
\caption{\label{fig:effective_area}Effective area for the X-IFU compared to the effective area of Hitomi/SXS. The requirements to be met by X-IFU are indicated by a red cross in the case of the large mirror configuration to be studied. The elements causing the main edges in the instrument response (e.g. from the mirror) are indicated with vertical blue arrows.}
\end{center}

\end{figure}

In the current design, the effective area of the X-IFU is shown in Figure \ref{fig:effective_area}, and compared to that of  Hitomi/SXS. The effective area of the mirror assumed an outer mirror module radius of 1469 mm and a 2.3 mm rib spacing (the so-called as proposed configuration), and on-axis case. A blanket factor of 0.9 for contingency and manufacturing errors in the mirror is also assumed. For the TES response, a detector with 245 \micron\ $\times 245$ \micron\ absorbers and a gap distance of 4 \micron\ (249 \micron\ pitch) are assumed corresponding to a pixel filling factor of 0.97. The absorber composition is 1.7 \micron\  of Au covered by 4.2 \micron\ of Bi. The low-energy response is determined by the aperture cylinder and focal plane filters. The current baseline filter setup for the X-IFU consists of five filters with a total thickness of 280 nm Polyimide and 210 nm Al. For each filter, an outer layer of 10 nm is assumed to consist of Al$_2$O$_3$, leaving an effective absorber thickness of 160 nm Al and 50 nm of Al$_2$O$_3$. Furthermore, the two largest filters have a support mesh of 10 \micron\ thick Polyimide with an open area fraction of 93\%. As can be seen, the anticipated effective areas at energies below 1 keV are close (and actually just below) to the requirements and therefore consolidation of the scientific requirements and/or optimization of the mirror response and/or thermal filter properties should be pursued (see section \ref{s:low_energy_response}). A multi-layer coating specifically designed to increase the effective area of the mirror (at energies between $\sim 4-7$ keV) is currently being studied by ESA and if adopted, may be used to relax the requirements on the absorber properties (if needed). 
\subsection{Spectral resolution}
\begin{figure}[!ht]
\begin{center}
\begin{tabular}{c}
\includegraphics[scale=0.225]{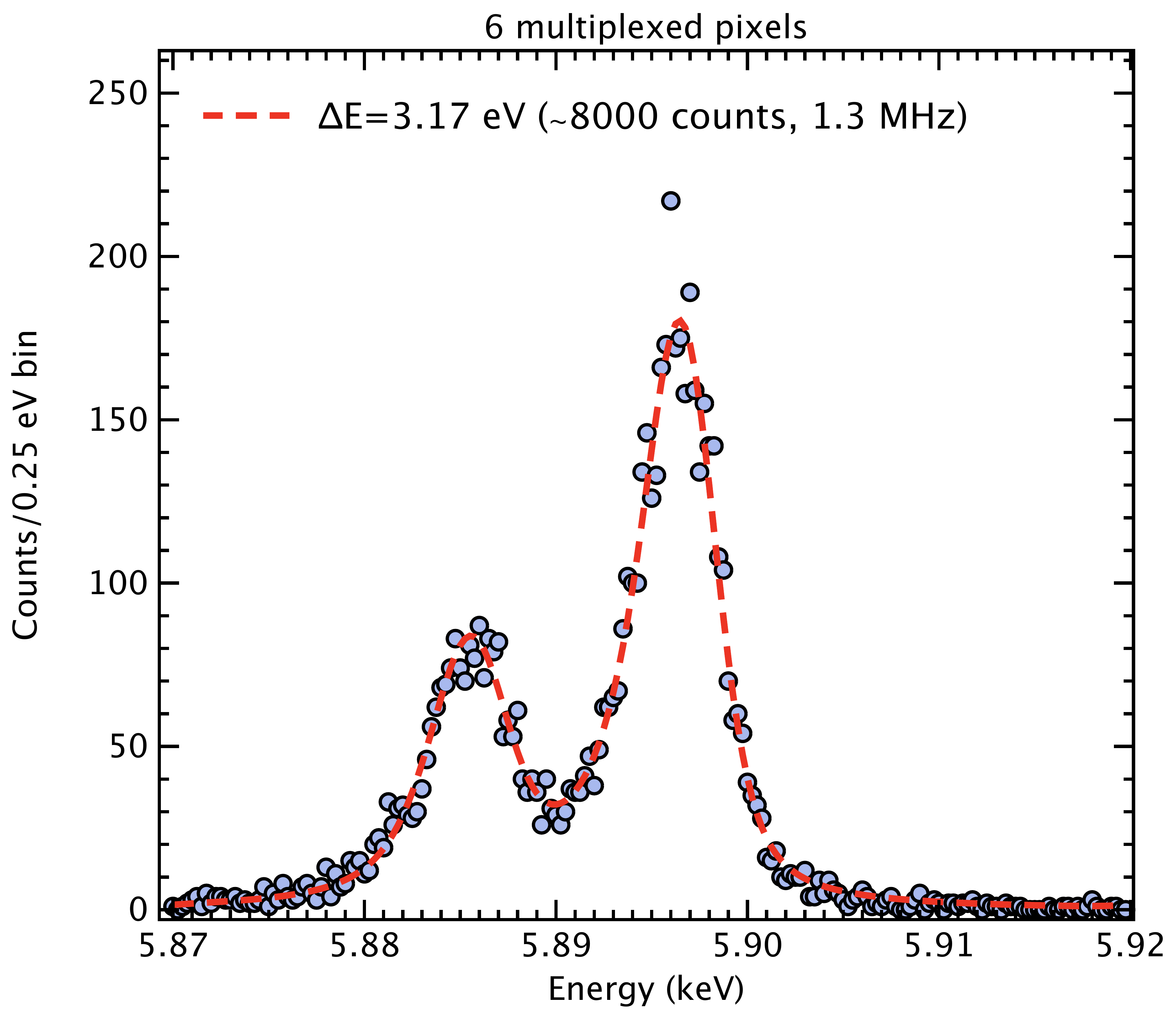}\includegraphics[scale=0.225]{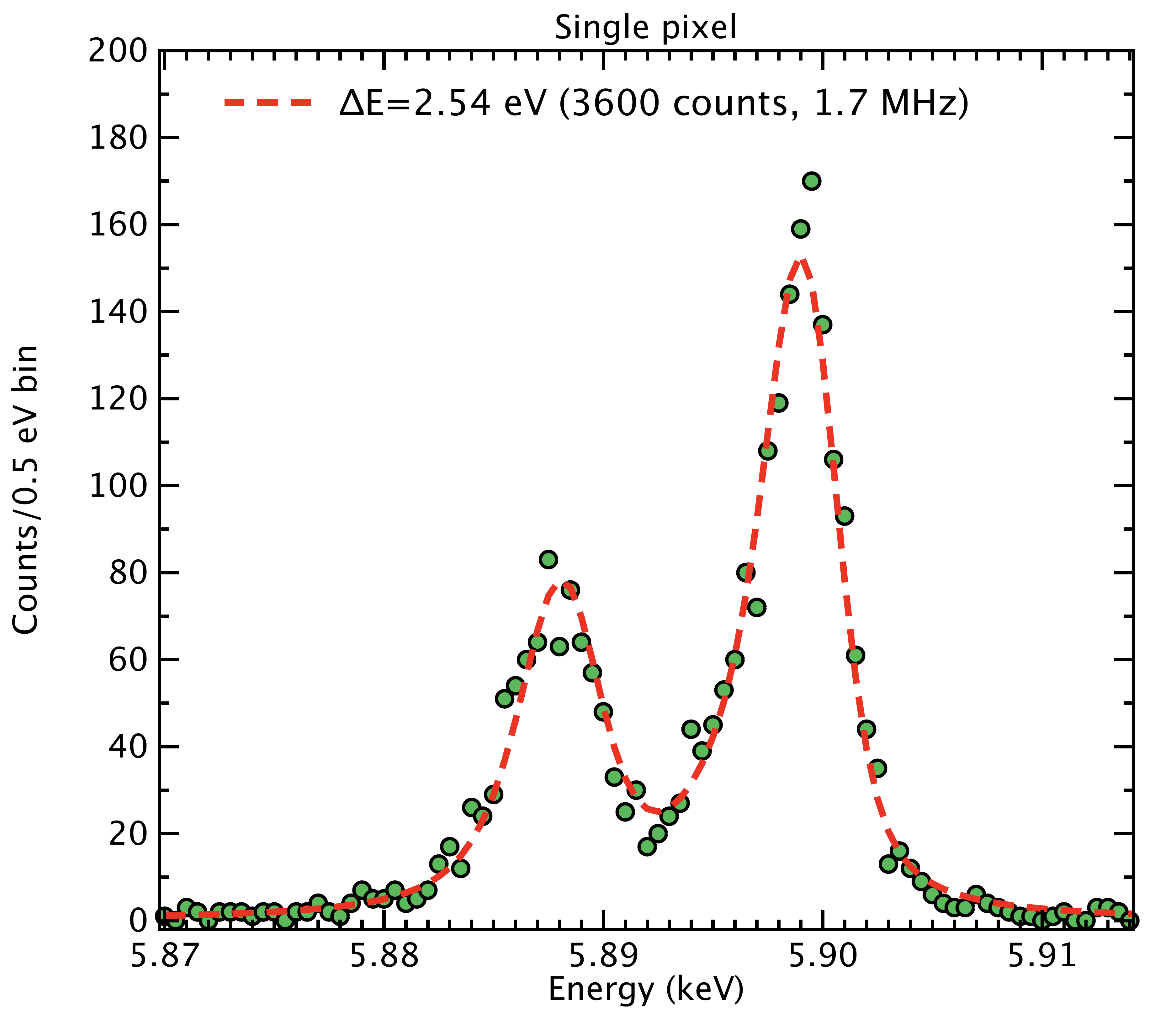}
\end{tabular}
\caption{\label{fig:fdm_spectral_resolution}  Left) Measured X-ray resolution ($^{55}$Fe: 5.9 keV) at 1.3 MHz under 6 pixel multiplexing mode.  The blue filled circles and red dashed line represent the measured count spectrum and the best fit model.  The data were taken with a  microcalorimeter array with 140$\mu$m GSFC TES (GSFC-1: see Ref. \citenum{akamatsu_spie_2016} for details). TESs are connected to 1.10, 1.27, 1.38, 1.55, 1.75, 2.05, 2.45, 2.55 MHz resonators.  Resonators at 1.10 and 1.38 MHz are excluded from multiplexing experiment due to undesired detector properties. Right) Measured X-ray resolution ($^{55}$Fe: 5.9 keV) under AC bias for a single pixel.  The green filled circles and red dashed line represent the measured spectrum and the best fit model.  The data was taken with 120 $\mu$m GSFC TES microcalorimeter array (GSFC-A2: see Ref. \citenum{akamatsu_spie_2016} for details). The TES was connected in series with the LC filter with resonance frequency of 1.7 MHz.}
\end{center}

\end{figure}

Progresses in FDM readout at SRON are presented in Refs. \citenum{akamatsu_spie_2016,vanderkuur_spie_2016}, demonstrating spectral resolution approaching the required 2.5 eV (see Fig. \ref{fig:fdm_spectral_resolution}, left panel). At single pixel level, the required spectral resolution is already achieved (see Fig. \ref{fig:fdm_spectral_resolution}, right panel). 

\begin{figure}[!t]
\begin{center}
\begin{tabular}{c}
\includegraphics[scale=0.3]{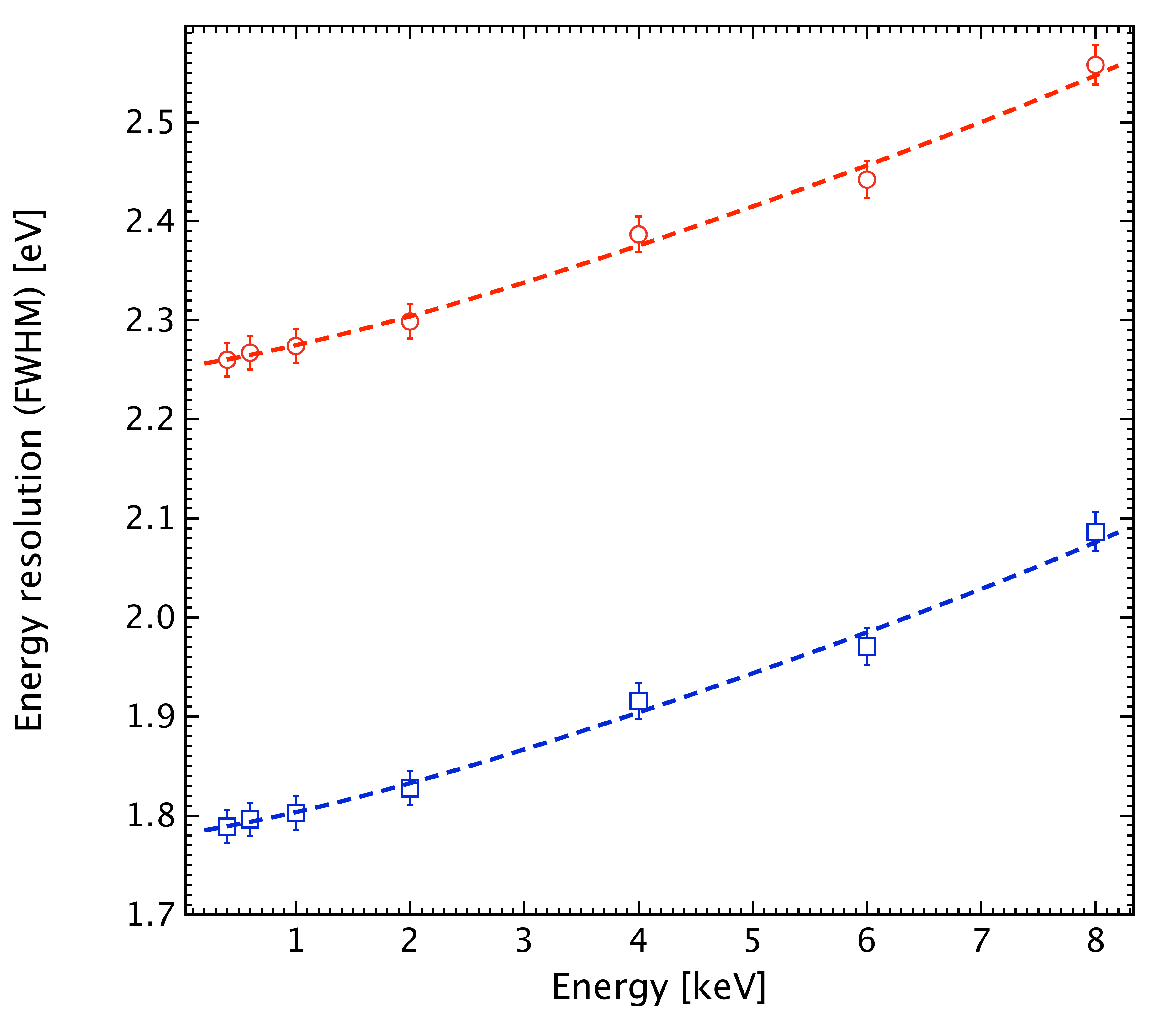}
\end{tabular}
\caption{\label{fig:spectral_resolution_versus_energy}In blue (squares), the energy resolution as a function of energy for baseline pixels as derived from {\it TESSIM} simulations (which include the contribution of the intrinsic detector resolution and the SQUID noise). The error bars show the remaining $1 \sigma$ statistical uncertainty from the simulation. The pulses were reconstructed with the optimal  filtering technique using a pulse template obtained at 1 keV and 16384 samples. All resolution values have been gain-scale corrected. Perfect knowledge was assumed by calibrating the method with a very large number of pulses. In red (circles), the data point have been scaled up to account for the additional noise contributions constrained such that the energy resolution at 7 keV is 2.5 eV. Polynomial fits between the measured data points are shown with dashed curves.}
\end{center}

\end{figure}

 The spectral resolution varies with energy, and depends on 1) the pulse reconstruction technique, 2) the duration of the record length, 3) the time difference between the previous and the next pulse (see Ref. \citenum{peille_spie_2016}). 
The energy resolution variation with energy was evaluated through simulations using {\it TESSIM} (Ref. \citenum{wilms_spie_2016}), with the pulses reconstructed with the standard optimal filter technique (see Ref. \citenum{peille_spie_2016}). The energy resolution as a function of energy is then simply estimated from the scatter of the reconstructed energies. A large record length of $\sim 104$ ms (16384 samples; the sampling frequency is 156 kHz) was used, as to produce a negligible contribution to the energy resolution degradation. As a consequence of the non-linearity of the TES, the spectral resolution degrades with energy, starting from an initial value of $\sim 1.8$ eV and reaching up 2.0 eV at 7 keV (see Fig. \ref{fig:spectral_resolution_versus_energy}).  Additional contributions to the spectral resolution budget to meet the 2.5 eV resolution required at 7 keV should therefore be limited (see red curve of Fig. \ref{fig:spectral_resolution_versus_energy}). 

\subsection{Bulk and turbulent velocity measurements}
\begin{figure}[!t]
\begin{center}
\begin{tabular}{c}
\includegraphics[scale=0.24]{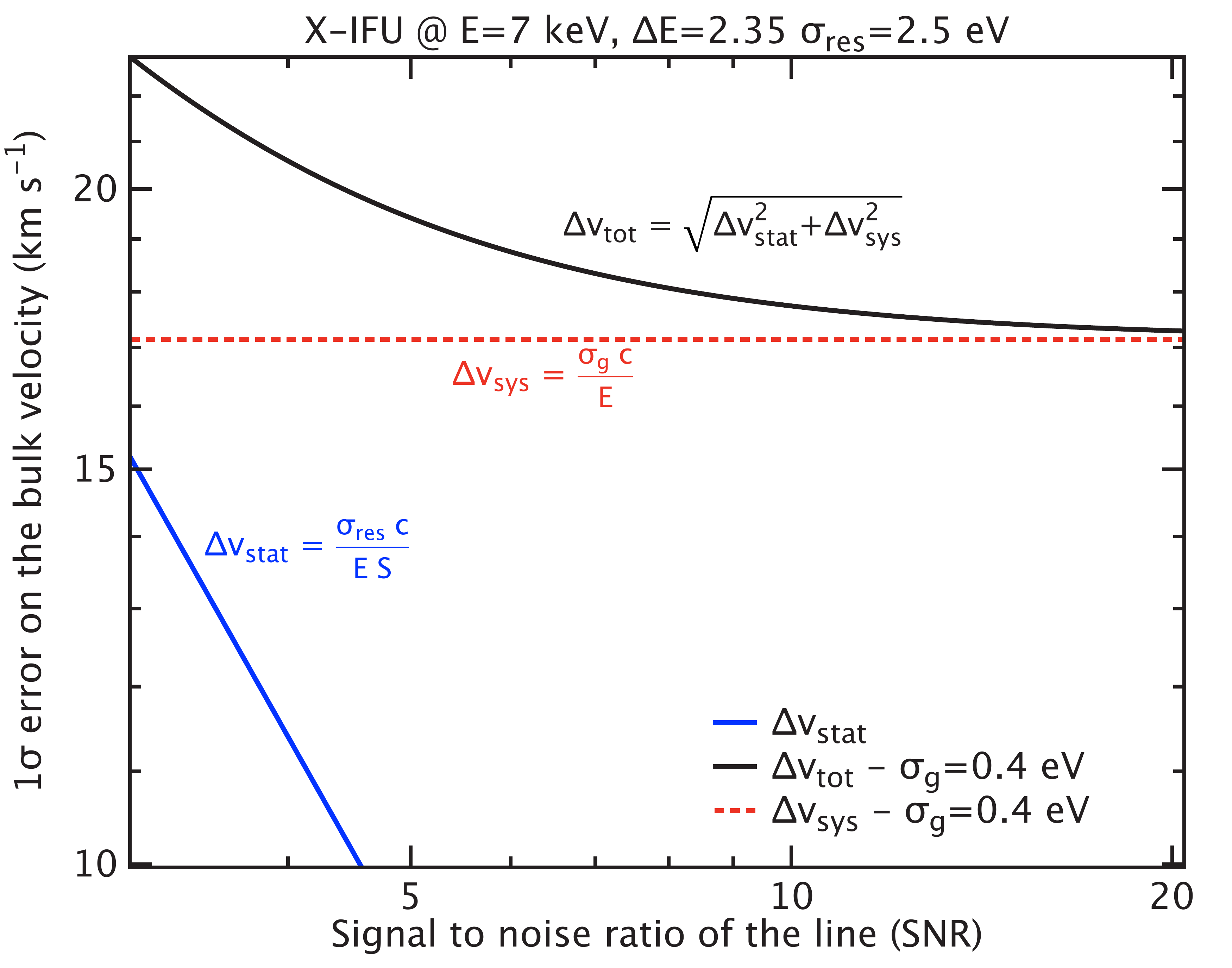}\includegraphics[scale=0.24]{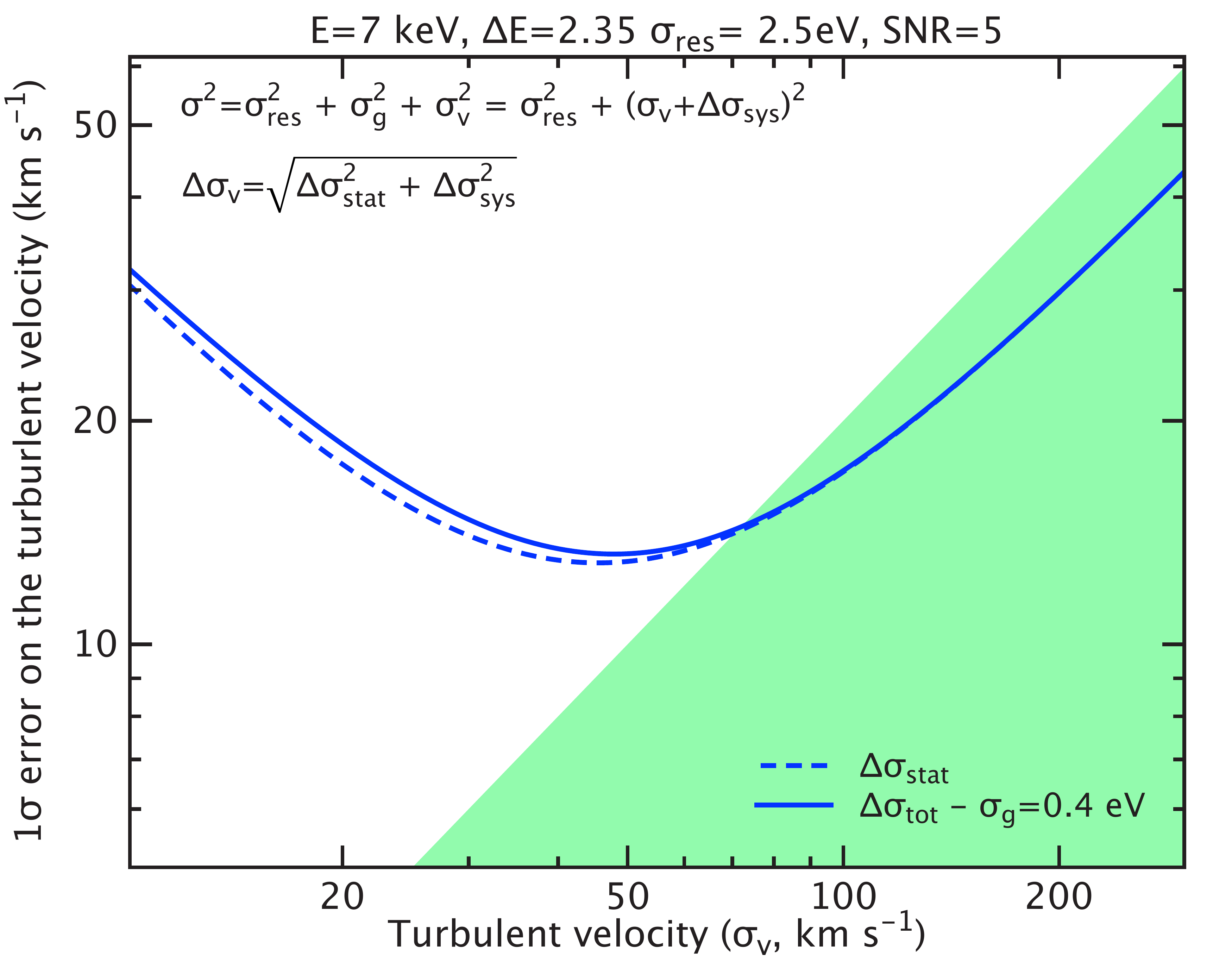}
\end{tabular}
\caption{\label{fig:bulk_turbulent_velocity}Bulk (left) and turbulent (right) velocity measurements expected from the X-IFU, assuming $\Delta E=2.5$ eV spectral resolution and a $\sigma_g=0.4$ eV gain calibration error (at 7 keV). Systematic and statistical components are shown in the error budgets. The green region defines the area where the $1\sigma$ error on the turbulent velocity is smaller than one fifth of the turbulent velocity (see Technical note by J. Kaastra, {\it Spectral diagnostics for IXO}, \url{http://space.mit.edu/home/nss/Jelle_Kaastra_ixo_spextroscopy.pdf}, November 18, 2008).}
\end{center}

\end{figure}

With the $\Delta E=2.5$ eV spectral resolution and a $\sigma_g=0.4$ eV gain calibration error, one can measure bulk and turbulent velocities from the line shift and line broadening. The expectations are shown in Figure \ref{fig:bulk_turbulent_velocity} at 7 keV. Note that the gain calibration error corresponds to a systematic error on the bulk velocities of about $17$ km/s. Similarly, for a line of signal to noise ratio of 5, turbulent velocities of $\sim 75$ km/s with an error of  15 km/s could be measured. 

\subsection{Gain calibration}
The impact of the gain drift due to fluctuations of the bath temperature and the AC bias point was investigated with {\it TESSIM} (Ref. \citenum{wilms_spie_2016}). Following the methodology described in Ref. \citenum{porter_2016}, several gain calibration curves will be determined from on-ground calibration for the in-flight calibration, and in order to keep the gain error below 0.4 eV (at 7 keV), the bath temperature should deviate at most by 1.5 mK from the reference temperature (55 mK). Similarly the effective bias voltage at TES level should not differ by more than 0.04 nV from the reference bias (assumed to be 68.06 nV). Other calibration issues (e.g. effective area and spectral resolution knowledges) and the way to address them with the X-IFU on-ground and in-flight measurements (e.g. with the MXS and astrophysical sources) will be the subject of a forthcoming paper (Ref. \citenum{cucchetti_2016}).

\subsection{Event grading and count rate capability}
Simulations of 7 keV pulse triplets were performed and the energy resolution (measured from the standard deviation of the reconstructed energies) and energy bias (measured as the mean of the energy offset between 7 keV and the reconstructed energies) of the middle pulse were determined using the standard optimal filtering technique for the pulse reconstruction (Ref. \citenum{peille_spie_2016}). When the preceding event is too close to the middle event, the reconstructed energy shows an offset. When the succeeding event is too close to the middle event, it limits the length for reconstructing the energy of the middle event, leading to a degradation of the energy resolution. One can then define high-resolution events as those which are properly reconstructed (not perturbed by the preceding and succeeding event) and thus corresponding to the 2.5 eV spectral resolution, mid-resolution events as those suffering from a modest decrease of spectral resolution (3 eV, equivalent to 2.5 eV before the additional noise contributions are added) and finally low-resolution events, as all the remaining valid events, for which the energy can be reconstructed. Invalid events are a significant fraction of events at high count rates and represent a measure of the TES array deadtime (see Ref. \citenum{macculi_spie_2016} for a discussion of the CryoAC induced deadtime). 

The baseline grading scheme is presented in Table \ref{tab:grades_baseline} for the baseline TES configuration (Ref. \citenum{smith_spie_2016}).

\begin{table}[ht]

\begin{center}       
\begin{tabular}{|l|l|l|l|} 
\hline 
Grade & Time until next pulse & Time since previous pulse & Resolution \\
\hline
High resolution & $> 1024$ samples (6.6 ms) & $> 400$ samples (2.6 ms) & 2.5 eV \\
Medium resolution & $> 256$ samples (1.6 ms) & $> 400$ samples (2.6 ms) & 3 eV  \\
Low resolution & $< 256$ samples (1.6 ms) & $> 400$ samples (2.6 ms) & $\sim 15$ eV \\
Invalid & Any values & $< 400$ samples (2.6 ms) & N/A \\
\hline
\end{tabular}
\end{center}
\caption{Definition of the event grades for the baseline configuration pixels (Ref. \citenum{peille_phd_2016}). The sampling frequency is 156 kHz. } 
\label{tab:grades_baseline}
\end{table} 

\begin{figure}[!t]
\begin{center}
\begin{tabular}{c}
\includegraphics[scale=0.3]{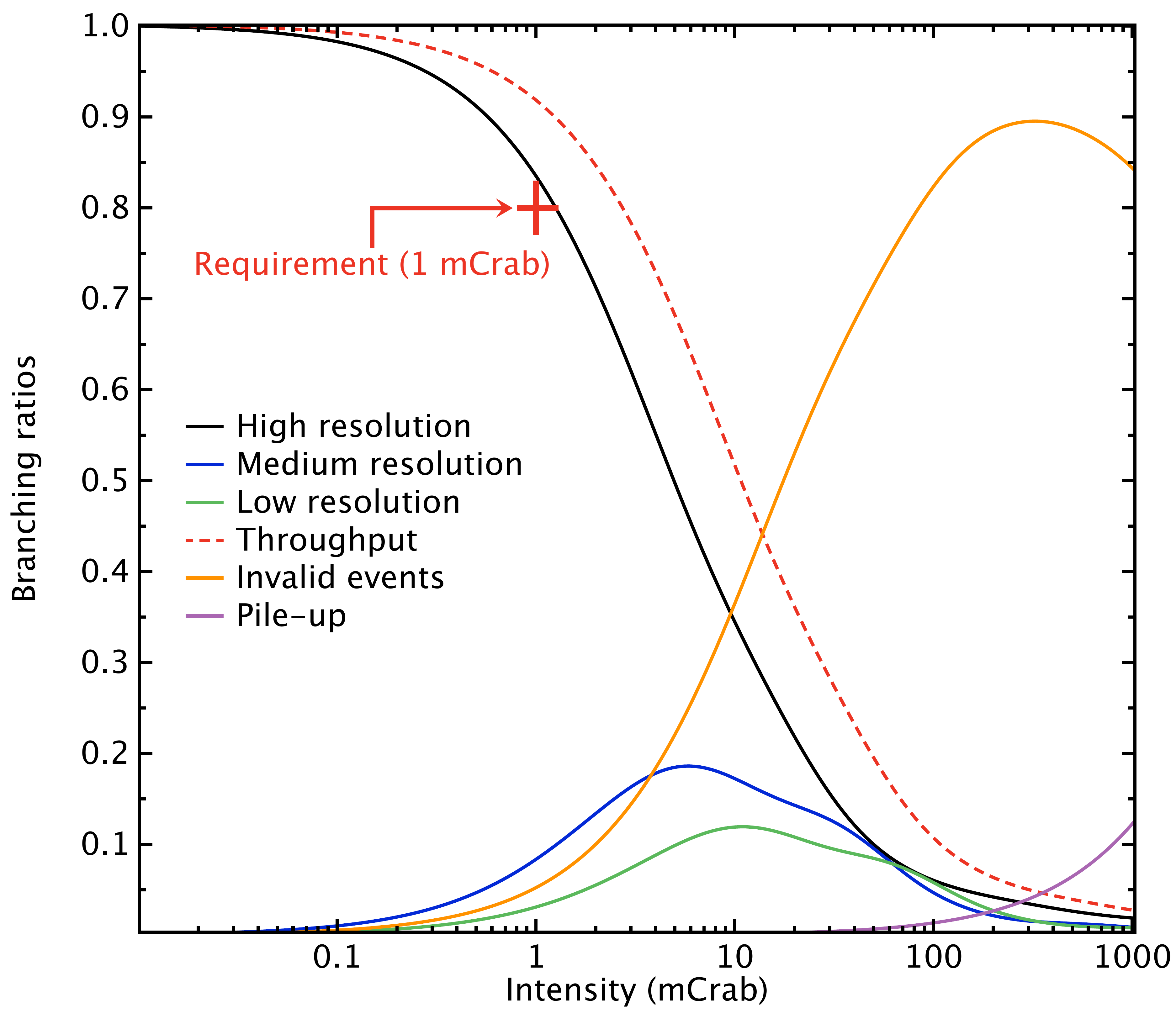}
\end{tabular}
\caption{\label{fig:lpa1_branching_ratios}The branching ratio of the different event grades for the X-IFU for the baseline configuration of the TES array. The throughput is given as the sum of all valid events. The invalid events are rejected. The pile-up fraction is also shown and starts increasing above 100 mCrab (the sampling frequency is 156 kHz). The pixel design was defined to match the count rate requirement of 80\% of high-resolution events at 1 mCrab (indicated by a red cross). The Crab spectrum is assumed to be a power law of index 2.1, a normalization of 9.5 photons/cm$^2$/s/keV and a column density of $4 \times 10^{21}$ cm$^{-2}$. Such a spectrum produces about 94000 counts/s on the TES array.}
\end{center}

\end{figure}

From this grading scheme, using Poisson statistics for the photon arrival times, it is possible to compute the branching ratios for the different grades at pixel level, and then sum up the ratios over the PSF array considering the PSF distribution of counts (e.g. in the baseline configuration, if the source is centered on a pixel, it receives $44$\% of the incident photon flux). One can also compute the pile-up fraction, as due to events falling in the same time bin. The branching ratios are plotted against the intensity of the source on the TES array expressed in units of mCrab (with the Crab spectrum being a power law of index 2.1, a normalization of 9.5 photons/cm$^2$/s/keV and a column density of $4 \times 10^{21}$ cm$^{-2}$ and producing about 94000 counts/s on the TES array) on Fig. \ref{fig:lpa1_branching_ratios}. As can be seen, while the 1 mCrab count rate requirement is met, the 1 Crab requirement for 30\% throughput is not simultaneously met. Consolidation of the latter requirement, which is very challenging, is on-going.

%
%In the current baseline configuration, consisting of an array of single size pixels, the count rate capability for a 5'' Point Spread Function  is shown in Figure \ref{event_grades}, and expressed in mCrab units, where the Crab is expected to produce 95000 counts/s over the full detector array. Accommodating faster pixels would drive further the read-out system (e.g. higher information bandwith) and keeping the same spectral resolution would be challenging due to the shortening of the available record length required to process the pulses. 
%
\subsection{Background}
The background requirement is one of the most challenging requirements to meet for the X-IFU, operating on a L2 orbit (a L1 orbit is also being assessed). Understanding the background necessitates a detailed design of the CryoAC system and the focal plane assembly, in addition to extensive GEANT-4 based Monte Carlo simulations that account for the materials surrounding the TES array and the physical processes at work in the low-energy domain of the X-IFU. It also puts requirements on the magnetic diverter screening out the protons focused by the Athena mirror and requires the focalization efficiency of the optics to be known. An update on the background expectations is provided in Ref. \citenum{lotti_spie_2016}, while the overview of the CyroAC development is presented in Ref. \citenum{macculi_spie_2016} (see also Ref. \citenum{biasotti_spie_2016,dandrea_spie_2016} for a discussion on the analysis of the CryoAC TES signal). As discussed in Ref. \citenum{lotti_spie_2016} the level of uncertainty of the current estimates is about a factor of 2 to 3, thus calling for optimization and consolidation of  the design of all the instrument components contributing to the background (e.g. FPA). It is anticipated that the fraction of the time the X-IFU observations will not be perturbed by soft protons is about 80\% and this is when the background requirement shall be met ($\le 5 \times 10^{-3}$ counts s$^{-1}$ cm$^{-2}$ keV$^{-1}$, in the 2--10 keV range). The non focused charged particle background should be known at the 2\% level, for the observations of low surface brightness sources (such as galaxy clusters). This may require $\sim 15$\% of blind ("out of the beam") pixels to the X-ray sky or the use of the closed position of the filter wheel to block the focal beam for $\sim 15$ \% of the time. How the latter requirement can be best met clearly requires further simulations, a better understanding of the L2 environment, and considering the use of external informations provided by the WFI instrument, which will be operated when the X-IFU is in focus. The requirement for a sensitive broad band particle monitor close to the instrument should also be defined (e.g. energy range, Ion species,\ldots).  
\subsection{Observation of bright optical sources}
Optical loading from bright UV and visible counterparts of X-IFU targets (massive stars, planets) would degrade the spectral resolution of the X-IFU by additional photon shot noise. Provision of two optical blocking filters (thin and thick OBF) on the filter wheel is therefore considered. The thin OBF would be made of 200 nm Polyimide + 40 nm Al and the thick one  of 200 nm Polyimide + 80 nm Al. It would allow to observe hot stars fainter than mV $\sim 7.5$ with no significant degradation of energy resolution, while the use of the thick OBF would allow to observe hot stars as bright as mV = 2. Note that the use of those two filters intercepting the focal beam will necessarily lead to a reduction of the effective area of the X-IFU at low energies.
\section{X-IFU performance optimization}
\label{performance_optimization}
Within the resource envelope allocated to the X-IFU, and without making the overall system more complex and demanding, several areas where the performance of the instrument could be improved are being investigated, both from the scientific and technical point of views. 
\begin{figure}[!t]
\begin{center}
\begin{tabular}{c}
\includegraphics[scale=0.2275]{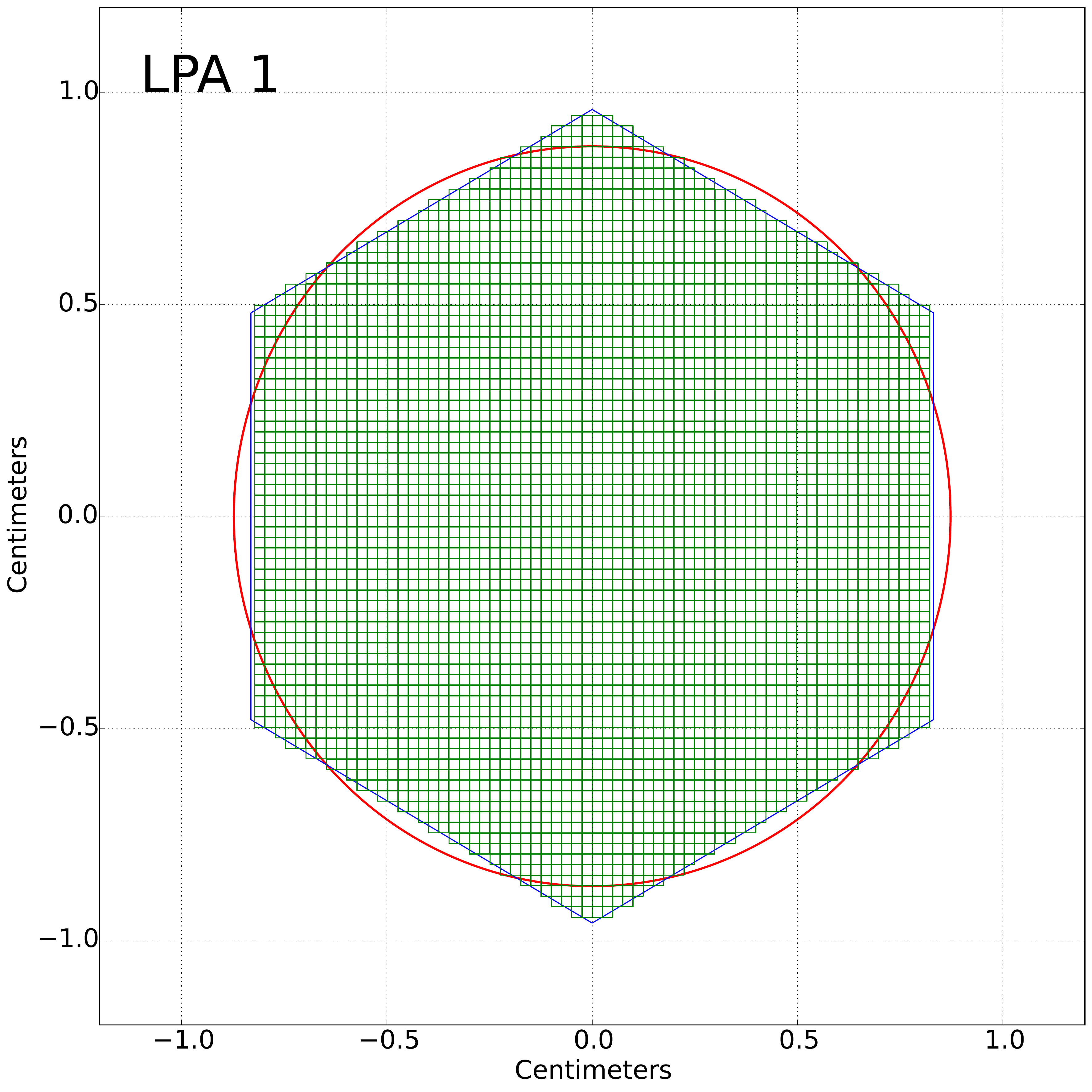}
\includegraphics[scale=0.2275]{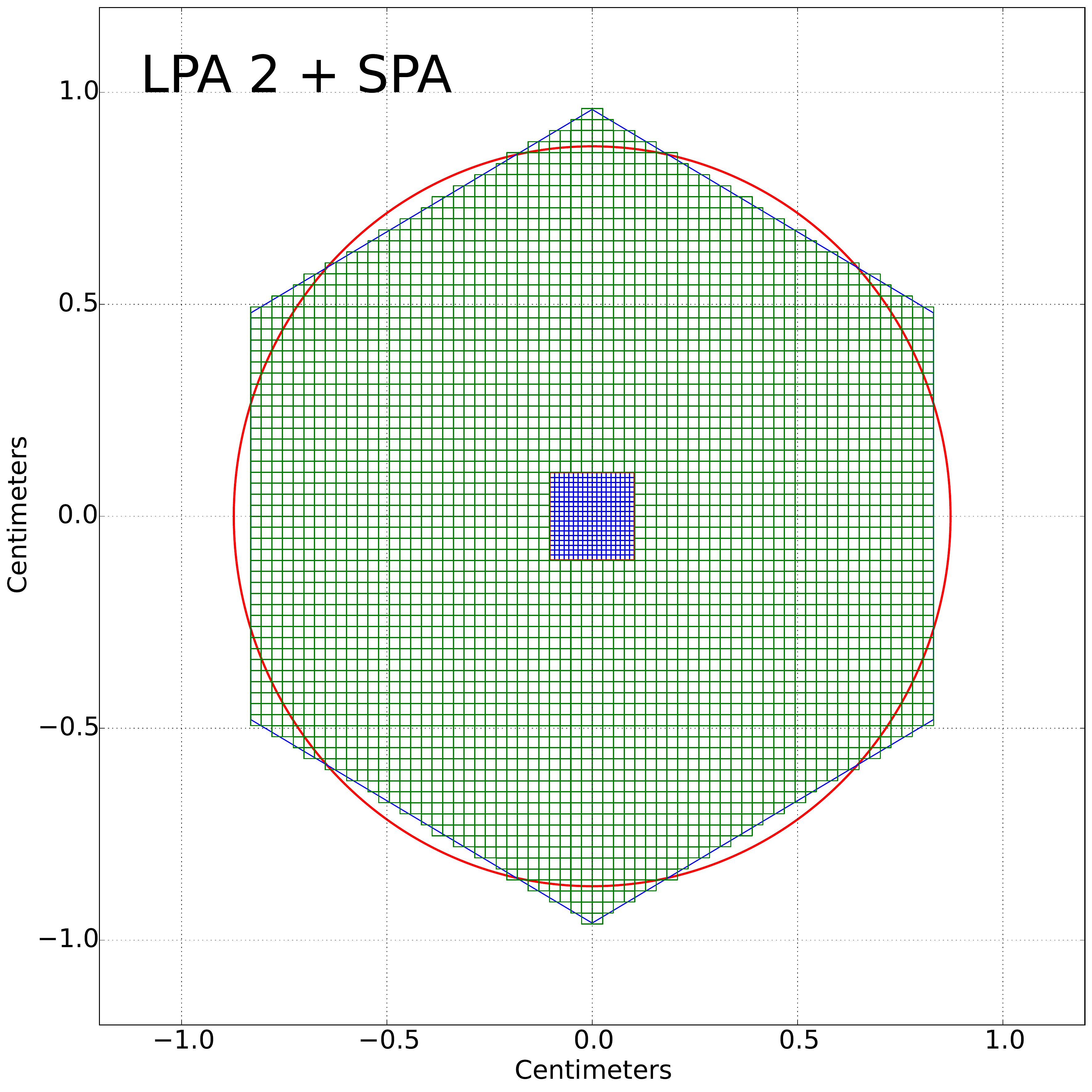}
\end{tabular}
\caption{\label{fig:array_configurations}Left) The baseline configuration of the X-IFU consisting of 3840 pixels of 249 \micron\ pitch (Large Pixel Array, LPA 1). Right) A hybrid TES array configuration considered for the X-IFU, consisting of a Large Pixel Array (LPA 2) and a Small Pixel Array (SPA). The SPA consists of 324 pixels of 110 \micron\ pitch ($\sim 1.9 ''$) to cover a field of view of $34 \times 34$ arc seconds. The pitch of the LPA 2 pixels has been increased to 260 \micron\ to keep the same overall field of view. The red circle indicates the 5' equivalent circular field of view. Further optimization of the LPA 2 + SPA configuration is on-going (e.g. size of the SPA, gap between the SPA and LPA 2...). The LPA 2 pixels are slower than the LPA 1 pixels whose parameters have been set to match both the spectral resolution, the count rate and field of view requirements. }
\end{center}
\end{figure}
\subsection{Count rate capability}
Improving the count rate capability of the X-IFU can be achieved by the addition of a small pixel array and by defocussing the mirror, which in both cases leads to spreading the PSF over a larger number of pixels. Improving the count rate capability of the X-IFU would be beneficial in several corners of the X-IFU science, including enlarging the number GRB afterglows usable for WHIM detection and enlarging the fraction of observable X-ray binaries with a larger throughput of high resolution events. 
\subsubsection{Introducing a Small Pixel Array (SPA)}
An alternative concept to the baseline configuration, in which two pixel geometries are combined to provide high count-rate capability for a small central field of view (Small Pixel Array) and reduced count-rate capability over the bulk of the array (Large Pixel Array) is thus being studied (see Figure \ref{fig:array_configurations}). The TES pixel parameter design of the microcalorimeter array and the implementation and fabrication considerations of an optimal hybrid array are presented in Ref. \citenum{smith_spie_2016}. The pitch of the LPA 2 pixel has been increased to 260 \micron\ to retain the 5 arc minute field of view in the presence of the SPA (while keeping a similar number of readout channels). They have also been made significantly slower (thermal conductance reduced by a factor of 2) as the count rate requirement applying to point sources only is now set to the SPA pixels (the SPA pixels are about a factor of 2 faster than the LPA 1 pixels, while preserving the same spectral resolution). The event grading schemes for the SPA and LPA 2 pixel configuration are presented in Table \ref{tab:grading_scheme_spa_lpa2}. The critically damped time constant for the LPA 2 pixels is $\sim 286~\mu$s and $\sim 79~\mu$s for the SPA pixels, compared to $\sim 160~\mu$s for the LPA 1 pixels (see Ref. \citenum{smith_spie_2016}). 

The increase in count rate capabilities moving from the baseline (LPA 1) configuration to the SPA is shown in Figure \ref{fig:count_rate_capability_different_configurations}. As it can be seen, the SPA provides about an order of magnitude improvement in count rate capability compared to the LPA 1 configuration. Beside relaxing the count rate requirement on the bulk of the array (and thus making it easier to build somehow, e.g. by reducing the number of readout chains (hence the instrument mass), and/or reducing the DAC dynamic range requirements and keeping the number of readout chains unchanged, the addition of an SPA may also provide the opportunity to improve the spectral resolution, by considering  pixel designs optimized for higher spectral resolution (lower heat capacity), while preserving adequate count rate capabilities (see \ref{optimization_spectral_resolution}). 
\begin{table}[ht]
\begin{center}       
\begin{tabular}{|l|l|l|l|l|} 
\hline 
Configuration & Grade & Time until next pulse & Time since previous pulse & Resolution \\
\hline
SPA & High resolution & $> 512$ samples (3.3 ms) & $> 190$ samples (1.2 ms) & 2.5 eV \\
 & Medium resolution & $> 128$ samples (0.8 ms) & $> 190$ samples (1.2 ms) & 3 eV  \\
 & Low resolution & $< 128$ samples (1.6 ms) & $> 190$ samples (1.2 ms) & $\sim 15$ eV \\
 & Invalid & Any values & $< 190$ samples (1.2 ms) & N/A \\
\hline
LPA 2 & High resolution & $> 16384$ samples (105 ms) & $> 700$ samples (4.5 ms) & 2.5 eV \\
 & Medium resolution & $> 512$ samples (3.3 ms) & $> 700$ samples (4.5 ms) & 3 eV  \\
& Low resolution & $< 512$ samples (3.3 ms) & $> 700$ samples (4.5 ms) & $\sim 30$ eV \\
& Invalid & Any values  & $< 700$ samples (4.5 ms) & N/A \\
\hline
\end{tabular}
\end{center}
\caption{Definition of the event grades for the SPA and LPA 2 pixel configurations (Ref. \citenum{peille_phd_2016}). The sampling frequency is 156 kHz.} 
\label{tab:grading_scheme_spa_lpa2}
\end{table} 

\subsubsection{Introducing defocussing}

\begin{figure}[!t]
\begin{center}
\begin{tabular}{c}
\includegraphics[scale=0.222]{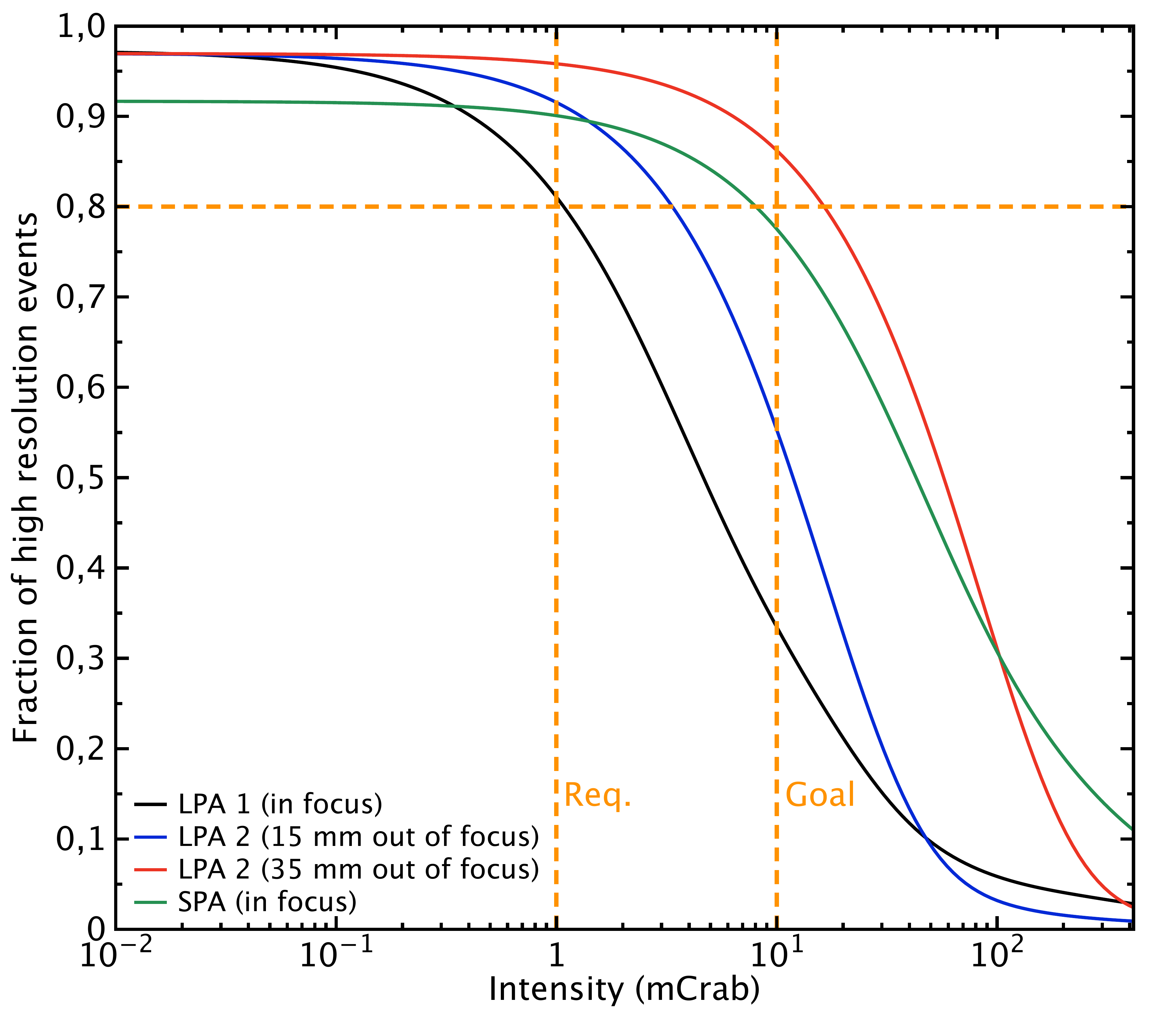}
\includegraphics[scale=0.222]{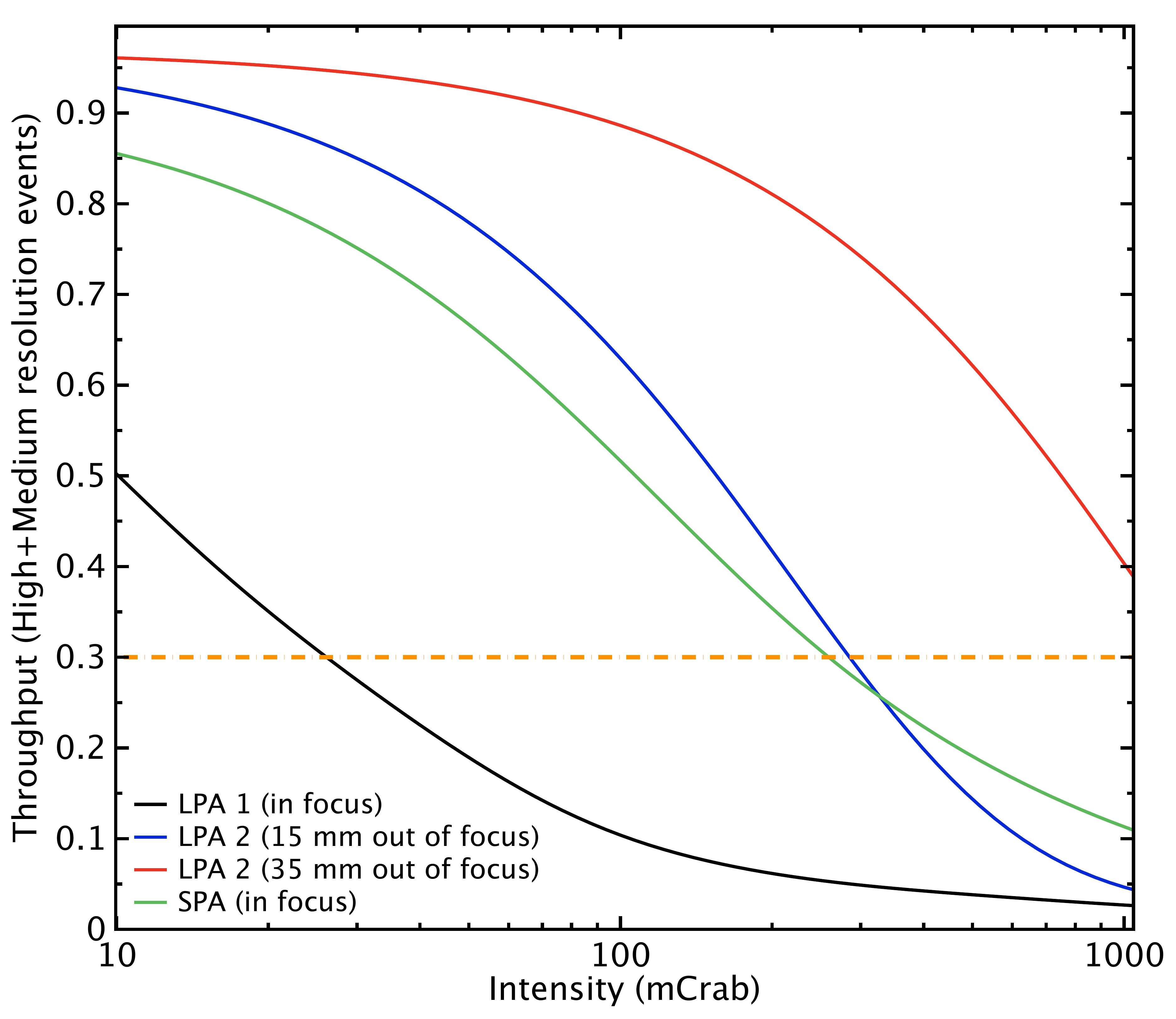}
\end{tabular}
\caption{\label{fig:count_rate_capability_different_configurations}Left) High resolution event fraction as a function of the point source intensity in mCrab units (1 mCrab is 95 cps/s) for different options compared to the baseline configuration (LPA 1 in focus). LPA 2 is made of slightly slower pixels but their count rate capabilities are recovered through defocussing. Two defocussing lengths are considered: 15 mm and 35 mm. For comparison, the performance of the SPA is also shown. The requirement of having 80\% of high-resolution events at 1 and 10 mCrab (goal) is also marked as a vertical orange line. The goal count rate capability can be achieved with safe margins through defocussing of the LPA 2 pixels by 35 mm. Right) Throughput of high-resolution and mid-resolution events in the four configurations for bright source observations. This figure shows that the throughput requirement of 30\% of events at 1 Crab is exceeded only with high resolution and mid resolution events, in the case of a strongly defocussed LPA 2. Adding the low-resolution events would make the SPA and LPA 2 defocussed at 15 mm compliant with the 30\% throughput requirement. Note that the addition of thick filter on the filter wheel (reducing the overall source flux by up to a factor of 5) could enable 1 Crab source to be observed with a throughput larger than 50\% with the use of an SPA and a defocussed LPA2. In these two figures, the ratios are normalized to the total number of photons hitting the absorbers and thus encode the different filling factors of the various configurations (this explains the differences in the low intensity part of the plots).}
\end{center}

\end{figure}

Another way of improving the count rate capability of the X-IFU is by using the defocussing option offered by the Movable Mirror Assembly (MMA), currently adopted as the baseline for the instrument switching mechanism. Improvement in the count rate capability of the X-IFU is shown in Figure \ref{fig:count_rate_capability_different_configurations}. As can be seen, a defocussing of 35 mm enables to improve the count rate capability by more than a factor of 10, even in the case of the slightly slower pixels of the LPA 2 configuration compared to the LPA 1 one. Thus defocussing appears a very promising route as it could enable the use of a single pixel design with relaxed and affordable count rate requirement.  Adopting defocussing as the baseline now would however make the X-IFU performance dependent on a parameter external to the instrument itself (i.e. the availability and performance of the defocussing option of the MMA). In addition, relatively large defocussing lengths are required to produce a meaningful improvement ($\ge 30$ mm): whether this will be achievable by the MMA remains to be demonstrated (maximum length is currently assumed to be 15 mm). On the other hand, the combination of a SPA and a short defocussing length ($<10$ mm) would still produce a very significant improvement in count rate capability. Shorter defocussing lengths may be less constraining for the MMA.

\subsection{Spectral resolution}
\label{optimization_spectral_resolution}
The spectral resolution requirement of 2.5 eV below 7 keV is achieved through a careful setting of the TES physical parameters (Ref. \citenum{smith_spie_2016}), which defines the intrinsic spectral resolution of the TES (1.8-1.9 eV) before any degradations due to finite record length, readout electronics, perturbations induced by the cooling system, and so on. While optimizing the count rate performance of the X-IFU, by introducing an SPA, it may be possible to take advantage of the small pixels to reduce their heat capacity (at equal stopping power a lower heat capacity can be achieved by smaller pixels), and thus improve their spectral resolution (especially if the count rate capability requirements are achieved through defocussing). Approaching a 1.5 eV  resolution would imply  the intrinsic spectral resolution of the TES to be of the order of 1 eV (provided that all the other contributors remain unchanged). Achieving this may however require mixing two different TES parameter settings on the same substrate (e.g. different absorber properties: all gold versus gold and bismuth) which may lead to fabrication issues, two different heat sinks, and possibly two dedicated readout and event processing chains. The benefit of increasing the spectral resolution, which would improve the weak line sensitivity (e.g. doubling the number of WHIM filaments detectable in a given exposure,  resolving the velocity and ionisation state of the various components of the AGN winds), lowering down the systematics, should thus be weighted against the likely added risk and complexity of the overall system. 

It is worth mentioning that having two different pixel sizes will also have an impact on the overall in-flight calibration, as small and large pixels will not see the same photon rate produced during the MXS pulses. Although the calibration requirements are difficult to anticipate before we have a consolidated design for the TES array, the characteristics of the MXS will have to be adjusted to minimize the on-board calibration time, while ensuring that all photons produced by the MXS are processed as high-resolution events (e.g. they must be well separated in time and they should not suffer from cross-talk issues).
 
\subsection{Low-energy response}
\label{s:low_energy_response}
The low energy effective area of the X-IFU is the convolution of the mirror effective area and the absorption of X-rays by any material placed between the mirror and the TES array. With the filter wheel in open position, absorption takes place predominantly in the thermal/optical blocking filters of the aperture cylinder and one possibility to improve the low-energy response (and match the 0.3 keV requirement, see Figure \ref{fig:effective_area}) would be to consider removing one filter or optimizing their thicknesses. Simulations indicate that removing one of the five filters would be acceptable in terms of heat load on the sensor array but not in terms of photon shot noise, exceeding the allocation in the spectral resolution budget. Before considering the removal of one filter, consolidation of the interface specifications (shield temperatures, distance between shields and TES array, required RF attenuation) is thus required. Alternatively a configuration consisting of 5 filters with a total thickness of 225 nm of Polyimide (instead of 280 nm in the baseline design)  and 150 nm of Aluminum (instead of 210 nm in the baseline design) with a few filters supported by metal meshes  may also be considered. The properties of the metal meshes  will be defined as to provide both mechanical strength and an attenuation of RF EMI from the uplink/downlink X-band telemetry and satellite operation signals. The drawback of using  metal meshes is that they reduce the instrument effective area across the full energy range, in proportion to the masked
area (a few \% per mesh). RF attenuation modeling and structural analysis are being carried out to optimize filter and mesh designs.

Note that as a way to reduce the background due to secondary electrons a thin filter just above the TES array may be needed (Ref. \citenum{lotti_spie_2016}). Those electrons hit the surface of the detector and bounce back releasing a small fraction of their energy. Half of these electrons have energies above 100 keV and are therefore difficult to block. Modeling of the backscattering process in GEANT-4 is rather uncertain, but in order to produce a significant effect on the background level (20\% level), filters such of 50 nm of Au would be required, reducing dramatically the low-energy effective area below 1 keV. Although the feasibility of implementing such a filter above the TES array has to be demonstrated, the benefit of the lower background should be weighted against the reduction of the effective area at low energies. 
 
\subsection{Pulse reconstruction}
Improving the spectral resolution and count rate capability of the X-IFU may be achieved through an optimization of the pulse reconstruction scheme. The baseline today is a method based on optimal filtering (Ref. \citenum{szymkowiak_1993}). Based on TESSIM (Ref. \citenum{wilms_spie_2016}), a thorough comparison of various pulse reconstruction techniques applicable to X-IFU is presented in Ref. \citenum{peille_spie_2016} (see also Ref. \citenum{vries_spie_2016} for a focus on a method based on principal component analysis). All techniques considered (optimal filtering, resistance space analysis, covariance based analysis) show relatively similar performance, with the covariance based analysis showing the larger improvement of the energy resolution compared to standard optimal filtering (improvement of $0.04$ eV at 7 keV). This method however requires significantly more demanding calibration resources than the others and this may prevent its use for the X-IFU. Resistance space analysis should then be considered as it indeed provides some performance improvement over optimal filtering (improvement of $0.02$ eV at 7 keV) at a very limited computational cost and no impact on the calibration needs. It is also the most robust technique at high count rates (see Ref. \citenum{peille_spie_2016} for more details). Before modifying the baseline for the pulse reconstruction technique, more work is needed to account for the environmental perturbations of the TES response (e.g. low frequency variations of the bath temperature or the voltage bias) and to study in greater details the sensitivity of the reconstruction techniques with the heat capacity of the TES and its non linearity (see Ref. \citenum{peille_spie_2016} for a comparison of the performance of the different reconstruction techniques for low heat capacity TES that may be envisaged for the small pixel array described above).  
\subsection{Pushing bright X-ray source observations}
In the current mission configuration, the telemetry allowance would enable to observe a 1 Crab source (the filter wheel in open position) for up to $\sim 3$ kilo-seconds and download the on-board processed and recorded data to the ground. Without requiring additional instrument complexity (e.g. on-board data compression, on-board memory for temporary data storage), we may consider pushing the bright source capability of the X-IFU. As discussed above, there are two ways of improving the count rate capability of the X-IFU, by the addition of an SPA and through defocussing of the optics. However, this may not be enough to observe sources at higher brightness, reaching Crab-like intensities. For this purpose, we are considering the addition of thick filters on the filter wheel as to suppress the more numerous low-energy photons, which for the bright source science are also scientifically less rewarding than the photons emitted above 5 keV (around the Iron line). Using a Be filter (100 $\mu$m) would reduce the overall Crab count rate by a factor of $\sim 5$, removing all photons below 1 keV (observations as long as 15 kilo-seconds could thus be feasible). Bright X-ray source science may not require the 2.5 eV spectral resolution but could make use of the mid-resolution events ($\sim 3$ eV resolution). The throughput expected by combining high resolution and medium resolution events is shown in Figure \ref{fig:count_rate_capability_different_configurations}. Reducing the 1-12 keV flux of a 1 Crab source by a factor of $\sim 5$ would correspond to a throughput of more than 50\% for a defocussed LPA 2 (without the SPA), thus providing 10000 cps/s with a spectral resolution of better than 3 eV.

Adding a gray filter, absorbing photons at all energies (it generally consists of tiny holes) may also be considered. The use of a gray filter would enable to reduce the overall flux by factors as high at 100-1000, and may be useful for keeping the observation of bright sources in the limit affordable by the on-board processing and telemetry allocation. The number of positions in the filter wheel will be subject to further optimization considering mechanical and mass constraints imposed to the instrument.

\begin{figure}[!t]
\begin{center}
\begin{tabular}{c}
\includegraphics[scale=0.2225]{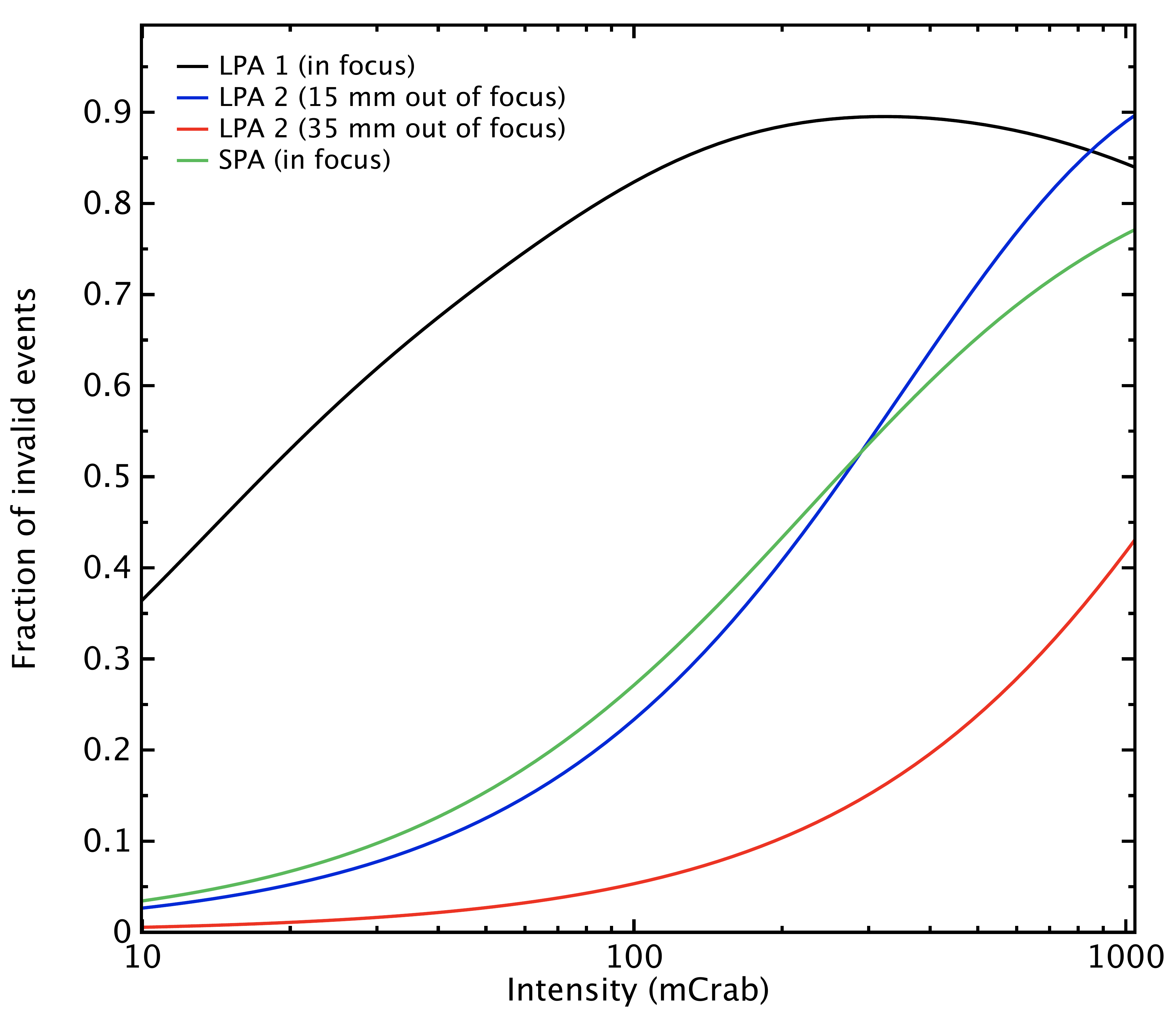}
\includegraphics[scale=0.2225]{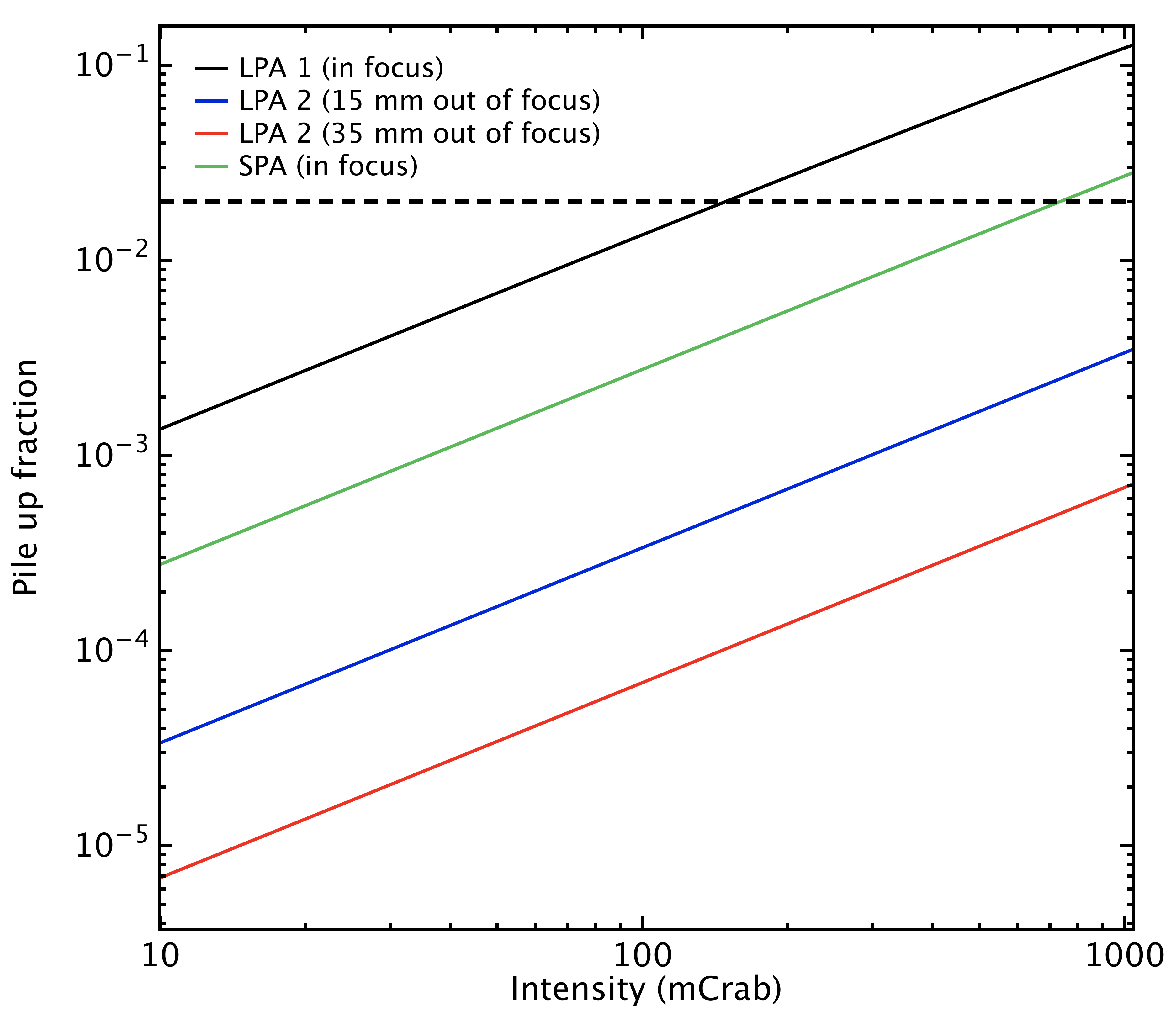}
\end{tabular}
\caption{\label{fig:deadtime_pileup}Left) The fraction of invalid events in the four configurations: LPA 1 in focus, LPA 2 pixels (15 mm out of focus), LPA 2 pixels (35 mm out of focus) and the SPA in focus. The fraction of invalid events traces the main TES array deadtime. As can be seen, large defocussing length reduces strongly the fraction of invalid events, even in the case of the slower LPA 2 pixels. Right) The pile-up fractions for the same four configurations. A value of 2\% of pile-up starts producing significant distortions in the recovered energy spectrum (indicated by a black dashed line). Pile-up in the defocussed LPA 2 configuration is negligible, but becomes significant for the baseline LPA 1 configuration around $\sim 100$ mCrab.}
\end{center}
\end{figure}

When observing bright sources, two instrumental processes have to be properly evaluated: pile-up and deadtime. Pile-up induces an artificial hardening of the input spectrum (as two photons arriving in the same time bin are reconstructed as one event with an energy that is the sum of the two, if no saturation occurs). Simulations indicate that deviations start to appear in the reconstruction of the input spectrum when the pile-up fraction exceeds a few percents. Rejected secondaries or invalid events trace the instrument deadtime. Deadtime level should be known accurately to properly recover the flux of the observed source. Deadtime complicates also time series analysis by causing significant departures of power spectra from those expected from simple Poisson statistics. In the case of the X-IFU, the deadtime is paralyzable in the sense that if an event arrives in the deadtime interval (during which no event can be reconstructed), the deadtime interval is extended (ultimately the sensor would not detect any events). Paralyzable deadtime in the X-IFU can be accurately modeled following the formalism in Ref. \citenum{zhang_apjl_1995}. The fraction of invalid events raises sharply with count rate, and only in the case of the LPA1 configuration, pile-up becomes significant above $\sim 200$ mCrab, and is negligible in all other configurations up to $\sim 1$ Crab intensities. This is shown in Figure \ref{fig:deadtime_pileup}, which demonstrates again the potential of the defocussing option for the LPA2, in particular by its almost negligible pile-up fractions (together with its lower deadtime).  

\subsection{Field of view}
When considering an hybrid TES array, while keeping the same multiplexing factor and the same number of readout channels, increasing the pixel pitch to 300 \micron\ (equivalent to 5.1 arc seconds) would increase the field of view of the X-IFU to $\sim 5.8$ arc minutes (equivalent diameter). 300 \micron\ was deemed to be the maximum pixel size able to retain the 2.5 eV spectral resolution requirement. A 5.8 arc minute diameter field of view would be 34\% more efficient in imaging than the smaller field of view, which translates into a 34\% saving in observing time for any observations that needs mosaicing. Similarly, the larger field of view would be 80\% more effective for WHIM autocorrelation studies: the reason being that the autocorrelation function signal scales as the power of 2 of the solid angle and thus as the power of 4 of the field of view diameter. Another benefit of the larger field of view could be that it would enable a 5 arc minute "on-sky" field of view while allowing 15\% of blind pixels to monitor the background, if the use of blind pixels is the solution retained for this purpose. Confusion effects in the larger pixel would be comparable to the LPA 1 and LPA 2 pixels. From the technical point of view, increasing the field of view would necessarily increase by geometric scaling the size of the focal plane assembly and its mass (e.g. larger shielding), with a snow ball effect on the overall instrument mass budget (which is currently above allocation). In addition, in the current Mock Observing Plan, going to a 5.8 arc minute field of view would lead to a saving of $\sim 1$ Ms of observing time. The benefits of the larger field of view do not currently appear substantial enough to drive further the resource demands of the X-IFU, but consolidation of the above findings is needed before a firm decision can be taken. 
\subsection{ToO efficiency}
The X-IFU cool time is currently baselined as 32 hours. The regeneration time is 8 hours, meaning that the cooling chain achieves a duty cycle of 80\%. The Athena ToO requirement states that the X-IFU should be able to observe for at least 50 \ksec, GRBs starting within 4 hours for at least 80\% of instances  (for a field of regard of 50\% of the sky): the latter defines the so-called ToO efficiency (and includes all delays due to the science ground segment operations). Assuming a larger field of regard of 60\%, the ToO efficiency required drops from 80 to 67\%. For any reasonable assumptions about the Athena ground segment and the ToO implementation scheme (alert reception and validation, generating and uploading commands, instrument setting...), meeting this requirement depends strongly on the way the X-IFU is operated (it depends less critically on the cooling chain duty cycle if restricted to plausible values below 90\%). The Mock Observing Plan indicates the X-IFU will observe for about 60\% of the time (i.e. the remaining 40\% will be for WFI observations). Assuming that the X-IFU performs observations during its entire cool time (32 hours), before WFI turns on to observe for 21.3 hours (to preserve the 60-40 ratio) and that regeneration takes place during the last 8 hours of the WFI observation, the ToO efficiency is only about 36\%, hence very far from the 67\% required. This low efficiency results from the fact that a successful ToO must happen within the first 18 hours of the 53.3 hour nominal operation cycle (to have 50 \ksec\ of remaining X-IFU observation time for the ToO). To improve significantly the ToO efficiency, a very effective solution is to split the X-IFU observations, as to always reserve 14 hours (50 \ksec) of cool time for ToO observations. The X-IFU observation would then stop after 18 hours and would be followed by a  WFI observation lasting 12 hours, with the cooler regenerating in the last 8 hours of the WFI observation. This compressed operational mode would raise the ToO efficiency up to 66\%, thus meeting the requirement. The technical implications and feasibility of this operational mode for the X-IFU need however to be assessed. 

Segmenting the observations into several shorter ones would imply a larger number of slews (driving the overall spacecraft agility), more reaction wheel off-loadings, more instrument switching actuation cycles, thus impacting on the overall spacecraft operational availability (set at 90\%). Preliminary analysis indicates that a large slew rate of at least 3 degrees/minute would be needed to meet the operational availability requirement in the compressed operational mode described above. Such a large slew rate may not be achievable with reaction wheels alone, but would require the addition of thrusters. For a slew rate of 1 degree/minute (achievable with reaction wheels alone), in the 18-12 hour compressed X-IFU/WFI cycle, the loss of operational availability is already 16\%, thus breaking the compliance with the 90\% availability requirement (this would correspond to an increase of $\sim 70$\% of instrument switches compared to the non-compressed operational cycle mode). A reduced operational availability could obviously be compensated by a longer mission duration, but one should first consolidate the ToO efficiency requirement which is driven in particular by the detection of the WHIM using GRB afterglows. This is being addressed in Ref. \citenum{brand_spie_2016b} which presents the extensive WHIM detection simulations being carried out,  considering realistic instrument set-up (spectral resolution, throughput, and cross-talk, see Ref. \citenum{denhartog_spie_2016}), afterglow and WHIM properties. As discussed above, recent cosmological simulations predict that the evolution of the WHIM with redshift is flat up to $z\sim0.8$ (Ref. \citenum{smith_apj_2011}), so that each distant GRB afterglow (at $z>0.8$) may probe multiple systems along the line of sight, with several of those predicted to be easily detectable against a not extremely bright GRB afterglow (see Figure \ref{fig:WHIM}). This mere fact will be folded in when consolidating the ToO efficiency requirement, derived from the need to detect 50 WHIM filaments with GRB afterglows. 

Athena (and the X-IFU) must offer a good ToO capability, as this is a top-level requirement for the mission, both for WHIM filament detection against GRB afterglows, high-z GRB studies and discovery science (e.g. follow-up of gravitational wave signals). This translates not simply into a requirement on ToO efficiency, but also on requirements on other mission performances (e.g. effective area) onto which virtually all Athena science objectives impose requirements. Consolidation of the requirements on each individual performance parameter has to be achieved from a global point of view taking into account overall constraints, such as mass and cost.  
\section{Short term plans and conclusions}
\label{conclusions}
X-IFU has now been studied as an integrated instrument for about 1 year. The next priorities are 1) to consolidate the mass budget in order to make it fit within the mass constraint derived from the spacecraft studies while preserving its scientific performance and 2) to define an optimized  accommodation for the X-IFU onto the focal plane module. This is achieved through a top-down approach, starting with a critical review of the thermal, mechanical and electrical budgets, and should lead to a consolidated design along the remaining part of the phase A study. Options to improve the X-IFU performance are being studied (e.g. count rate capability, spectral resolution, low-energy response,\ldots) within the constraint of not increasing its resource demands and complexity. Some of these options have however, system level impacts and therefore will require more extensive and longer analysis (e.g. the hybrid array). 

In parallel to the phase A study, technology developments are also being pursued in order to increase the Technology Readiness Level of the X-IFU, as required for the adoption of the mission. This covers the TES array, the focal plane assembly, the cold front end electronics, warm readout electronics, thermal/optical blocking filters, aperture assembly, cryo-coolers,\ldots Some of those activities will converge in the development of the Detector Cooling System, a Core Technology Program (CTP) contract issued by ESA to develop and characterize a 50 mK cryogenic chain with which functional testing of a TES array will be performed. 

%Finally, following the Mission Consolidation Review, ESA now considers transferring the Focal Plane Module (FPM) responsibility to the instrument teams. A short term activity is thus to define a baseline architecture for the FPM, able to accommodate the WFI and X-IFU. This will involve the two Primes (Airbus Defense and Space \& Thales Alenia Space). The management and procurement scheme of the FPM is currently being discussed between the two instrument teams and ESA. Defining the exact perimeter of the instrument team contributions to Athena is mandatory for ESA to issue the AO for the payload and science ground segment contributions. 

The outstanding Hitomi SXS results have demonstrated the transformational character of high-resolution X-ray spectroscopy. The X-IFU consortium, which includes all key members of the SXS team, has gained additional motivation for delivering to the community the most ambitious X-IFU, while recognizing plainly the challenge to be faced for building it.
\section{Acknowledgments}
We acknowledge support from the Athena Science Study Team, the Athena Working  Group Chairs, the Athena Topical Panel Chairs and the Topical Panel members in strengthening the X-IFU top level performance requirements. Particular thanks go to: E. Rasia, V. Biffi, S. Borgani and K. Dolag for providing cosmological hydrodynamic simulations of a cluster used to produce simulation of X-IFU observation presented in Fig.~\ref{fig:yields}; P.T. O'Brien for assistance with Sect.~\ref{s:GRB}, A.C. Fabian and C. Pinto for providing inputs for Figure \ref{fig:perseus}.  We also thank the ESA project team, and in particular Mark Ayre and Ivo Ferreira, for their work on the assessment of the ToO efficiency requirement. The Italian contribution to X-IFU is supported through the ASI contract n. 2015-046-R.0. 
XB, MTC and BC  acknowledge financial support by MINECO through grant ESP2014-53672-C3-1-P. A.R., P.O, and A.J. were supported by Polish NSC grants: 2015/17/B/ST9/03422 and 2015/18/M/ST9/00541. GR, ER, YN, and PJ acknowledges support by FNRS and Prodex (Belspo). This work was supported by the French Space Agency (CNES). 
\bibliographystyle{spiebib} % makes bibtex use spiebib.bst
\bibliography{barret_etal_spie_2016-VS_ArXiv} % bibliography data in report.bib

\end{document}